\documentclass[aps,prb,twocolumn,preprintnumbers,amsmath,amssymb,superscriptaddress,longbibliography,nofootinbib]{revtex4-2}

\usepackage{graphicx}
\usepackage{dcolumn}
\usepackage{bm}
\usepackage{color}
\usepackage{ulem}
\usepackage[usenames,dvipsnames]{xcolor}
\usepackage{hyperref}

\hypersetup{
    bookmarks=false,        
    colorlinks=true,        
    linkcolor=blue,        
    citecolor=blue,        
    filecolor=blue,      
    urlcolor=blue
}

\newcommand{\GF}{\mathcal{G}}

\newcommand{\be}{\begin{equation}}
\newcommand{\ee}{\end{equation}}
\newcommand{\bea}{\begin{eqnarray}}
\newcommand{\eea}{\end{eqnarray}}
\newcommand{\eq}[1]{Eq.~(\ref{#1})}
\newcommand{\fig}[1]{Fig.~\ref{fig:#1}}

\newcommand{\ie}{\textit{i.e.}}
                             
\newcommand{\e}{\varepsilon}
\newcommand{\w}{\omega}
\newcommand{\s}{\sigma}

\newcommand{\up}{\uparrow}
\newcommand{\down}{\downarrow}

\newcommand{\Pol}{\mathcal{P}}
\newcommand{\T}{\mathcal{T}}

\newcommand{\Lor}{\mathcal{L}}
\newcommand{\tkap}{\widetilde{\kappa}}
\newcommand{\Geff}{\Gamma_{\rm eff}}

\newcommand{\es}{& = &}

\newcommand{\nn}{\nonumber}

\newcommand{\new}[1]{{\color{ForestGreen}#1}}

\usepackage{ulem}
\usepackage{verbatim} 

\begin{document}

\title{Spin-resolved thermal signatures of Majorana-Kondo interplay
	\\in double quantum dots}

\author{Piotr Majek}
\email{pmajek@amu.edu.pl}
\affiliation{Institute of Spintronics and Quantum Information,
	Faculty of Physics, Adam Mickiewicz University in Pozna{\'n}, 
	ul.~Uniwersytetu Poznańskiego 2, 61-614 Pozna{\'n}, Poland}

\author{Krzysztof P. W\'ojcik}
\affiliation{Institute of Molecular Physics, Polish Academy of Sciences, 
	ul.~Smoluchowskiego 17, 60-179 Pozna{\'n}, Poland}
\affiliation{Physikalisches Institut, Universit\"at Bonn, Nussallee 12, D-53115 
	Bonn, Germany}

\author{Ireneusz Weymann}
\affiliation{Institute of Spintronics and Quantum Information,
	Faculty of Physics, Adam Mickiewicz University in Pozna{\'n}, 
	ul.~Uniwersytetu Poznańskiego 2, 61-614 Pozna{\'n}, Poland}

\date{\today}

\begin{abstract}
We investigate theoretically the thermoelectric transport properties of a 
T-shaped double quantum dot	side-coupled to a topological 
superconducting nanowire hosting Majorana zero-energy modes.
The calculations are performed using the numerical renormalization group method
focusing on the transport regime, where the system exhibits the two-stage Kondo effect.
It is shown that the leakage of Majorana quasiparticles into the double dot system
results in a half-suppression of the second stage of the Kondo effect,
which is revealed through fractional values of the charge and heat conductances
and gives rise to new resonances in the Seebeck coefficient. 
The heat conductance is found to satisfy a modified Wiedemann-Franz law.
Finally, the interplay of Majorana-induced interference 
with strong electron correlations is discussed in the behavior
of the spin Seebeck effect, a unique feature of the considered setup.
\end{abstract}

\maketitle

\section{Introduction}

At the time of its publication, Ettore Majorana's rederiving of the Dirac equation 
in terms of real wave functions \cite{Majorana1937Apr},
the solution of which pointed at a particle that is its own anti-particle, has not attracted much attention.
The situation has changed after the Kitaev's proposal to realize Majorana fermions as quasiparticles in 
quantum wires, and to utilize their non-Abelian statistics in quantum computation protocols
\mbox{\cite{Kitaev2001,Kitaev2003,Fu2008Mar,Nayak2008Sep,BraidingReview}}. 
The enormous interest in this field stems from the topological protection of Majorana
quasiparticles, allowing for fault-tolerant computation 
\cite{Kitaev2003,Fu2008Mar, Hasan2010Nov, Qi2011Oct, Wang2017Nov, Sato2017Rev}.
It exploded after the first experimental observation of Majorana signatures 
\cite{Mourik2012May}, which followed earlier theoretical proposals \cite{Lutchyn2010Aug,Oreg2010Oct}.
Since then, fabrication of the so-called Majorana wires has been reported by numerous groups,
with still improving quality of the characteristic Majorana features in increasingly sophisticated 
nanostructures \cite{Sato2017Rev, Lutchyn2018May, Prada2020Rev}.
Up to recently, the Majorana signatures were in fact reduced to detection of zero-bias peaks corresponding 
to single-electron conductance through the superconducting wire, which is not possible in 
a trivial Cooper-pair dominated medium. However, this laid ground to a controversy, 
since such an anomaly can also be a fingerprint of zero-energy Andreev states
\cite{Lee2013Dec,Kells2012Sep,Wang2021Feb} or weak anti-localization effects \cite{Pikulin2012Dec}.
With recent interferometric experiments \cite{Borsoi2020Jul,Whiticar2020Jun},
unified theoretical picture \cite{Avila2019Oct,Prada2020Rev} and prediction of Majorana 
oscillations in the microwave spectra as further candidates for signature of the presence 
of Majorana bound states \cite{Avila2020Sep}, the controversy may seem close to the conclusion
supporting the existence of localized Majorana zero-energy modes (MZMs) at the ends of 
topological superconductor wires, but the discussion among the community continues
\cite{Saldana2021Jan}.
Meanwhile, the tremendous improvement of experimental techniques may lead to realization
of Majorana modes also in strongly correlated systems involving quantum dots (QDs) \cite{Deng2016Dec}.
Here, we examine unconventional (magnetic and thermoelectric)
signatures of MZM presence in an example of such a device.

\begin{figure}[t!]
	\includegraphics[width=0.95\columnwidth]{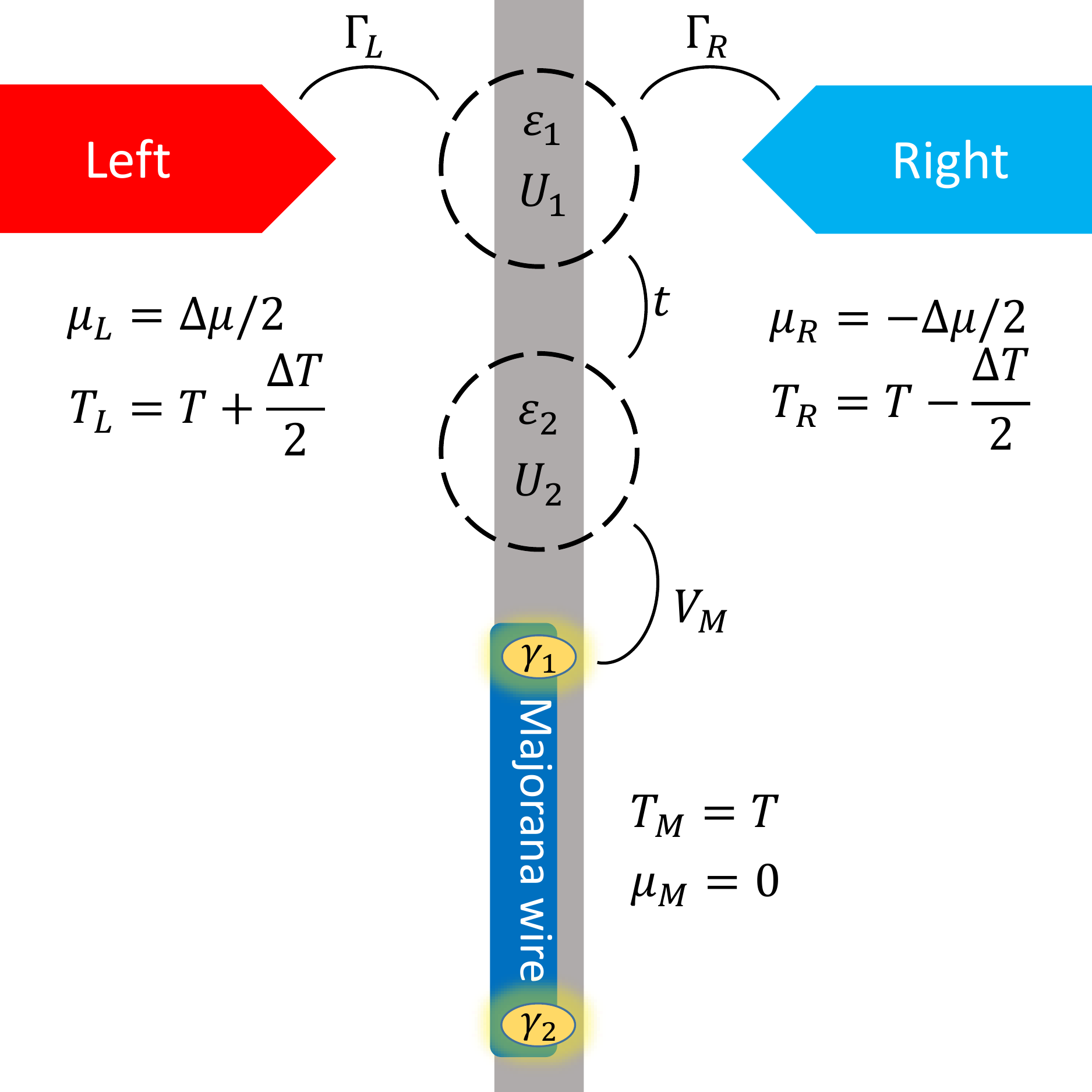}
	\caption{\label{fig:model}
		The schematic of the considered system.
		The first quantum dot (of on-site energy $\e_1$ and Coulomb correlations $U_1$)
		is connected to metallic leads with the strength $\Gamma_r$,
		with $r = L(R)$, for the left (right) contact.
		The second quantum dot (of on-site energy $\e_2$ and Coulomb correlations $U_2$)
		is coupled to the topological superconducting nanowire,
		hosting Majorana quasiparticles denoted by $\gamma_1$ and $\gamma_2$,
		with coupling matrix elements $V_M$.
		The hopping between the dots is described by $t$.
		The nanowire temperature is assigned as $T$ and
		there is a temperature $\Delta T$ and chemical potential $\Delta\mu$
		gradient applied between the leads.
	}
\end{figure}

Generally, strong electron correlations can lead to 
various forms of the Kondo effect \cite{Kondo1964Jul,hewson_1993},
different types of superconductivity and magnetism,
or non-Fermi-liquid phases \cite{hewson_1993,Lohneysen2007Aug,Si2016Mar,Paschen2021Jan},
to name just a few.
Only very recently have this field reached into the realms of topological materials,
such as topological Kondo insulators \cite{Dzero2016Mar}, 
Weyl-Kondo semi-metals \cite{Lai2018Jan},
and---most relevant for our study---Kondo-Majorana interplay 
\cite{Cheng2014Sep,Vernek2019}.
In this paper we thus seek for interesting features at the crossroads between
strong electron correlations and topologically protected Majorana modes,
exploiting a tunable playground of quantum dot systems \cite{Deng2016Dec}.
It is important to note that there exists a number of works
concerning the Majorana-Kondo physics.
A study of a nanowire in-between two leads revealed a fixed point distinct from the 
conventional Kondo one \cite{vanBeek2016Sep}.
In setups with multiple Majorana wires the topological Kondo effect may appear \cite{Beri2012Oct},
and in some cases the mapping onto multi-channel Kondo model is possible \cite{Herviou2016Dec}.
Very recently, a novel correlation-fueled mechanism has been proposed to obtain mobile 
Majorana modes in exotic two-channel Kondo insulators \cite{Kornjaca2021Apr}.
Moreover, in various T-shaped configurations, very similar to existing experimental setups for MZM detection 
\cite{Deng2016Dec,Prada2020Rev}, perturbative RG analysis indicates yet another distinct 
fixed point \cite{Cheng2014Sep}, with robust strong-coupling nature and independence of
the Kondo temperature $T_K$ on the coupling to Majorana wire $V_M$ \cite{Cheng2014Sep,Vernek2019}.
Numerical studies confirm the former \cite{Lee2013Jun},
and even show that $T_K$ increases with enhancing the coupling to the Majorana mode in the case of
single quantum dots \cite{Ruiz-Tijerina2015Mar,Weymann2017Apr}.
\new{
In addition to the modification of the relevant energy scales
\cite{Flensberg2010Nov,Lee2013Jun,Vernek2014Apr,Deng2016Dec,Weymann2020Jan},
the side-coupling to Majorana wire also gives rise to new fractional values of the conductance.
In particular, for single dots a $25\%$ reduction of the conductance
to $3e^2/2h$ has been found \cite{Lee2013Jun,Weymann2017Apr}.
However, in this case the relative change is not large
and may be difficult to be detected.
Much more interesting in this regard is a double quantum dot (DQD) scenario,
where the two-stage Kondo effect develops \cite{Pustilnik2001Nov,Cornaglia2005Feb,Sasaki2009Dec,Wojcik2015Apr,Guo2021Mar},
in which at low temperatures the conductance becomes fully suppressed.
The presence of Majorana mode results then in
an increase of the conductance to some new fractional value \cite{Weymann2020Jan}.
Consequently, the relative change induced by the Majorana mode is much more
pronounced in the case of double dots instead of single quantum dot setups.
Besides, the considered system allows us to address
more fundamental questions regarding the Majorana-Kondo
interplay in the presence of more exotic Kondo states.
While such questions have already been partly addressed in the case
of electronic transport, their thermoelectric signatures remain still rather unexplored.
Therefore, the main focus of this paper is on the caloritronic transport properties
of a T-shaped double quantum dot side-attached to Majorana wire,
as schematically depicted in \fig{model}.}

Studying the response of the system subject to thermal gradient
gives an important information about the interplay between the Majorana and Kondo physics.
Signatures of such interplay have been identified 
for hybrid devices comprising single quantum dot coupled to metallic leads
and to MZM \cite{Lopez2014May,Weymann2017Jan}.
In this regard, however, the thermoelectric way of exploring Majorana modes seems particularly unexplored,
especially as far as more complex structures are considered.
This is despite the fact that thermopower can provide additional insight into the properties of
the system, which has been achieved for Kondo QDs both theoretically \cite{CostiZlatic}
and experimentally \cite{Dutta2018Dec,Svilans2018Nov}, and even more could be taken out from 
the spin-resolved caloritronics \cite{Bauer2012May,Weymann2017Jan}.
It has been also predicted that half-fermionic nature of Majorana quasiparticles
may give rise to the violation of the Wiedemann-Franz law 
\cite{Gong2014Jun,Ramos-Andrade2016Oct,Buccheri2021Aug}
and it should leave various signatures in QDs weakly coupled 
with Majorana modes \cite{Hou2013Aug,Leijnse2014Jan,Valentini2015Jan,Smirnov2018Apr,Wang2019May,Smirnov2020Mar}.
Here, we especially show that a generalized Wiedemann-Franz law
holds for strongly correlated DQD-Majorana system,
as in other Kondo scenarios where topology does not play any role \cite{CostiZlatic,Wojcik2016Feb}. 
Furthermore, the presence of the second Kondo scale (introduced by the second QD) in our setup allows
for tuning the system into the regime where the Seebeck coefficient changes sign in the presence 
of even very small coupling to the Majorana mode.

Finally, it should be noted that there exist other considerations on DQD Majorana systems
\cite{Gong2014Jun,Sherman2017Mar,Vayrynen2020Oct}, however, 
these works do not capture the Kondo regime.
We believe that with rapid improvements in fabrication techniques 
\cite{Carrad2020Jun,Heedt2020Jul,Kanne2020Feb,Munning2021Feb}, in particular reducing 
the necessary magnetic fields \cite{Desjardins2019Oct,Delfanazari2020Jul},
construction of such devices would be possible soon,
and we hope that our study will stimulate further efforts in this direction.

The paper is organized as follows.
In Sec. II we present the model of the studied system,
define the thermoelectric coefficients and 
describe the method used in calculations.
To gain an intuitive understanding of the transport behavior, 
in the next section we discuss the non-interacting case.
The fully interacting case is analyzed in great detail in Sec. IV.
Finally, the paper is summarized in Sec. V.

\section{Theoretical framework}
\label{sec:modelandmethod}

\subsection{Model}
\label{subsec:model}

The schematic of the considered DQD-Majorana setup is presented in \fig{model}.
It consists of two single-level quantum dots coupled to each other
through the hopping $t$, with one of the dots attached to the external leads.
The second dot is directly interacting with superconducting nanowire
hosting Majorana zero-energy modes described by  the operators $\gamma_1$ and $\gamma_2$.
The system can be modeled by the following Hamiltonian $H = H_{\rm leads} + H_{\rm tun} + H_{\rm DD-Maj}$.
The first term stands for the left and right metallic leads,
denoted respectively by $r=L$ and $r=R$, for the left and right lead,
which are modeled as reservoirs of noninteracting quasiparticles
\be
H_{\rm leads} = \sum_{r=L,R}\sum_{\mathbf{k}\sigma}
\e_{r\mathbf{k}} c^\dag_{r\mathbf{k}\sigma} c_{r\mathbf{k}\sigma}.
\ee
Here, $c_{r\mathbf{k}\sigma}^\dag$ is the creation operator for an electron with
spin $\s$, momentum $\mathbf{k}$ and the energy $\e_{r\mathbf{k}}$ in the lead $r$.
The second term describes the tunneling processes between the first 
quantum dot and the leads. It is given by
\be
H_{\rm tun} = \sum_{r=L,R}\sum_{\mathbf{k}\sigma} v_{r} \left(d^\dag_{1\s}
c_{r\mathbf{k}\sigma} + c^\dag_{r\mathbf{k}\sigma} d_{1\s} \right),
\ee
with $v_r$ denoting momentum-independent tunnel matrix elements. The operator 
$d^\dag_{1\s}$ creates an electron with spin $\s$ in the first quantum dot.
The tunnel coupling results in the broadening of the first dot level,
which is given by $\Gamma = \Gamma_L + \Gamma_R$, where
$\Gamma_r=\pi\rho_r v_r^2$, with $\rho_r$ being the 
density of states of lead $r$. In the following we assume $\Gamma_L = \Gamma_R \equiv \Gamma/2$.

The last term of the total Hamiltonian models the subsytem consisting of double quantum 
dot and Majorana wire, which can be described by the following effective Hamiltonian
\bea \label{Eq:HDDM}
H_{\rm DD-Maj} &=& \sum_{j=1,2}\sum_{\s} \e_j d_{j\s}^\dag d_{j\s}
+ \sum_{j=1,2} U_j d_{j\uparrow}^\dag d_{j\uparrow} d_{j\downarrow}^\dag 
d_{j\downarrow}
\nonumber\\
&&+ \sum_\s t (d_{1\s}^\dag d_{2\s} +  d_{2\s}^\dag d_{1\s})
\nonumber\\
&&+ \sqrt{2} V_M (d^\dag_{2\downarrow} \gamma_1 + \gamma_1 d_{2\downarrow}) ,
\eea
where the first three parts model the double quantum dot.
$d_{j\s}^\dag$ stands for the creation operator for an electron
with spin $\s$ on the $j$-th dot with the energy $\e_j$,
$U_j$ is the corresponding Coulomb correlation energy
and the dots are coupled by the hopping matrix element $t$.
The second quantum dot is coupled to Majorana wire with tunnel matrix elements given by $V_M$ \cite{Flensberg2010Nov,Liu2011Nov,Lee2013Jun,Weymann2017Apr}.
\new{Note that since the bare double dot Hamiltonian  ($V_M=0$)
has a full spin symmetry, one can choose the quantization axis
in such a way that it coincides with the Majorana mode.
Because of that, the Majorana wire couples
only to the spin-down electrons on the double dot
\cite{Flensberg2010Nov,Liu2011Nov,Lee2013Jun}.}
The Majorana operators, $\gamma_1$ and $\gamma_2$, describe
Majorana zero-energy modes at the ends of topological superconducting wire.
These operators can be expressed in terms of an auxiliary fermion operator $f$
as  $\gamma_1 = (f^\dag+f)/\sqrt{2}$ and $\gamma_2 = i(f^\dag-f)/\sqrt{2}$.
In our considerations we assume that the wire is much longer than the
superconducting coherence length, such that 
the Majorana modes do not overlap \cite{Albrecht2016Mar}, unless stated otherwise.

\subsection{Transport coefficients}
\label{subsec:TransCoef}

In this paper we are interested in the linear-response 
thermoelectric transport properties of the setup presented in \fig{model}.
The temperature and voltage gradients, $\Delta T$ and $\Delta V$, respectively, are symmetrically
applied to the left and right contacts,
while the topological superconductor is assumed to be grounded
and kept at temperature $T$.
\new{
Under such assumptions,
in the linear response regime,
the average electric and heat currents
flowing between the normal contacts and the wire
vanish, provided the system is left-right symmetric
\cite{Benenti2017Jun}.}
On the other hand, the electric $I$ and heat $I_h$ currents
flowing between the left and right leads in the linear response regime
can be expressed as \cite{barnard1972thermoelectricity}
\begin{equation}
\label{eq:current_matrix}
\begin{pmatrix}
I \\
I_{h}
\end{pmatrix}
= \sum_{\s}
\begin{pmatrix}
e^2 L_{0\s}	&	-\frac{e}{T} L_{1\s} \\
-e L_{1\s}	&	\frac{1}{T} L_{2\s}
\end{pmatrix}
\begin{pmatrix}
\Delta V \\
\Delta T
\end{pmatrix},
\end{equation}
where the functions introduced above, $L_{n\s}$, are given by
\begin{equation}
\label{eq:onsager_integral}
L_{n\s} = - \frac{1}{h} \int \w^n ~\frac{\partial f(\w)}{\partial \w} \T_\s (\w) d\w .
\end{equation}
Here, $f(\w)$ denotes the Fermi-Dirac distribution function for $\Delta \mu =  \Delta T = 0$
and $\T_\s (\w)$ is the transmission coefficient through the double-dot-Majorana setup for spin $\s$.
For the considered system's geometry, it can be related to the spectral function
of the first quantum dot, $\T_\s (\w) = \pi\Gamma A_\s (\w)$,
with $A_\s (\w) = - \frac{1}{\pi} \rm{Im}~\GF_\s^R (\w)$,
where $\GF_\s^R (\w)$ is the Fourier transform
of the retarded Green's function of the first dot,
$\GF^R_\s(t) = -i \Theta (t) \langle\{ d^{}_{1\s}(t), d_{1\s}^\dagger(0)\}\rangle$.

Because in our considerations it is assumed
that the Majorana mode is coupled to the spin-down electrons,
it is interesting to investigate the behavior
of spin-resolved transport coefficients
\cite{DiasdaSilva2013May, Wojcik2014Sep,Wojcik2015Apr}.
The linear conductance, Seebeck coefficient and heat conductance
in the spin channel $\sigma$ can be expressed as
\begin{align}
\label{eq:conductance}
	G_\s = \left( \frac{\partial I_\sigma}{\partial \Delta V}\right)_{\Delta T = 0} &= e^2 L_{0\s}, \\
\label{eq:seebeck}
	S_\s = \frac{1}{G_\s} \left( \frac{\partial I_\sigma}{\partial \Delta T} \right)_{\Delta V = 0} &= - \frac{1}{e T} \frac{L_{1 \s}}{L_{0 \s}}, \\
	\kappa_\s = \left( \frac{\partial I_{h\sigma}}{\partial \Delta T} \right)_{I_\sigma = 0} &= \frac{1}{T} \left( L_{2\s} - \frac{L_{1\s}^2}{L_{0\s}} \right),
\end{align}
where $I_\sigma$ and $I_{h\sigma}$ denote the spin-resolved electric and heat currents.
Considering both spin channels, the total conductance can be simply calculated with 
the formula, $G = G_\uparrow + G_\downarrow$, whereas 
the total thermopower is defined as \cite{barnard1972thermoelectricity},
$S = -(1/eT)(L_1/L_0)$, where $L_n = \sum_\s L_{n\s}$.
On the other hand, for the heat conductance one finds,
$\kappa=T^{-1}(L_2-L_1^2/L_0)$.
We note that in this analysis we only consider the electronic contribution
to the heat conductance, with the contribution due to phonons
is assumed to be negligible. This is justified in the low-temperature
regime considered here.

As can be seen from the above formulas, the main task is to find 
the transmission coefficient through the device and determine the integrals $L_{n\sigma}$
in the most accurate way. We achieve this goal by using the numerical renormalization 
group (NRG) procedure \cite{Wilson1975Oct, Bulla2008Apr, NRG_code}, which allows for 
construction of full density matrix \cite{Anders2005} and calculation of the relevant 
Green functions directly from their Lehmann representation.

To perform the calculations, it is convenient to re-express the last term of 
the double dot-Majorana wire Hamiltonian in terms of an auxiliary fermionic operator $f$.
One then finds, ${\sqrt{2} V_M (d^\dag_{2\downarrow} \gamma_1 + \gamma_1 d_{2\downarrow})} =
{V_M (d^\dag_{2\downarrow}-d_{2\downarrow}) (f^\dag+f)}$.

\section{Non-interacting case}
\label{sec:analytics}

\begin{figure*}[t]
\includegraphics[width=0.85\linewidth]{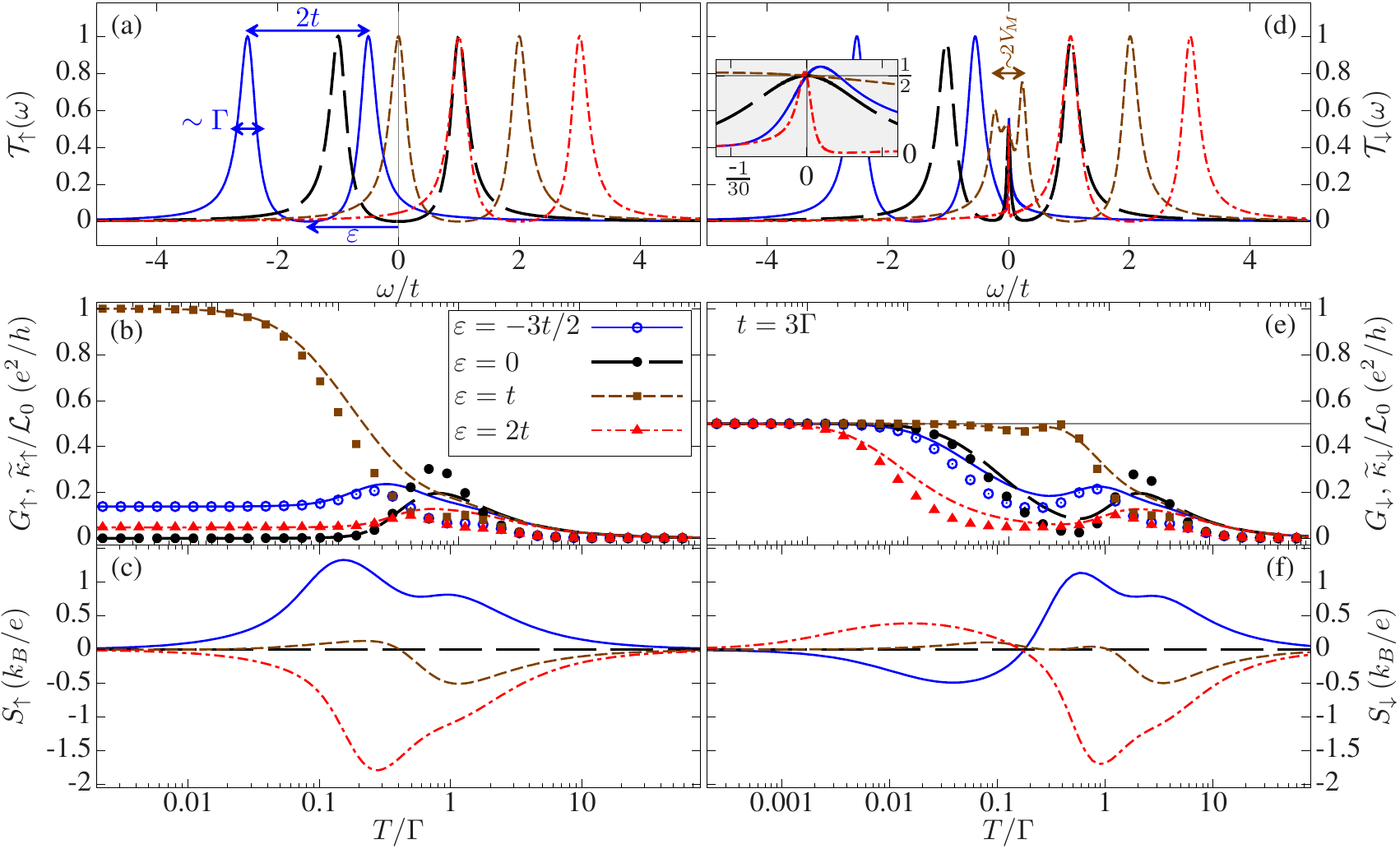}
\caption{Left column: 
			   (a) $T=0$ transmission coefficient $\T_{\up}(\w)$
			   for $\e=\e_1=\e_2$ indicated in the legend and $t=3\Gamma$,
			   (b) the conductance $G_{\up}$ as a function of temperature $T$ (lines)
			   and the shifted thermal conductance $\tkap_\up$ (points, see text for details),
			   and (c) spin-up thermopower $S_{\up}$. 
  		 	  The right column shows the corresponding
		 	  plots for the spin-down component. 
		       The interaction with Majorana wire is assumed to be $V_M=\Gamma/2$.
		       The inset in (d) zooms the $\w\approx 0$ region.
			   }
\label{fig:A}
\end{figure*}
\new{To make our theoretical discourse complete, we would like to begin with 
the description of the non-interacting case.
This allows us to obtain analytical formulas
and generate exact results that will serve as a starting point for the further analysis
when correlations are relevant.}
In the absence of electron correlations it is possible to get an exact solution
for the transmission coefficient by the equation of motion
for the retarded Green's function. Moreover, the two spin channels 
become in practice independent and one can analyze them separately.
Even though in realistic case relaxation processes would come into play 
and blur the picture, we nevertheless find it instructive to perform
such an idealized analysis, which serves as a starting point for the discussion 
of numerical results presented in the next section.

The spin-up transmission $\T_{\up}(\w)$
is the same as in the model without the Majorana wire and equals
\begin{equation}
\T_{\up}(\w) = \frac{\Gamma^2}{  \Gamma^2  + \left( \w-\e_1 + \frac{t^2}{\w-\e_2}\right)^2}.
\label{Tup}
\end{equation}
We see that it exhibits two peaks reaching $\T_\up(\w)=1$ at zeros of the term in the bracket,
and $\T_\up(\w)=0$ at $\w=\e_2$. The width of the peaks is of the order of $\Gamma$. 
A representative illustration of these features for $\e_1=\e_2=\e$ is shown in \fig{A}(a). 
The maxima of $\T(\w)$ correspond to single-particle excitations,
in this case associated with additional spin-up electron occupying or missing in the double quantum dot.

Whenever the peak of $\T_\up(\w)$ occurs close to $\w=0$, this mean there are low-energy 
charge fluctuations, which give rise to good conductance, as illustrated in \fig{A}(b).
This also applies to normalized thermal conductance $\kappa/T$. Actually, for a Fermi liquid at low $T$,
such that the Sommerfeld expansion is valid, the Wiedemann-Franz (WF) law is fulfilled, \ie{} 
$\Lor_0^{-1} \kappa(T)/T = G(T)$, with $\Lor_0 \equiv \pi^2/3$. More generally, it 
has been observed that some interacting systems exhibit a modified WF law
\cite{Wojcik2016Feb}, where the shifted and rescaled heat conductance 
\begin{equation}
\tkap(T) \equiv \kappa(\alpha T)/(\alpha T)
\end{equation}
fulfills $\tkap/\Lor_0=G$ for $\alpha\approx 2$.
%

In general, narrow peaks of $\T(\w)$ close to the Fermi level can also lead 
to large thermopower \cite{MahanSofo}. Its sign depends on the slope of 
$\T(\w)$ and reveals the type of majority charge carriers (electrons or holes). 
Moreover, the Seebeck coefficient vanishes at the particle-hole symmetry point.
For low temperatures, via the Sommerfeld expansion, one finds
$S_\up\sim T$. On the other hand, at high $T$ from \eq{eq:seebeck} follows $S_\up\sim T^{-1}$.
Such behavior can be seen in \fig{A}(c).

Furthermore, the thermopower may also change sign between different transport regimes. 
This behavior is visualized e.g. by a short-dashed
curve in \fig{A}(c) and the corresponding curve in \fig{A}(a).
For $T\lesssim 2t$ (peaks separation), most of the relevant spectral weight 
is above $\w=0$ due to the peak asymmetry, such that 
electrons overtake holes in transport and, consequently, $S_\up>0$.
On the contrary, at high $T$, the peak deep below
the Fermi energy becomes relevant (\ie{} holes can be excited there by thermal fluctuations),
which facilitates the hole transport and the sign of $S_\up$ is flipped at $T\sim 2t$.

The results concerning $\T_\up(\w)$ would be also valid in the second spin channel,
if not for the coupling to the Majorana mode. From the equation of motion technique,
the Green's function in the spin-down channel is conveniently found as a continued fraction,
\begin{eqnarray}
\label{Adown}
\GF_\down\! (\w) \es [\w - \e_1 + i\Gamma - t^2/A_1(\w)]^{-1} ,
	\\&&
	A_1(\w) = \w - \e_2 - 2V_M^2/A_2(\w) ,\nn\\&&
	A_2(\w) = \w - \e_M^2/\w - 2V_M^2 / A_3(\w) ,\nn\\&&
	A_3(\w) = \w + \e_2 - t^2 / [\w + \e_1 + i\Gamma] , \nn
\end{eqnarray}
in agreement with Ref.~\cite{Gong2014Jun}.
Note that the above formula is quite general and it also takes into account
a finite overlap between Majorana modes \cite{Albrecht2016Mar}, which can be included
in the effective Hamiltonian via the term $i \e_M \gamma_1 \gamma_2$.
This overlap is determined by the ratio of the superconducting coherence length to the length of the wire:
if the coherence length is much shorter than the wire,
the Majorana modes do not overlap and, consequently, $\e_M = 0$.
Although the expression for the transmission coefficient is quite cumbersome,
it follows from \eq{Adown} via simple algebra and one can give the 
exact formulas for $\T_\down(\w)$ and its derivative at $\w=0$,
\begin{align}
\T_\down(0) &= \left\{ \begin{array}{l @{\;\mathrm{for}\;}l} \left[ 1 + (\e_1-t^2/\e_2)^2/\Gamma_{\down}^2 \right]^{-1} & \e_M \neq 0  \\[6pt]
										\frac{1}{2} & \e_M = 0 \end{array} \right. ,
\label{Adown0}
\\
\T'_\down(0) &= \left\{ \begin{array}{l @{\;\mathrm{for}\;}l} 
					\frac{\left(\e_1 - t^2/\e_2\right)\left[V_M^2 t^2 + 2\e_M^2(t^2+\e_2^2)\right]}{
					\Gamma_\down^2 \e_2^2 \e_M^2 \left[ 1 + (\e_1-t^2/\e_2)^2/\Gamma_{\down}^2 \right]^{2}} & \e_M \neq 0  \\[6pt]
					\frac{\e_2}{4V_M^2}-\frac{\e_1\e_2^2}{4V_M^2t^2} -\frac{\e_1}{t^2} & \e_M = 0 \end{array} \right. .
\label{dAdown0}
\end{align}
Quite strikingly, for $\e_M=0$ the transmission becomes completely universal at low energies, 
$\T_{\down}(0)=1/2$ \cite{Gong2014Jun,Weymann2020Jan}, 
which is a clear manifestation of half-fermionic 
nature of Majorana quasiparticles. Moreover, $\T'(0)$ becomes independent of $\Gamma$,
as the width of the spectral features close to $\w=0$ is set by the strength of the coupling
to the Majorana mode, $V_M$. 
It seems noteworthy that $\T_{\down}(\e_M)=1/2+\mathcal{O}(\e_M)$, so even though 
$\T_{\down}(0)$ is not universal for finite $\e_M$, at non-zero temperatures
$T\sim \e_M$ one should expect results obtained for $T=\e_M=0$ to be relevant.
From now on, let us focus on the $\e_M=0$ case, i.e. long Majorana wire case.

The features discussed above for the spin-down component
are visible in \fig{A}(d), and especially in the inset there.
However, as is clear from the main plot, far from the Fermi level,
the behavior of $\T_\down(\w)$ very much resembles that of $\T_\up(\w)$,
compare \fig{A}(a) with \fig{A}(d).
Whenever $\T_\down(\w)$ has no peak close to $\w=0$, just an additional 
(in general asymmetric) peak appears, such that $\T_{\down}(0)=1/2$. 
Yet the peak of $\T_\up(\w)$ at $\w=0$ visible in \fig{A}(a) for a short-dashed curve
splits into three peaks visible in $\T_\down(\w)$, separated approximately by $V_M$. 
The peaks at $\w\approx \pm V_M$ somewhat resemble the case
one would get if Majorana wire was replaced by usual resonant level,
while the third peak remains a unique Majorana signature. 

The properties of $\T_\down(\w)$ reveal in the corresponding transport characteristics of the system.
${G_\down(T\to 0) = e^2/2h}$, irrespective of $\e$, as can be seen 
in \fig{A}(e), where the low temperature limit is reached at $T\sim \Gamma_M \approx V_M^2/t$.
The modified WF law is still well satisfied at low temperatures, compare the curves and points in \fig{A}(e).
\new{On the other hand, the Seebeck coefficient $S_{\down}(T)$
exhibits an additional sign change for $\e$ away from the resonance, what cannot be 
seen for the spin-up component of the thermopower, cf. Figs.~\ref{fig:A}(c) and (f).
This sign change occurs for $T/\Gamma \approx 0.1$,
and can be related to the corresponding change of slope of 
the linear conductance, as described by the Mott's formula
\be
\label{eq:S_Sommerfeld}
S \approx - \frac{\pi^2}{3} \frac{k_B^2}{e} \frac{T}{\T(0)} \frac{\partial 
	\T(\omega)}{\partial \omega}.
\ee
}

\section{Numerical results and discussion}
\label{sec:Num_results}
The considered system reveals interesting effects
associated with the interplay between the Kondo and Majorana physics, 
as compared to bare double quantum dot.
In the following sections we present and discuss the numerical results 
obtained with the aid of the numerical renormalization group method for the fully interacting case.
At the beginning, we analyze the spin-resolved charge and thermal conductance in the context of the Wiedemann-Franz law 
and its persistence in the presence of topological superconductor.
We also examine the Majorana-induced current spin polarization.
Then, we study the thermopower $S$, where Majorana-induced sign change is observed. 
In the discussion we focus on the two regimes 
of hopping between quantum dots: the weak and strong ones.
The weak interdot hopping regime is characteristic
of the two-stage Kondo effect, whereas when the hopping
is strong, a molecular singlet state forms between the dots
and the Kondo correlations are less important.

In NRG computations we take the discretization parameter $\Lambda = 2-2.5$
and keep at least $3000$ states during the procedure of iterative diagonalization.
All energies are expressed in terms of band halfwidth $D$, which is used
as energy unit $D\equiv 1$.
We set the Coulomb interaction $U_1=U_2\equiv U = 0.2$,
while the first quantum dot is coupled to metallic leads with
strength $\Gamma_L = \Gamma_R \equiv \Gamma = U/10$.
Unless otherwise stated, the energy levels of quantum dots
are set as $\e_1 = \e_2 \equiv \e =  -U/3$.
\new{We note that although we focus on the case of symmetric dots,
	the results presented here are also valid for slightly asymmetric systems,
	i.e. dots having different charging energies,
	couplings to the contacts and detuned levels,
	as long as the system is in the Kondo regime.}

\subsection{Persistence of the modified Wiedemann-Franz law}
\label{sec:WFlaw}

\begin{figure}[ht!]
	\includegraphics[width=0.9\columnwidth]{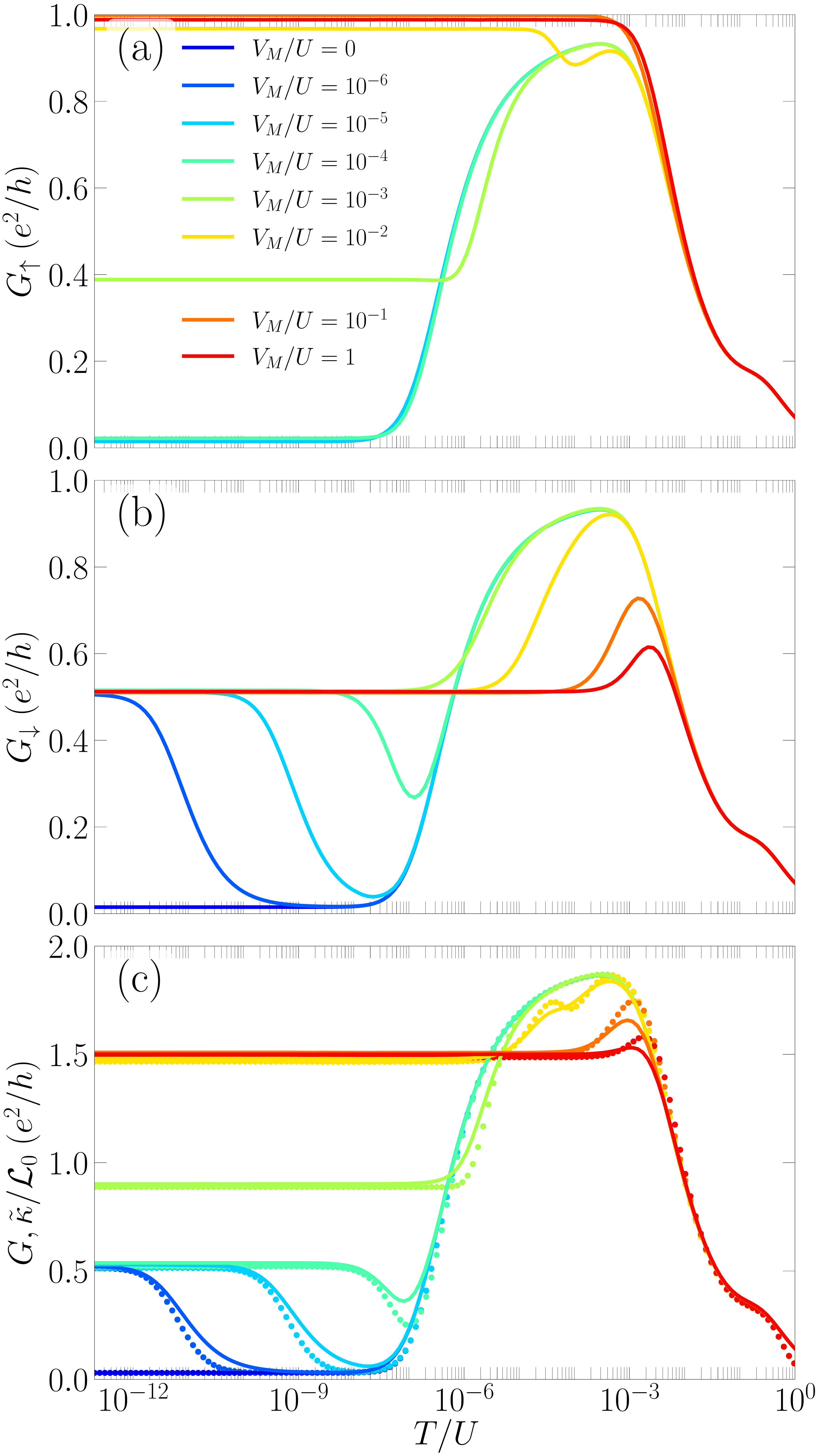}
	\caption{\label{fig:WFlaw}
		(a) The spin-up $G_\up$, (b) spin-down $G_\down$ and (c) total conductance $G$
		plotted as a function of temperature.
		The points in (c) present the rescaled heat conductance 
		$\tilde \kappa \equiv \kappa(\alpha T)/(\alpha T)$ with $\alpha =2$.
		The parameters are: $U=0.2$, $\Gamma=0.1U$, $t=0.02U$ and 
		$\e_1=\e_2=-U/3$. \new{Please note that the curves 
		for $G_\up$ overlap when $V_M/U \lesssim 10^{-4}$,
		since the energy scale associated with the coupling 
		to Majorana wire is smaller than the Kondo energy scale.
		This is contrary to $G_\down$, where due to direct coupling
		to Majorana wire quantum interference
		results in suppression of the second stage of the Kondo screening,
		lifting the conductance,
		already for small coupling $V_M/U \approx 10^{-6}$.}
		}
\end{figure} 

\begin{figure}[ht!]
	\includegraphics[width=0.9\columnwidth]{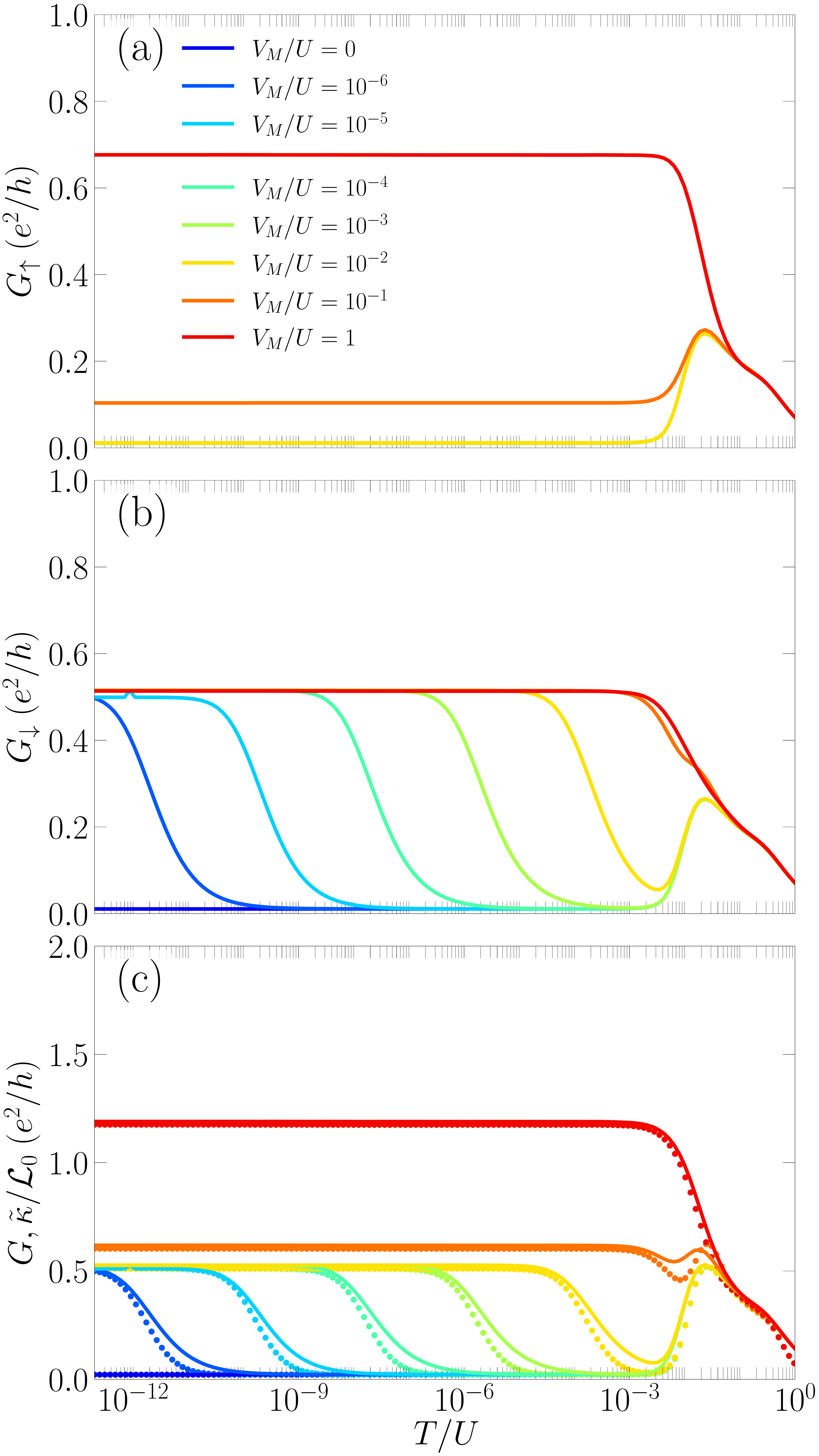}
	\caption{\label{fig:WFlaw2}
		The same as in \fig{WFlaw} calculated for $t=0.1U$.}
\end{figure} 

To begin with, in Figs.~\ref{fig:WFlaw} and \ref{fig:WFlaw2}
we present the spin-resolved and total linear conductance as a
function of temperature, calculated for different values of the coupling to the 
Majorana zero mode in the case of weak (\fig{WFlaw}) and strong (\fig{WFlaw2})
hopping $t$ between the quantum dots.
For the weak value of hopping, lack of the coupling to superconducting 
nanowire results in the typical two-stage Kondo behavior, where besides the Kondo 
temperature $T_K$, defined as \cite{Haldane_Phys.Rev.Lett.40/1978}
\be
\label{eq:TK}
T_K = \sqrt{\frac{\Gamma U}{2}} \exp{\left[ \frac{\pi}{2} \frac{\e_1 
(\e_1 + 
U)}{\Gamma U} \right]}
\ee
one can define another energy scale $T^*$, the so-called second-stage Kondo temperature,
at which the total conductance drops to zero
\cite{Cornaglia2005Feb}
\be
\label{eq:Tstar}
T^* \approx \alpha T_K \exp{\left( -\beta T_K/J_{\rm{eff}} \right)},
\ee
with $J_{\rm{eff}} = 4 t^2 / U$ and $\alpha, \beta$ dimensionless 
constants of the order of unity. This drop is 
the result of spin screening of the second quantum dot \cite{Cornaglia2005Feb, 
Wojcik2016Feb}. As can be seen in the figures,
for $t=0.02U$, $T^*\approx 10^{-6}U$,
whereas for $t=0.1U$, there is only a small 
resonance due to the Kondo effect at the first quantum dot, which becomes
quickly suppressed as T is lowered. 
One can then estimate $T^*\approx T_K$.
Consequently, $T^*$ is a few orders of magnitude larger 
in the case of $t=0.1U$ compared to the case of $t=0.02U$.
When the coupling to the Majorana wire is turned on, it starts to play a 
main role in the electronic transport through the system.
In the low-temperature regime one observes competitive interplay between the Kondo and Majorana physics, 
causing the reduction of the second dot screening, lifting conductance to the 
value of $e^2/2h$, which comes mainly from the 
spin-down contribution. Higher values of $V_M$ affect also the spin-up 
electrons, rising the conductance over $e^2/2h$
to reach its maximum value for weak hopping between the dots, see \fig{WFlaw}(a).
Consequently, the existence of Majorana mode in the system
has a great impact on the spin-resolved transport channels,
utterly destroying the second stage of the Kondo effect, which
can be seen as the rise of the total conductance from $e^2/2h$ to $3e^2/2h$ in the low 
temperature regime,
\new{
which happens for $V_M/U \gtrsim 10^{-3}$, see \fig{WFlaw}(c).
}%
The maximum of the conductance develops for $T^* \lesssim T 
\lesssim T_K$, and is being reduced due to spin-down electrons contribution, which 
are more affected within the whole energy spectrum.

Enhancing the inter-dot interaction (see \fig{WFlaw2}), one can observe that the 
competition between the coupling to the Majorana wire and the hopping becomes 
more fierce, which is especially visible
in the spin-down component of $G$ and the total conductance of the system.
In this regime $G_\down$ and, thus, $G$ are
being raised to $e^2/2h$ at the characteristic energy scale
$\omega \sim \Gamma_M$ \cite{Weymann2020Jan}.
For larger values of $V_M$, contrary to the case of $t=0.02U$,
the Kondo effect does not develop fully
even when $V_M$ is on the order of $U$, which is 
due to the formation of a molecular singlet state
between the dots with binding energy
of the order of the Kondo temperature, see \fig{WFlaw2}.

Let us now focus on Figs.~\ref{fig:WFlaw}(c) and \ref{fig:WFlaw2}(c)
where the rescaled thermal conductance in the units 
of temperature, $\tilde \kappa \equiv \kappa(\alpha T)/(\alpha T)$,
is plotted (dotted lines) with colors corresponding 
to the values of $V_M/U$ of the charge conductance.
Comparing the rescaled thermal conductance and the charge conductance allows us to analyze the 
Wiedemann-Franz law \cite{Franz1853Jan}, which states that $\kappa/T = G 
\Lor_0$, with $\Lor_0 = \pi^2 /3$. 

\new{%
It has been shown that 
in transport through the Majorana wire in different configurations
the Wiedemann-Franz law is violated \cite{Ramos-Andrade2016Oct}. 
Also, in a more general case of a multi-terminal superconducting island,
the violation of the Wiedemann-Franz law has been proposed as a means to detect
topological character of the superconductor \cite{Buccheri2021Aug}.
On the contrary, it has been demonstrated that for single \cite{CostiZlatic} 
and double quantum dot setup without Majoranas \cite{Wojcik2016Feb}, a modified 
Wiedemann-Franz law can be introduced, $\Lor_0^{-1} \kappa(\alpha T)/(\alpha T) = G$,
which is satisfied with good accuracy up to temperatures $T\sim U$.
The key result we show here is that such modified law, quite surprisingly,
also persists in the presence of coupling to topological superconducting wire.
As can be seen in Figs.~\ref{fig:WFlaw}(c) and \ref{fig:WFlaw2}(c),
with $\alpha \approx 2$, the rescaled thermal conductance $\tilde \kappa$ 
retraces the charge conductance even for higher values of $V_M$. 
In particular, the rescaled heat conductance in the second-stage Kondo regime rises 
to the quarter of its maximal value. This proves that whatever fractionalization 
of the effective particles takes place in the DQD structure, it is the same for
charge and heat carriers, {\it i.e.} no separation of charge and 
heat transport takes place. This is in contrast to results obtained 
in Refs.~\cite{Buccheri2021Aug}~and~\cite{Giuliano2021Aug}, where excitations 
of the superconducting island in the topologically non-trivial phase 
are composed partially of electrons and partially of Andreev-reflected holes, 
such that the heat and charge currents are no longer simply proportional. 
This apparent difference is a consequence of the heat and charge current 
\textit{not} flowing through the topological superconductor in the setup proposed here.
Instead, heat and charge carriers in DQD structure have a Fermi-liquid character, 
and behave qualitatively similar to the non-interacting case discussed in Sec.~\ref{sec:analytics}.
The coupling to the Majorana mode leads to interferometeric effects, in particular,
universal fractional values of both conductances, yet the local spectrum structure of DQD
remains qualitatively different than inside the topological superconductor,
in particular, the mutual coupling of heat and charge transfers is not altered.

With the Wiedeman-Franz law qualitatively correct in the proposed setup,
in the following sections we proceed to pinpoint the signatures of the Majorana modes presence 
in other spin-resolved transport characteristics of the device,
such as the current spin polarization and the (spin) Seebeck effect.
}

\subsection{Majorana-induced current polarization}
\label{sec:Pol}

\begin{figure}[ht!]
	\includegraphics[width=0.9\columnwidth]{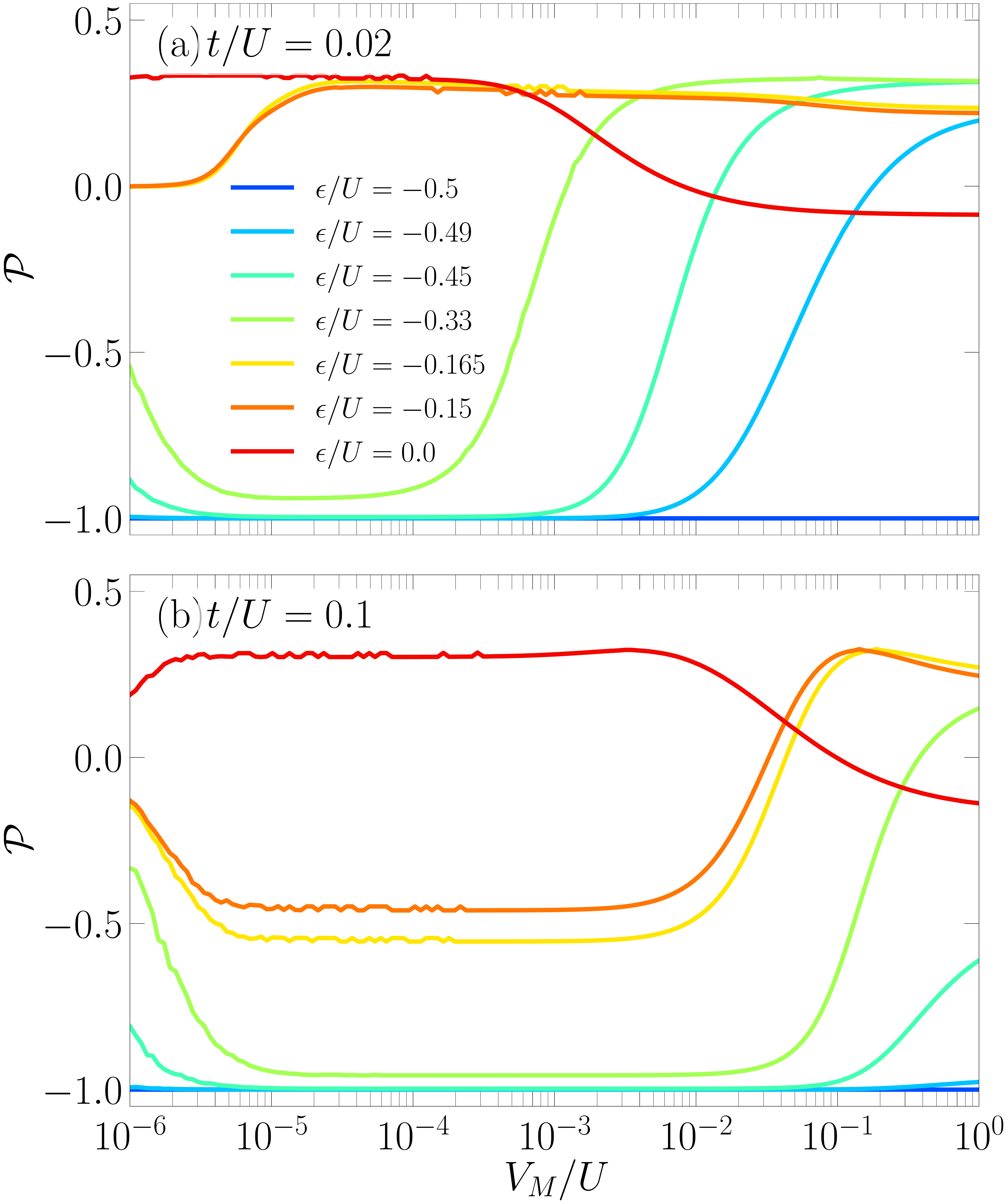}
	\caption{\label{fig:Pol}
		The current spin polarization $\Pol$ as a function of $V_M$ (note the 
		logarithmic scale) for multiple values of DQD level positions
		$\e_1=\e_2\equiv \e$
		out of the particle-hole symmetry point, calculated for (a) $t=0.02U$ and (b) 
		$t=0.1U$.
		The other parameters are the same as in \fig{WFlaw} with $T = 10^{-10}U$.
		Note that due to very low values of integrals $L_{n\s}$ there are 
		oscillations in $\Pol$ for $V_M \lesssim 10^{-3}U$ that should be 
		considered as numerical error, which does not affect qualitative behavior of presented results. 
	}
\end{figure} 

Maintaining our focus on the conductance, 
before the behavior of the Seebeck coefficient is discussed, let us 
take a gander on Fig.~\ref{fig:Pol}, where the current spin polarization is plotted as a function of 
$V_M$ for multiple values of the DQD level positions for weak and strong hopping between 
quantum dots in the low-temperature regime with $T = 10^{-10}U$.
The spin polarization of the current can be defined as
\be
\Pol = \frac{G_{\uparrow}-G_{\downarrow}}{G_{\uparrow}+G_{\downarrow}}.
\ee
Usually, in systems such as quantum dots, finite spin 
polarization of the current is accompanied with suppression of the Kondo effect
\cite{Weymann2011Mar}. External magnetic field \cite{Costi2000Aug} or 
ferromagnetic leads \cite{MartinekEx}, which usually introduce the spin imbalance of 
current carriers, split the Kondo peak and eventually cause it to disappear. 
In the case of the system studied here, the spin polarization is associated with the
coupling to topological superconductor and it is not necessarily
accompanied with suppression of the Kondo state.

\new{In \fig{Pol}(a) the current spin 
polarization for weak hopping $t/U = 0.02$ is shown.
First, we note that $\Pol = -1 $ when the system is at the particle-hole
symmetry point, $\e=-U/2$, within the whole range of $V_M$.
This is due to the fact that $G_\up$ is then suppressed due to the 
two-stage Kondo effect, while $G_\down = e^2/2h$, due to the quantum interference
with the Majorana wire.}
Let us now analyze what happens when detuning from $\e=-U/2$ is turned on.
Starting with the case of $\e = -0.49U$, for low values of $V_M$,
$G_\uparrow$ is suppressed due to the second-stage Kondo effect (note that $T=10^{-10}U$),
but finite value of $V_M$ affects the Kondo state
giving rise to small but finite $G_\downarrow$.
It results in a perfect negative spin polarization $\Pol \approx -1$,
quite a remarkable feature in the fully screened Kondo regime,
stressing its unusual character.
This result holds up to $V_M \approx  10^{-2} U$, when
further increase of $V_M$ eventually affects the spin-up conductance
restoring the first-stage Kondo effect when $V_M \gtrsim 10^{-2} U$.
This leads to an increase of spin polarization to a moderate positive value for $V_M \approx U$.

Shifting the DQD levels further from the particle-hole symmetry point,
the effect of perfect spin polarization with $\Pol\approx -1$ becomes distorted,
which is due to Majorana-induced splitting, which grows with detuning the levels
\cite{Lee2013Jun,Weymann2017Apr}, as well as the fact that the low-temperature
second-stage Kondo conductance attains a finite value.
As a consequence, the spin-up conductance is being lifted,
while the spin-down channel remains approximately intact. 
Moreover, smaller values of $V_M$ are now needed to increase $G_\up$.
This manifests itself as a shift in sign change of $\Pol$ toward lower values of 
the Majorana interaction.
Such behavior is maintained until $\e$ crosses the 
point where $T^* \approx \Gamma_M$, which appears at $\e \approx -0.25 U$.
For $\e$ above this value, 
the spin polarization becomes suppressed, $\Pol \approx 0$, for low coupling to
the superconducting nanowire. Increasing $V_M$ leads to the spin polarization sign change, 
setting it positive at $\Pol \approx 0.25$. This abrupt change is due to 
entering the regime, at which both spin-up and spin-down channels take part in 
transport, with $G_\downarrow = e^2/2h$ for a significant range of 
$V_M$. Finally, reaching the resonance point $\e \approx 0$, low Majorana 
coupling limit exhibits a rather strong positive spin polarization, while for larger 
values of $V_M$, $\Pol$ becomes almost suppressed due to comparable contribution of both spin channels.

Considering the case of strong hopping between quantum dots shown in \fig{Pol}(b),
one can see that the spin polarization takes values similar to the ones
already discussed, however, the range of $V_M$ where e.g. $\Pol<0$ develops is now extended.
Strong enough $t$ blocks the conductance in both spin channels, nevertheless,
the interplay of Kondo and Majorana physics can result in an increase of conductance
and the corresponding behavior of $\Pol$.
While $G_\uparrow$ is damped for almost whole energy level 
spectrum, $G_\downarrow$ behaves similarly as in the Kondo regime, being 
half-suppressed for $V_M \gtrsim 10^{-5} U$. It leads to almost full negative 
spin polarization for dot level position between $\e \approx -0.66 U$ and $\e  
\approx -0.33 U$ ($\pm U/6$ around the particle-symmetry point).
Above (below) the aforementioned range, $G_\uparrow$ starts to increase,
lifting $\Pol$ towards zero, finally changing the sign of spin polarization at 
resonance.

\new{To sum this section up, we have shown that the finite coupling to the 
Majorana wire results in nonzero spin polarization. Furthermore, this spin polarization
can have large negative values and it changes sign depending on the
strength of coupling to topological superconductor and the position
of the double dot levels.}

\subsection{Majorana-induced thermopower sign change}
\label{sec:Seebeck}

\new{
The following section will expand the discussion to another important 
thermoelectric effect---the thermopower (or the Seebeck coefficient).}
In general, for a Fermi liquid (FL) at sufficiently low temperature $T$, $S$ follows the Mott 
formula based on the Sommerfeld expansion,
with its characteristic $S \sim T$ dependence 
\cite{barnard1972thermoelectricity}, cf. Eq.~(\ref{eq:S_Sommerfeld}).
This is valid when $T$ is smaller than all energy scales characterizing given 
FL. Typically, Fermi liquid features a peak in $\T(\w)$ at $\w=\w_0$ corresponding to 
quasi-particle energy level, of width $\Geff$ inversely proportional to the 
quasi-particle life time. For $T\ll \w_0, \Geff$, the linear dependence of 
thermopower follows from the Mott formula (\ref{eq:S_Sommerfeld}).
On the other hand, for $T\gg \w_0,\Geff$, the integrals in 
\eq{eq:onsager_integral} in fact average all the $\w$-dependence out and the 
$T^{-1}$ term from \eq{eq:seebeck} remains \cite{BulkaZitko}.
This leaves $S(T)$ no choice, but to exhibit a peak 
at $T\sim \w_0,\Geff$, with a sign depending on which slope of $\T(\w)$ is 
present at $\w=0$.

This picture repeats itself in a series of peaks of alternating sings 
in T-shaped double quantum dots, where the two-stage Kondo effect is present 
\cite{Wojcik2016Feb}.
Three FLs are involved there: at elevated $T\sim U$, even QD levels 
deep below Fermi level contribute to the transport. Transport is suppressed in 
the Coulomb blockade regime $T_K \ll T \ll U$, where relevant excitations 
do not form a FL, only to rise again at $T$ below $T_K$, when FL character 
of relevant degrees of freedom is restored and another peak appears in $S(T)$.
Similar scenario happens for $T<T^*$, then only the spectral function of the second dot
exhibits a peak, while $\T(\w)$ shows the corresponding dip,
leading to a peak in $S(T)$ of opposite sign.

\subsubsection{Temperature dependence}

\begin{figure}[t]
	\centering
	\includegraphics[width=\columnwidth]{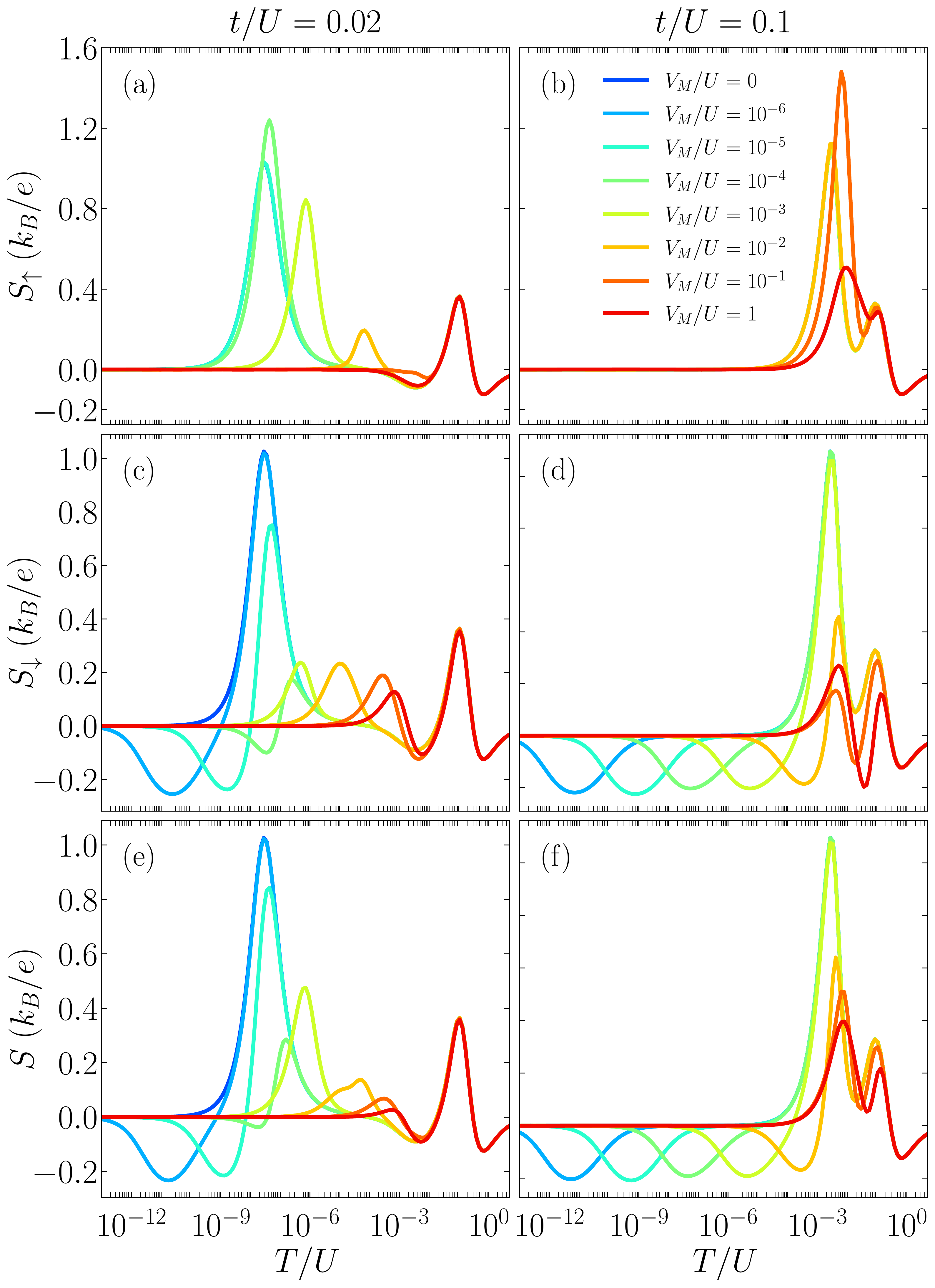}
	\caption{The Seebeck coefficient $S$ as a function of temperature for (left 
	column) $t = 0.02 U$, (right column) $t = 0.1 U$ and multiple values of 
	$V_M$ as indicated in the legend.
	The other parameters are the same as in \fig{WFlaw}.}
	\label{fig:S}
\end{figure}


The situation becomes even more interesting in the presence of topological 
superconductor. In \fig{S} the spin-resolved Seebeck coefficient for 
multiple values of $V_M$ and two selected values of $t$ is presented.
For relatively weak hopping $t = 0.02 U$ (see the left column of \fig{S}),
coupling the system with Majorana mode causes the low-temperature sign change of 
the spin-down and total thermopower, see Figs.~\ref{fig:S}(c) and (e).
This picture directly reflects the behavior of the conductance 
presented in \fig{WFlaw}, where an increase of $G$ with lowering $T$ is observed.
For $10^{-6} U \lesssim V_M \lesssim 10^{-4} U$, i.e. when $\Gamma_M$ is smaller than 
Kondo-related energy scales, there is the corresponding minimum in $S_\down$ and (thus in)
$S$ occurring at $T\sim \Gamma_M$.
At this temperature, the resurgence of conductance 
$G$ is observed, and according to the Mott formula, a sudden change of the 
thermopower can be expected. Stronger coupling with the superconducting 
nanowire results in shallowing of the minimum, which one can relate to 
stronger Majorana-Kondo competition in the system.
For values of $V_M$ up to about $10^{-4} U$, the low-temperature transport is dependent 
only on the spin-down electron channel.
However, increasing the interaction between the Majorana mode 
and the second quantum dot, additional contribution from spin-up 
electrons emerges, see \fig{S}(a).
The minimum of thermopower is then being reduced and only positive peaks of the Seebeck coefficient,
combined from the spin-up and spin-down electron channels, remain.
Moreover, stronger coupling to the Majorana wire increases
the second-stage Kondo temperature $T^*$, which results in shifting of thermopower peak
toward higher temperatures. We also note that the behavior of the Seebeck
coefficient at higher temperatures, i.e. $T\gtrsim T_K$, is consistent
with that predicted for single quantum dots \cite{CostiZlatic,Weymann2013Aug}.

The quantum interference with Majorana zero-energy mode is 
much better visible in the case of strong hopping between the dots, which is 
presented in the right column of \fig{S}.
In this regime one can notice a series of minima of the Seebeck coefficient
corresponding to $T\sim\Gamma_M$. As long as $V_M$ 
is not strong enough to dominate the transport in the system,
the prevailing spin-down electron contribution is observed, where the coupling to 
the superconducting nanowire results in increased conductance to the total value of $G = e^2/2h$. 
Such destructive behavior relative to the second-stage Kondo effect plays a 
main role in the sign changes of the thermopower. Similar to the previous case, 
when the coupling to the Majorana wire is of the order of or above the Kondo energy scales,
it also affects the spin-up electrons, resulting in a large maximum, which is
visible both in $S_\up$ and $S_\down$. Consequently, the total thermopower
features a maximum around $T\sim T_K$, see \fig{S}(f).

\subsubsection{Gate voltage dependence}

\begin{figure}[t]
	\includegraphics[width=\columnwidth]{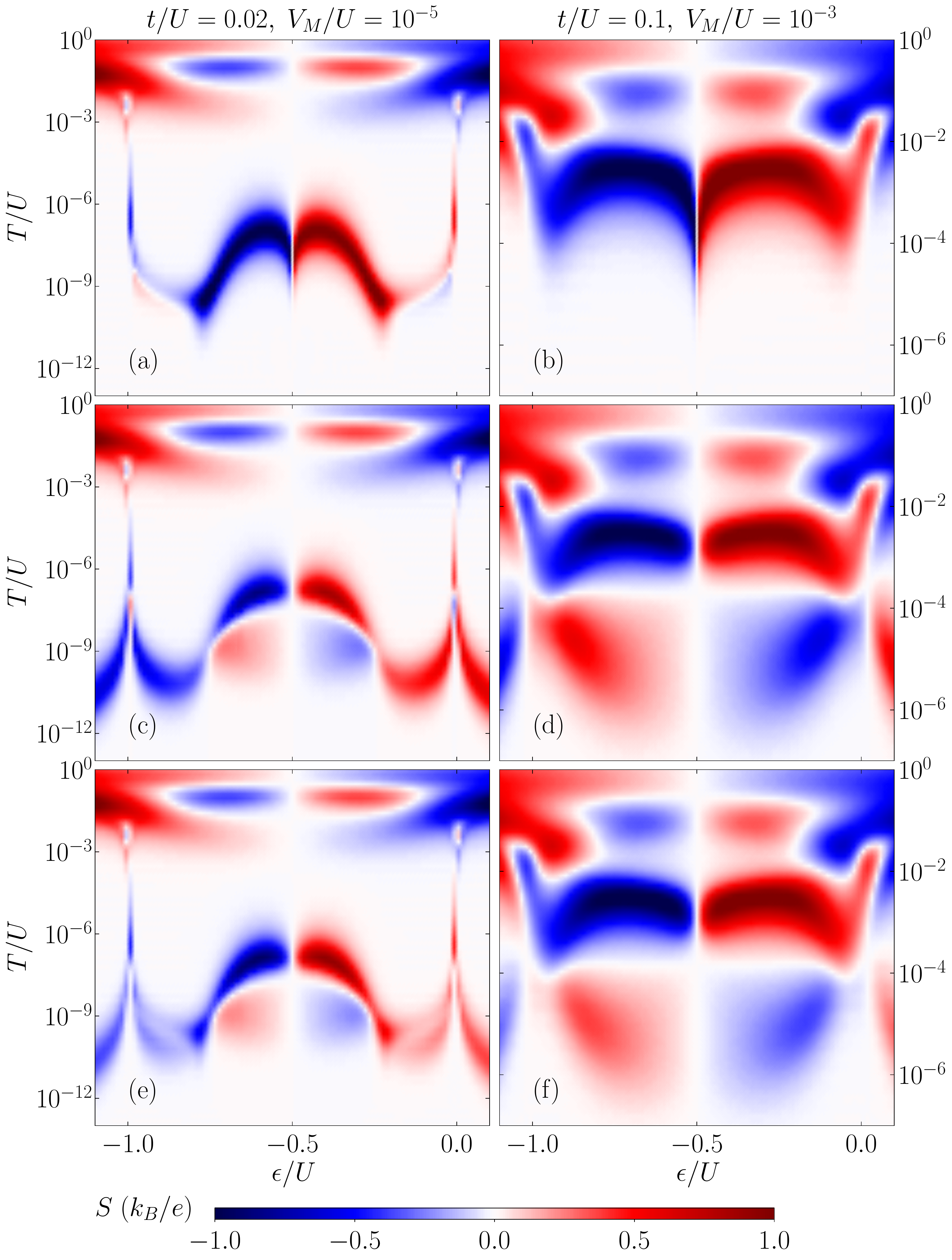}
	\caption{
The (a,b) spin-up, (c,d) spin-down, and (e,f) total Seebeck coefficient for 
$t = 0.02 U$ and $V_M/U=10^{-5}$ (left column) and $t = 0.1 U$ and $V_M/U=10^{-3}$ (right column) as a function of 
temperature $T$ and position of the quantum dot levels $\e_1 = \e_2 = \e$. Note different 
temperature scale for each column. The other parameters are the same as in 
\fig{WFlaw}.}
    \label{fig:S2D}
\end{figure} 

Figure \ref{fig:S2D} presents the Seebeck coefficient as a function of temperature $T$ and 
double dot energy levels, with $\e_1 = \e_2 = \e$. Left column of \fig{S2D}
shows the results corresponding to the weak interdot hopping, while the right column
is for the case of strong $t$. The first (second) row depicts the spin-up (spin-down)
thermopower, while the bottom row presents the total Seebeck coefficient.
First of all, one can note that thermopower changes sign when crossing
the particle-hole symmetry point, which is a natural consequence of the fact
that the type of majority charge carriers changes from electrons to holes
as $\e$ crosses $-U/2$.
Moreover, in the behavior of thermopower one can generally distinguish four regimes
determined by the relevant energy scales: $\Gamma$, $T_K$, $T^*$ and 
$\Gamma_M$. Within $-0.75 U \lesssim \e \lesssim -0.25 U$, they outline an interesting 
behavior in total thermopower of the system. 
Note that because $T^*$ strongly depends on $t$, the corresponding low
temperature behavior in the case of weak hopping is much more extended 
in the case when the hopping is strong, cf. the left and right column of \fig{S2D}.
Let us start the discussion with the former case.
One can see that in the Coulomb blockade regime there is a characteristic peak at $T \approx \Gamma$,
which does not depend on $V_M$ [cf. \fig{S}], but changes sign near the resonance. 
Then, with lowering the temperature, the first stage of the Kondo effect occurs,
where the first Fermi liquid forms and the charge transport 
through the first quantum dot is relatively strong. For this energy range, the
thermopower exhibits a small sign change with the minimum around $T\approx T_K$. 
Shifting to the lower temperature regime, where $T^* \lesssim T \lesssim T_K$, the total 
thermopower gets suppressed, while the total conductance reaches a maximum
signaling the full development of the first-stage Kondo regime.

An interesting behavior is revealed when the temperature drops below $T^*$. In 
that case one can distinguish the regime limited by $T^*$ and
$\Gamma_M$ (for assumed parameters $\Gamma_M\approx 10^{-9}U$),
where a sign change, not visible for $V_M = 0$ [cf. \fig{S}(e)], is observed.
In this parameter space one can observe consecutive thermopower minima and maxima. It 
can be related to charge conductance which is affected by the coupling to the 
Majorana mode, and considering the low-temperature limit, one can understand that 
with the Mott formula. Therefore, these peaks are related to $\Gamma_M$ and 
$T^*$, at which the change in transmission coefficient $\T(\w)$ is observed, and the 
sign change of total thermopower stems from the derivative of $\T(\w)$.
When the temperature drops below $\Gamma_M$, one can see the result of 
Majorana-induced half-destruction of the second-stage of the Kondo effect, hence 
the thermopower becomes suppressed. 

When the hopping between the dots is stronger [see Figs.~\ref{fig:S2D}(b,d,f)
for $S_\up$, $S_\down$ and $S$, respectively],
the Kondo effect hardly develops when the system is not coupled with the Majorana zero mode.
However, strong enough coupling to the superconducting nanowire 
induces additional minimum in thermopower for $\e<-U/2$
(maximum for $\e>-U/2$), which forms at $T \approx \Gamma_M$.
The energy at which a sign change occurs with further lowering the temperature
is equal in almost the entire range of DQD level position, see \fig{S2D}(f).
The parameter space at which the sign change is observed
is related to the minimum of conductance and Majorana-induced
increase of $G_\downarrow$ at low temperatures, cf. \fig{WFlaw2}.

We also note that the colormaps shown in \fig{S2D} allow one to easily
identify the new behavior associated with the presence of Majorana modes,
which is mostly revealed in the low-temperature behavior of $S_\down$
and then, consequently, in the corresponding dependence of $S$.

\new{The sign change of the thermopower is undoubtedly a notable result of this 
paper, showing unconventional effect arising from the coupling of the double quantum 
dot to topological superconductor.}

\subsection{Spin Seebeck effect due to Majorana proximity}
\label{sec:sSeebeck}

Since it is assumed that the Majorana mode couples to only one of the spin components
in the second quantum dot, it breaks the spin symmetry of the system
and may thus give rise to interesting spin-resolved thermoelectric phenomena, 
\new{such as nonzero spin thermopower (or spin Seebeck effect)}.
If the electrodes are characterized by long spin relaxation time,
spin accumulation may build up in the contacts,
such that the voltage generated by the temperature
gradient could become spin dependent $\Delta V_\sigma$.
In this case, the linear response current in the spin channel $\sigma$ is given by
\cite{Swirkowicz2009Nov}
\be
I_\sigma = e^2 L_{0\s} \Delta V +\eta e^2 L_{0\s} \Delta V_S - \frac{e}{T} L_{1\s} \Delta T,
\ee
where $\Delta V_S = (\Delta V_\up - \Delta V_\down)/2$
and $\eta = +1$ for $\sigma=\up$ and $\eta = -1$ for $\sigma=\down$.
There exist two definitions of the spin thermopower in the literature \cite{Swirkowicz2009Nov},
depending on the experimental implementation.
Defining the spin thermopower $S_S$ assuming open circuit conditions (vanishing of both
charge and spin currents), one gets \cite{Swirkowicz2009Nov,Weymann2013Aug}
$S_S = (S_\up - S_\down)/2$, while for $S$ one finds $S=(S_\up + S_\down)/2$.
On the other hand, if one requires that only the spin current vanishes
on the condition that the voltage gradient is zero $\Delta V=0$,
the spin thermopower is given by \cite{Misiorny2015Apr}
\be
\label{eq:SS}
S_S = -\frac{1}{eT} \frac{L_{1\up} - L_{1\down}}{L_0}.
\ee
In the following we study the behavior of the spin thermopower 
defined in the latter formula. We also note that since we have already discussed
in detail the spin-resolved components of thermopower $S_\s$ in previous sections,
the behavior of thermopower determined assuming open circuit conditions,
which is given either by a sum or difference of these components,
can be anticipated from previous results.

\subsubsection{Temperature dependence}

\begin{figure}[tb!]
	\includegraphics[width=0.9\columnwidth]{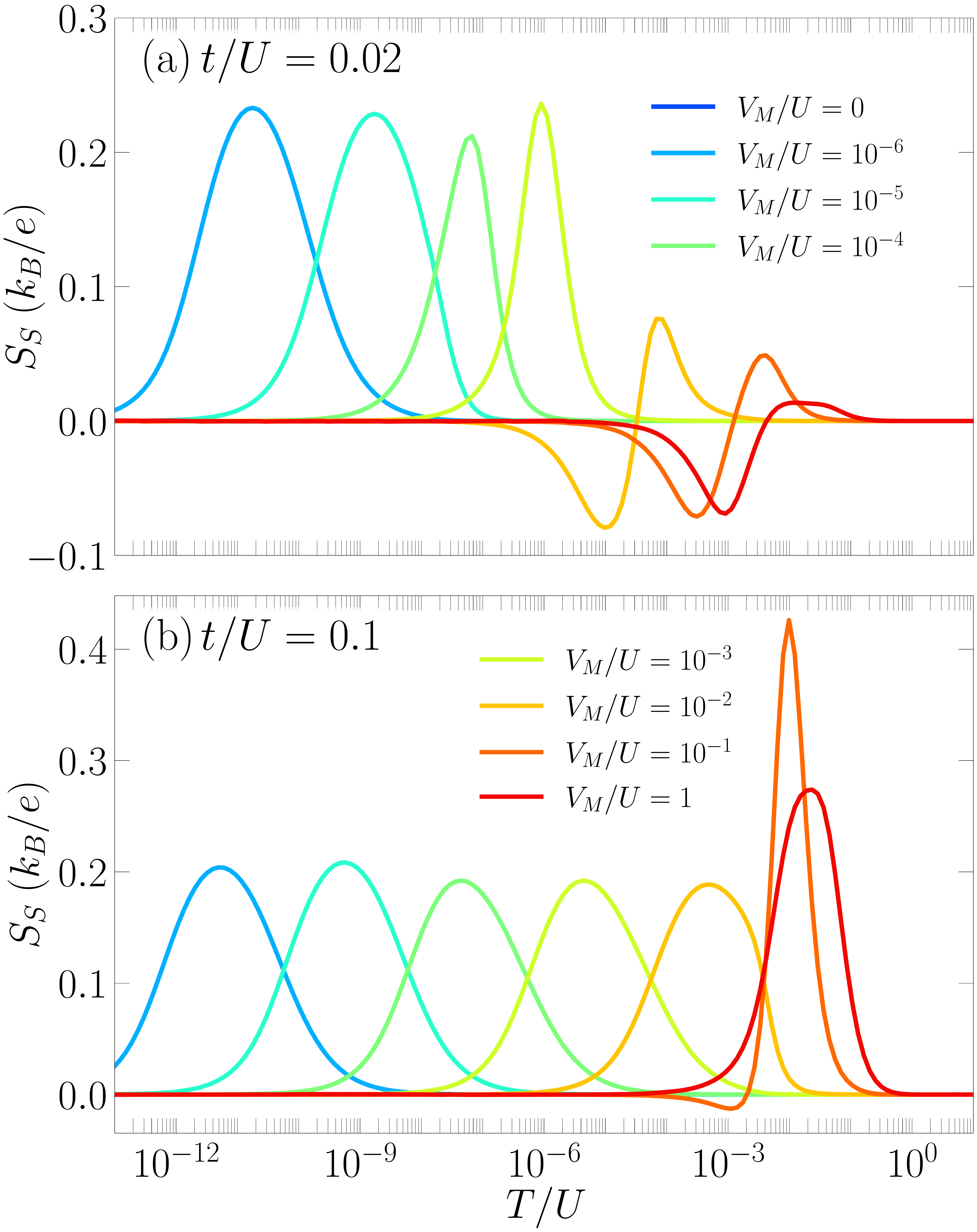}
	\caption{\label{fig:SS}
		The spin Seebeck coefficient $S_S$ plotted as a function of $T$ 
		for different values of $V_M$ as indicated and
		for (a) $t = 0.02U$ and (b) $t = 0.1 U$.
		The other parameters are the same as in \fig{WFlaw}.
	}
\end{figure}

Figure~\ref{fig:SS} presents the spin thermopower $S_S$
plotted as a function of temperature $T$ for multiple values of $V_M$, for 
both weak and strong value of hopping $t$ between the quantum dots.
Consider first the case where $t = 0.02 U$ [see \fig{SS}(a)].
Starting with $V_M = 0$, the spin Seebeck effect does not 
develop. Increasing the coupling to Majorana zero mode, a single positive peak 
appears at $T\approx \Gamma_M$, and remains when increasing $V_M$ until $\Gamma_M \approx T^*$.
Moreover, while the height of this maximum is qualitatively similar
for different values of $V_M$, its width becomes sharper (on logarithmic scale)
as $\Gamma_M$ becomes of the order of $T^*$. 
We can associate this phenomenon with the interplay of
the Majorana physics and the two-stage Kondo effect, where $\Gamma_M$ 
and $T^*$ energy scales are relevant. They limit the conductance gap due 
to the second quantum dot screening, what quantitatively affects the width of the peak. 
Increasing $V_M$, the gap becomes narrower as the 
difference between $T^*$ and $\Gamma_M$ decreases.
As a result, one observes the maximum of spin thermopower
shifting toward higher temperatures.
When $V_M$ is so large that $\Gamma_M$ reaches and exceeds $T^*$,
see the curves for $V_M\gtrsim 10^{-2}U$ in \fig{SS},
one observes a sudden sign change in $S_S$.
In this regime, the spin-up contribution becomes stronger, which is also
visible in the spin polarization of the current,
which changes sign to positive values as $V_M$ grows, cf. \fig{Pol}(a).
The height of minima (maxima) in $S_S$
strongly depends on the spin-down carriers of which the Majorana-Kondo 
interplay is visible with its characteristic half-suppression of $G_\downarrow$.

The case of strong interdot hopping, where $t = 0.1U$,
is shown in \fig{SS}(b). Now, the situation is rather different.
In this case, when the double quantum dot is decoupled from the topological nanowire,
the Kondo effect hardly develops due to strong singlet state, which forms between the dots.
Nonetheless, as in the situation of weak hopping discussed above,
turning on the coupling to Majorana zero mode changes this behavior significantly.
Finite $V_M$ gives rise to quantum interference in the spin-down channel,
resulting in $G_\downarrow = e^2/2h$.
This leads to a peak visible in spin thermopower.
The position of this peak follows the increase of $V_M$,
with approximately equal height and width (on the logarithmic scale)
until $\Gamma_M \approx 10^{-2}U$,
where both the width and height become changed, see \fig{SS}(b).

\subsubsection{Gate voltage dependence}

\begin{figure}[t]
	\includegraphics[width=\columnwidth]{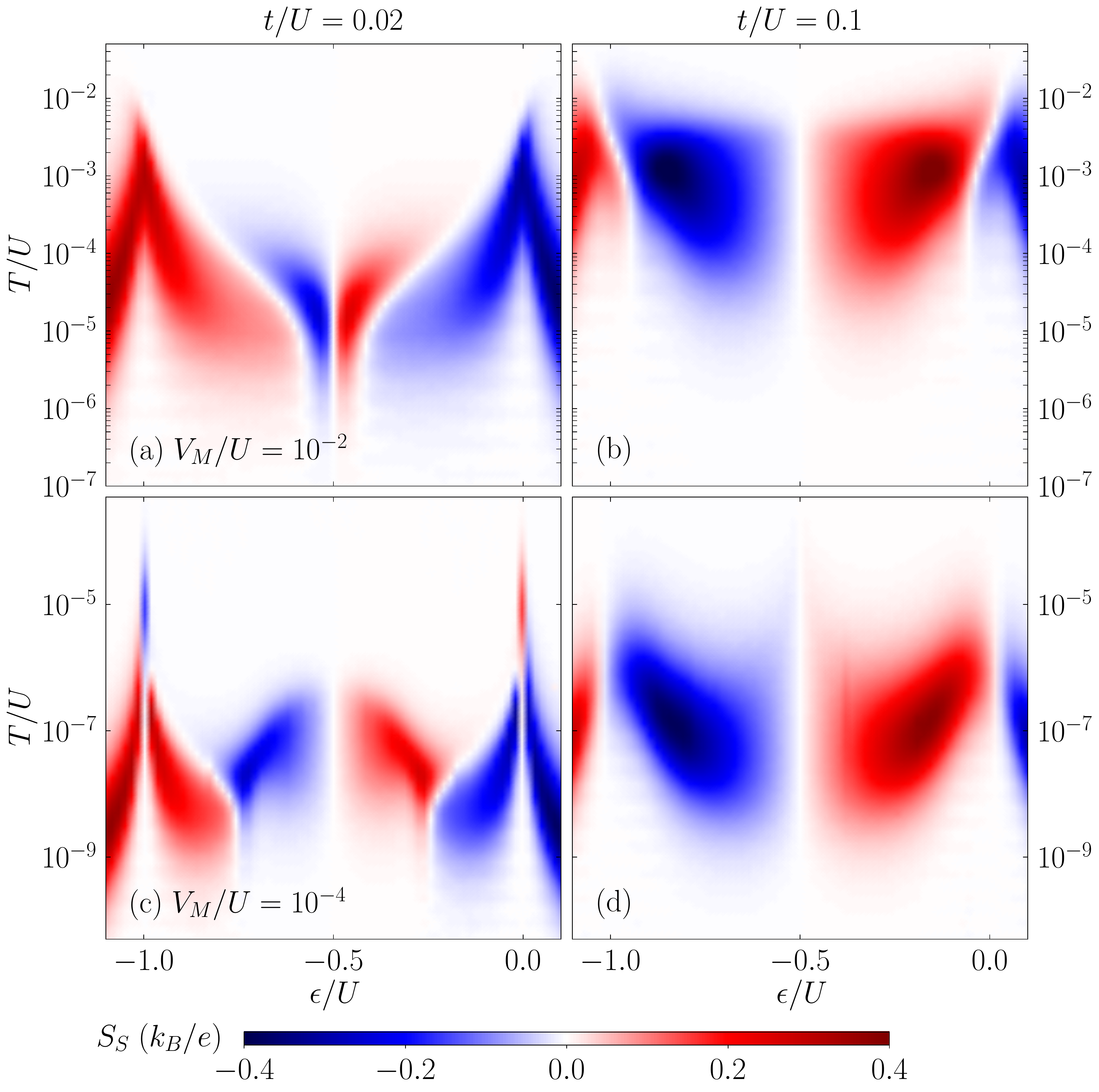}
	\caption{\label{fig:SS2D}
		The spin Seebeck effect as a function of $T$ and $\e$ for (a,c) $t = 0.02U$ and (b,d) $t = 0.1U$.
		The first row (a,b) is calculated for $V_M = 10^{-2}U$,
		while the second row (c,d) presents $S_S$ for $V_M = 10^{-4}U$. 
		The other parameters are the same as in \fig{WFlaw}.
	}
\end{figure} 

Figure~\ref{fig:SS2D} presents the behavior of the spin Seebeck effect as a function of the double dot levels $\e = 
\e_1 = \e_2$ and temperature $T$. The left (right) column is calculated in the case
of weak (strong) hopping between the dots, while the first (second) row
presents the data for $V_M=10^{-2}U$ ($V_M=10^{-4}U$).
Let us start the discussion with the case of weak interdot hopping.
When $V_M=10^{-2}U$, the spin-resolved conductance [cf. \fig{WFlaw}]
exhibits changes for $T\gtrsim 10^{-5}U$, while at lower temperatures it takes
constant value, with $G_\down\approx e^2/2h$ and $G_\up\approx e^2/h$.
Revoking the Mott formula, this helps to understand the temperature range where
non-zero spin Seebeck effect can emerge.
As one can see in the figure, around the particle-hole symmetry point
$S_S$ is either positive or negative depending on the sign of detuning
from $\e=-U/2$. In the case of $\e>-U/2$, $S_S$ exhibits a maximum 
which moves to higher temperatures as the detuning grows.
Moreover, for $\e\gtrsim -0.4U$, a sign change in the
$T$-dependence of the spin Seebeck effect develops, cf. also \fig{SS}(a).
When the detuning from the particle-hole symmetry point grows, the contributions
from the spin-up and spin-down channels become comparable,
which results in suppression and the corresponding reversal 
of the spin thermopower. Such a reversal is visible
in a narrow range of DQD level position, $-0.4U\lesssim \e \lesssim -0.2 U$
and $-0.8U\lesssim \e \lesssim -0.6 U$,
see \fig{SS2D}(a). Otherwise $S_S$ as a function of $T$
exhibits a large negative (positive) peak for $\e\gtrsim-0.2U$ ($\e\lesssim-0.8U$),
whose position strongly depends on $\e$.
When the coupling to Majorana wire is weaker,
see the case of $V_M=10^{-4}U$ presented in \fig{SS2D}(c),
the behavior of $S_S$ becomes greatly modified.
First of all, the characteristic temperature range where
spin thermopower emerges is shifted to lower $T$,
which is due to the fact that the characteristic energy scale
associated with the presence of Majorana mode is now reduced.
Moreover, the behavior around the particle-hole symmetry point
is extended over a wider range of $\e$.
This results from the fact that the Majorana
energy scale is smaller and the relevant interplay between
Majorana-induced interference and the Kondo correlations now takes 
place in larger parameter space of the system.

The situation when the interdot hopping is strong is shown in the right column of \fig{SS2D}.
First of all, one can see that the relevant behavior of $S_S$ is shifted to larger temperatures,
since now the second-stage Kondo temperature is of the order of the Kondo temperature.
Furthermore, contrary to the weak hopping limit, there is no effect where $S_S$ changes sign within 
the Coulomb valley. Instead, one can observe a single positive (or negative for $\e<-U/2$) 
peak of width depending on $\e$, whose extremum shifts toward higher temperatures
with detuning from the particle-hole symmetry point.
Such behavior results from the corresponding dependence
of the spin-resolved conductance [cf. \fig{WFlaw2}], which changes on the scale of $T\sim T^*$
and reaches a constant value for lower temperatures.
This peak for chosen parameter space reaches its maximum 
(minimum) for $\e \approx -0.15 U$ ($\e \approx -0.85 U$). Shifting away from 
this point toward the symmetry point, $S_S$ diminishes and the peak widens. 
A similar behavior can be observed when the Majorana coupling
is smaller, see the case of $V_M=10^{-4}U$ in \fig{SS2D}(d),
but now the temperature range where the spin thermopower
can be observed is shifted to lower temperatures.
Otherwise, the qualitative behavior is similar.

\new{To conclude, the generation of the spin thermopower is another important 
result of this paper. Together with the spin polarization, these effects provide
insight into the spin-dependent properties, characteristic of the model presented in this paper
and stemming from the coupling to topological superconductor.}

\begin{figure}[t]
	\includegraphics[width=0.9\columnwidth]{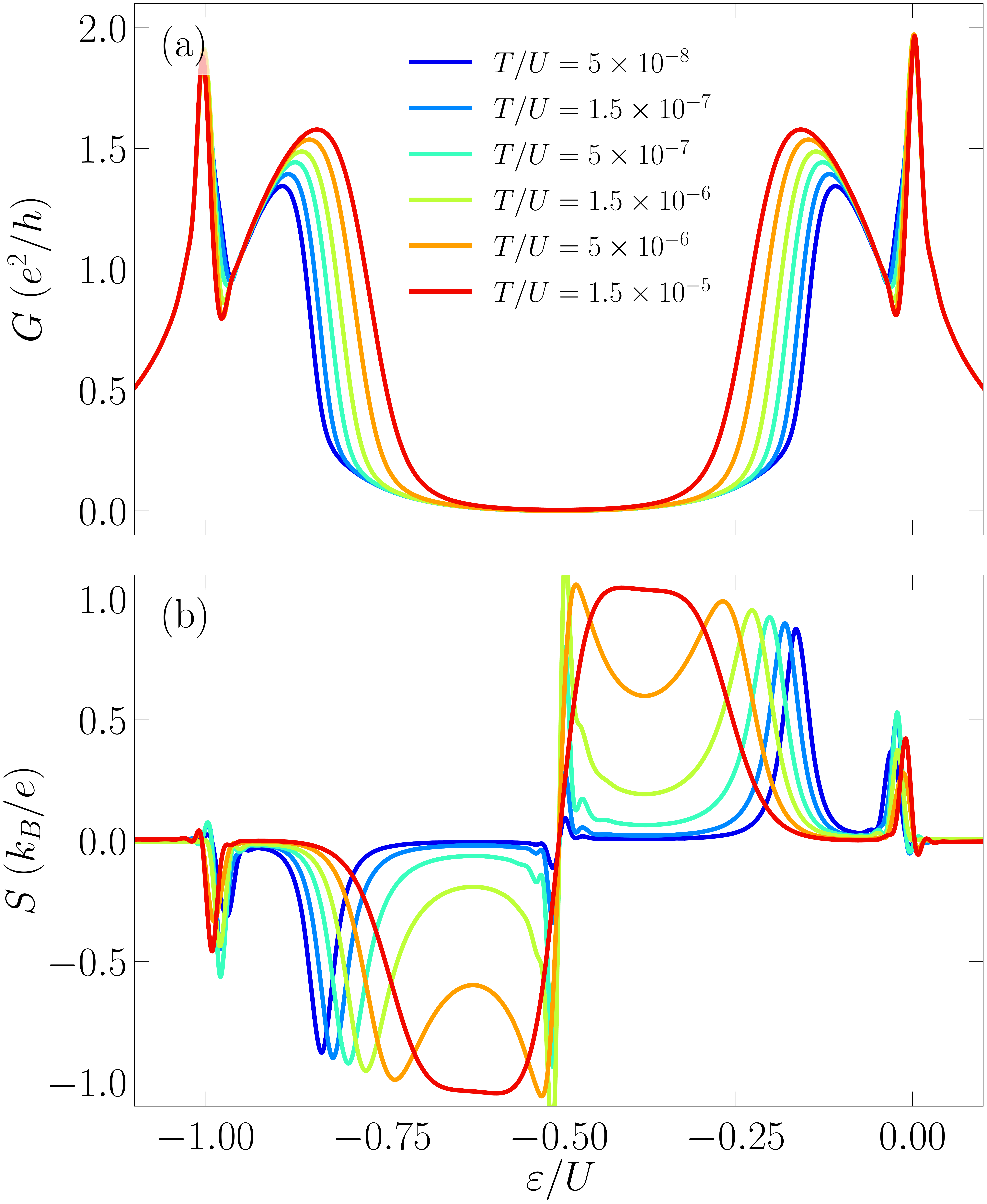}
	\caption{\label{fig:summaryVM0}
		The conductance and thermopower as a function of 
		$\e = \e_1 = \e_2$ for different values of $T$ and hopping 
		between the dots $t/U = 0.033$ in the case when the system is not coupled to
		the superconducting nanowire, $V_M= 0$.
	}
\end{figure}
\begin{figure}[t]
	\includegraphics[width=0.9\columnwidth]{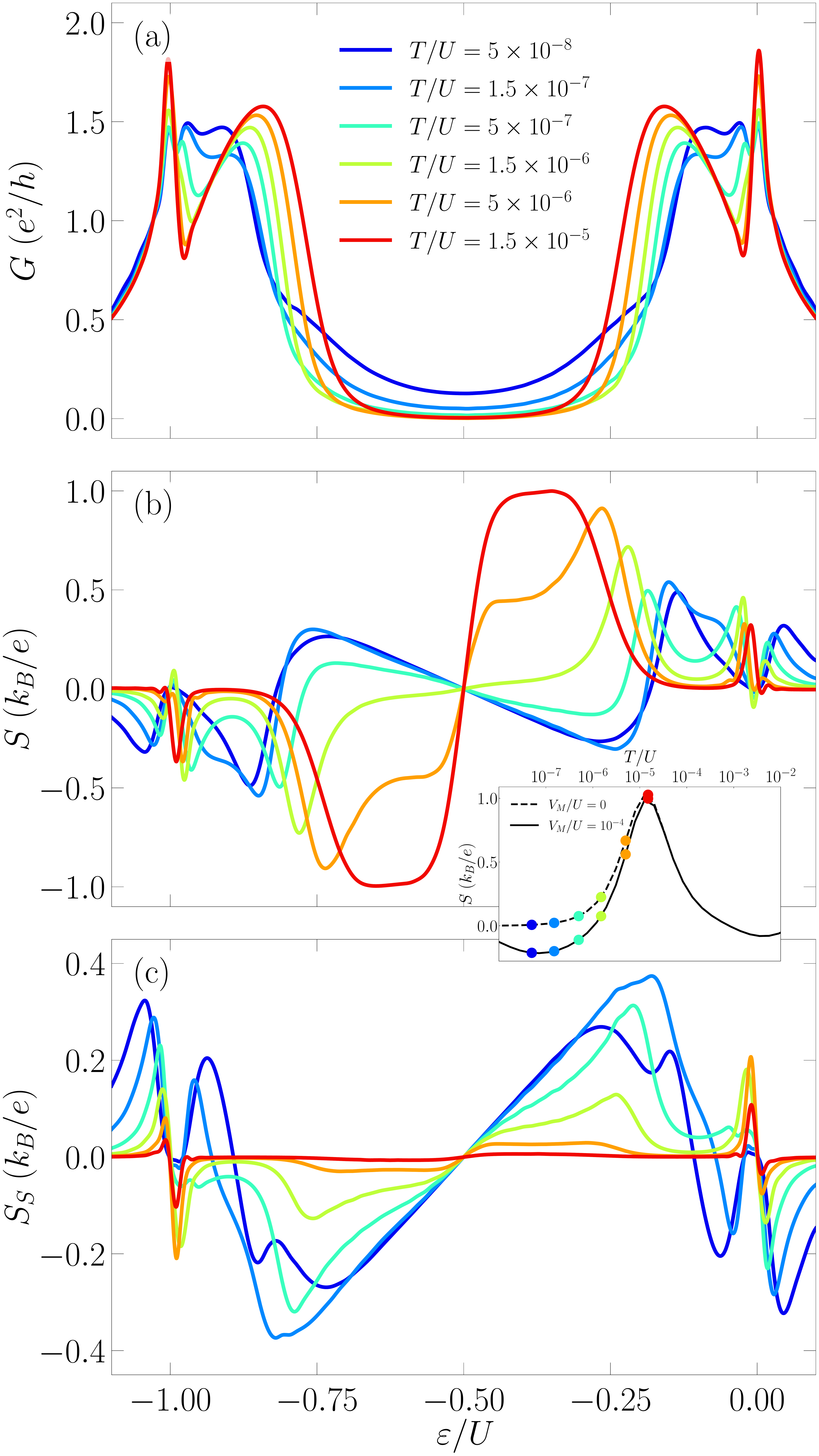}
	\caption{\label{fig:summaryVM}
		The conductance, thermopower and spin thermopower with the same 
		parameters as in \fig{summaryVM0}, but with $V_M/U = 10^{-4}$. The 
		inset shows the thermopower as a function of temperature, calculated for 
		$V_M/U = 0$ and $V_M/U = 10^{-4}$. The dots' colors correspond to the 
		temperatures shown in the legend of both figures. As can be seen, for 
		$T/U \gtrsim 10^{-5}$, $S$ hardly depends on $V_M$, and these values 
		of $T/U$ are not shown. Panel (c) displays the spin thermopower, which is 
		induced by the coupling to Majorana mode, thus this panel is not shown 
		for $V_M = 0$ in \fig{summaryVM0}.
	}
\end{figure} 

\section{Summary}
\label{sec:summmary}

We have studied the spin-resolved electrical and thermal signatures
of the interplay between the Majorana and Kondo physics in the case
of T-shaped double quantum dots attached to normal contacts
and to topological superconductor hosting Majorana zero energy modes.
To accurately resolve the transport behavior in the full parameter
space we have used the density-matrix numerical renormalization group method.
We have determined the behavior of the conductance, current spin polarization,
heat conductance as well as the Seebeck and spin Seebeck coefficients, focusing on the
parameter space where the system exhibits the Kondo correlations.
In our analysis, we have considered two specific cases of interdot hopping:
the weak hopping regime, where the two-stage Kondo effect develops,
and the strong hopping regime, where a molecular singlet state forms between the dots
allowing for only an onset of Kondo state in the system.

Analyzing the temperature dependence of the linear conductance,
we have shown that quantum interference with Majorana wire
breaks the second-stage of Kondo screening raising the conductance
to a fractional value of $G=e^2/2h$. We have also shown
that the presence of Majorana mode gives rise to finite spin polarization
of the current, which strongly depends on the magnitude of coupling to Majorana wire
and the position of the double dot energy levels.

As far as thermometric properties are concerned,
we have demonstrated that the system exhibits a modified Wiedemann-Franz law,
which is satisfied by the heat conductance {\it at a rescaled temperature}, also
in the presence of coupling to topological superconductor. 
Furthermore, we have analyzed the temperature
dependence of the spin-resolved Seebeck effect and shown that 
the thermopower exhibits additional sign change associated
with the presence of Majorana zero energy mode in the system.
Finally, assuming long spin relaxation time in the contacts,
we have examined the behavior of the spin Seebeck coefficient,
which emerges due to the presence of coupling to topological superconductor
and is a unique property of the considered Majorana-double quantum dot setup.
We have demonstrated that the spin thermopower nicely reveals the features 
resulting from the interplay of strong electron correlations
and Majorana-induced interference.

\new{
We would like to note that although the spin-resolved results presented in this paper
are challenging from experimental point of view, 
they have a clear fundamental relevance and are crucial to fully understand the system’s behavior,
especially how the coupling to Majorana wire affects the different spin components.
However, as far as experimental aspects are concerned,
we believe that the spin-resolved measurements could be performed by using e.g. spin-polarized STM \cite{Wiesendanger2009Nov}
or by attaching spin filters made by e.g. ferromagnetic electrodes \cite{Zutic2004Apr}.
We hope that our work will foster further efforts in spin-resolved
transport spectroscopy of hybrid nanostructures.

Finally, to emphasize the effects stemming from the presence of Majorana quasiparticles
and allow for their clear identification, in Figs.~\ref{fig:summaryVM0} and \ref{fig:summaryVM}
we present the dependence of the conductance and the (spin) Seebeck coefficient
on the position of dots levels for different values of temperatures.
Figure \ref{fig:summaryVM0} presents the data in the absence of coupling to Majorana wire,
while Fig. \ref{fig:summaryVM} shows the results in the presence of coupling to topological superconductor.
Indeed, the sign change of the thermopower due to finite $V_M$
is now clearly visible. Moreover, nonzero spin Seebeck coefficient emerges
only when the Majorana mode is present in the system.
}


\begin{acknowledgments}
This work was supported by the National Science Centre
in Poland through the Project No. 2018/29/B/ST3/00937.
The computing time at the Pozna\'n Supercomputing 
and Networking Center is acknowledged.
KPW acknowledges support from A. von Humboldt Foundation.
\end{acknowledgments}


\begin{thebibliography}{96}%
	\makeatletter
	\providecommand \@ifxundefined [1]{%
		\@ifx{#1\undefined}
	}%
	\providecommand \@ifnum [1]{%
		\ifnum #1\expandafter \@firstoftwo
		\else \expandafter \@secondoftwo
		\fi
	}%
	\providecommand \@ifx [1]{%
		\ifx #1\expandafter \@firstoftwo
		\else \expandafter \@secondoftwo
		\fi
	}%
	\providecommand \natexlab [1]{#1}%
	\providecommand \enquote  [1]{``#1''}%
	\providecommand \bibnamefont  [1]{#1}%
	\providecommand \bibfnamefont [1]{#1}%
	\providecommand \citenamefont [1]{#1}%
	\providecommand \href@noop [0]{\@secondoftwo}%
	\providecommand \href [0]{\begingroup \@sanitize@url \@href}%
	\providecommand \@href[1]{\@@startlink{#1}\@@href}%
	\providecommand \@@href[1]{\endgroup#1\@@endlink}%
	\providecommand \@sanitize@url [0]{\catcode `\\12\catcode `\$12\catcode
		`\&12\catcode `\#12\catcode `\^12\catcode `\_12\catcode `\%12\relax}%
	\providecommand \@@startlink[1]{}%
	\providecommand \@@endlink[0]{}%
	\providecommand \url  [0]{\begingroup\@sanitize@url \@url }%
	\providecommand \@url [1]{\endgroup\@href {#1}{\urlprefix }}%
	\providecommand \urlprefix  [0]{URL }%
	\providecommand \Eprint [0]{\href }%
	\providecommand \doibase [0]{https://doi.org/}%
	\providecommand \selectlanguage [0]{\@gobble}%
	\providecommand \bibinfo  [0]{\@secondoftwo}%
	\providecommand \bibfield  [0]{\@secondoftwo}%
	\providecommand \translation [1]{[#1]}%
	\providecommand \BibitemOpen [0]{}%
	\providecommand \bibitemStop [0]{}%
	\providecommand \bibitemNoStop [0]{.\EOS\space}%
	\providecommand \EOS [0]{\spacefactor3000\relax}%
	\providecommand \BibitemShut  [1]{\csname bibitem#1\endcsname}%
	\let\auto@bib@innerbib\@empty
	\bibitem [{\citenamefont {Majorana}(1937)}]{Majorana1937Apr}%
	\BibitemOpen
	\bibfield  {author} {\bibinfo {author} {\bibfnamefont {E.}~\bibnamefont
			{Majorana}},\ }\bibfield  {title} {\bibinfo {title} {{Teoria 
			simmetrica
				dell{'}elettrone e del positrone}},\ }\href
	{https://doi.org/10.1007/BF02961314} {\bibfield  {journal} {\bibinfo
			{journal} {Nuovo Cim.}\ }\textbf {\bibinfo {volume} {14}},\ 
			\bibinfo {pages}
		{171} (\bibinfo {year} {1937})}\BibitemShut {NoStop}%
	\bibitem [{\citenamefont {Kitaev}(2001)}]{Kitaev2001}%
	\BibitemOpen
	\bibfield  {author} {\bibinfo {author} {\bibfnamefont {A.~Y.}\ \bibnamefont
			{Kitaev}},\ }\bibfield  {title} {\bibinfo {title} {{Unpaired 
			Majorana
				fermions in quantum wires}},\ }\href
	{https://doi.org/10.1070/1063-7869/44/10s/s29} {\bibfield  {journal}
		{\bibinfo  {journal} {Phys. Usp.}\ }\textbf {\bibinfo {volume} {44}},\
		\bibinfo {pages} {131} (\bibinfo {year} {2001})}\BibitemShut {NoStop}%
	\bibitem [{\citenamefont {Kitaev}(2003)}]{Kitaev2003}%
	\BibitemOpen
	\bibfield  {author} {\bibinfo {author} {\bibfnamefont {A.~{\relax Yu}.}\
			\bibnamefont {Kitaev}},\ }\bibfield  {title} {\bibinfo {title}
		{{Fault-tolerant quantum computation by anyons}},\ }\href
	{https://doi.org/10.1016/S0003-4916(02)00018-0} {\bibfield  {journal}
		{\bibinfo  {journal} {Ann. Phys.}\ }\textbf {\bibinfo {volume} {303}},\
		\bibinfo {pages} {2} (\bibinfo {year} {2003})}\BibitemShut {NoStop}%
	\bibitem [{\citenamefont {Fu}\ and\ \citenamefont {Kane}(2008)}]{Fu2008Mar}%
	\BibitemOpen
	\bibfield  {author} {\bibinfo {author} {\bibfnamefont {L.}~\bibnamefont
			{Fu}}\ and\ \bibinfo {author} {\bibfnamefont {C.~L.}\ \bibnamefont 
			{Kane}},\
	}\bibfield  {title} {\bibinfo {title} {{Superconducting Proximity Effect and
				Majorana Fermions at the Surface of a Topological Insulator}},\ 
				}\href
	{https://doi.org/10.1103/PhysRevLett.100.096407} {\bibfield  {journal}
		{\bibinfo  {journal} {Phys. Rev. Lett.}\ }\textbf {\bibinfo {volume} 
		{100}},\
		\bibinfo {pages} {096407} (\bibinfo {year} {2008})}\BibitemShut 
		{NoStop}%
	\bibitem [{\citenamefont {Nayak}\ \emph {et~al.}(2008)\citenamefont {Nayak},
		\citenamefont {Simon}, \citenamefont {Stern}, \citenamefont 
		{Freedman},\ and\
		\citenamefont {Das~Sarma}}]{Nayak2008Sep}%
	\BibitemOpen
	\bibfield  {author} {\bibinfo {author} {\bibfnamefont {C.}~\bibnamefont
			{Nayak}}, \bibinfo {author} {\bibfnamefont {S.~H.}\ \bibnamefont 
			{Simon}},
		\bibinfo {author} {\bibfnamefont {A.}~\bibnamefont {Stern}}, \bibinfo
		{author} {\bibfnamefont {M.}~\bibnamefont {Freedman}},\ and\ \bibinfo
		{author} {\bibfnamefont {S.}~\bibnamefont {Das~Sarma}},\ }\bibfield  
		{title}
	{\bibinfo {title} {{Non-Abelian anyons and topological quantum
				computation}},\ }\href 
				{https://doi.org/10.1103/RevModPhys.80.1083}
	{\bibfield  {journal} {\bibinfo  {journal} {Rev. Mod. Phys.}\ }\textbf
		{\bibinfo {volume} {80}},\ \bibinfo {pages} {1083} (\bibinfo {year}
		{2008})}\BibitemShut {NoStop}%
	\bibitem [{\citenamefont {Beenakker}(2020)}]{BraidingReview}%
	\BibitemOpen
	\bibfield  {author} {\bibinfo {author} {\bibfnamefont {C.~W.~J.}\
			\bibnamefont {Beenakker}},\ }\bibfield  {title} {\bibinfo {title} 
			{{Search
				for non-Abelian Majorana braiding statistics in 
				superconductors}},\ }\href
	{https://doi.org/10.21468/SciPostPhysLectNotes.15} {\bibfield  {journal}
		{\bibinfo  {journal} {SciPost Phys. Lect. Notes}\ ,\ \bibinfo {pages} 
		{15}}
		(\bibinfo {year} {2020})}\BibitemShut {NoStop}%
	\bibitem [{\citenamefont {Hasan}\ and\ \citenamefont
		{Kane}(2010)}]{Hasan2010Nov}%
	\BibitemOpen
	\bibfield  {author} {\bibinfo {author} {\bibfnamefont {M.~Z.}\ \bibnamefont
			{Hasan}}\ and\ \bibinfo {author} {\bibfnamefont {C.~L.}\ 
			\bibnamefont
			{Kane}},\ }\bibfield  {title} {\bibinfo {title} {{Colloquium: 
			Topological
				insulators}},\ }\href 
				{https://doi.org/10.1103/RevModPhys.82.3045} {\bibfield
		{journal} {\bibinfo  {journal} {Rev. Mod. Phys.}\ }\textbf {\bibinfo
			{volume} {82}},\ \bibinfo {pages} {3045} (\bibinfo {year}
		{2010})}\BibitemShut {NoStop}%
	\bibitem [{\citenamefont {Qi}\ and\ \citenamefont 
	{Zhang}(2011)}]{Qi2011Oct}%
	\BibitemOpen
	\bibfield  {author} {\bibinfo {author} {\bibfnamefont {X.-L.}\ \bibnamefont
			{Qi}}\ and\ \bibinfo {author} {\bibfnamefont {S.-C.}\ \bibnamefont 
			{Zhang}},\
	}\bibfield  {title} {\bibinfo {title} {{Topological insulators and
				superconductors}},\ }\href 
				{https://doi.org/10.1103/RevModPhys.83.1057}
	{\bibfield  {journal} {\bibinfo  {journal} {Rev. Mod. Phys.}\ }\textbf
		{\bibinfo {volume} {83}},\ \bibinfo {pages} {1057} (\bibinfo {year}
		{2011})}\BibitemShut {NoStop}%
	\bibitem [{\citenamefont {Wang}\ and\ \citenamefont
		{Zhang}(2017)}]{Wang2017Nov}%
	\BibitemOpen
	\bibfield  {author} {\bibinfo {author} {\bibfnamefont {J.}~\bibnamefont
			{Wang}}\ and\ \bibinfo {author} {\bibfnamefont {S.-C.}\ \bibnamefont
			{Zhang}},\ }\bibfield  {title} {\bibinfo {title} {{Topological 
			states of
				condensed matter}},\ }\href {https://doi.org/10.1038/nmat5012} 
				{\bibfield
		{journal} {\bibinfo  {journal} {Nat. Mater.}\ }\textbf {\bibinfo 
		{volume}
			{16}},\ \bibinfo {pages} {1062} (\bibinfo {year} 
			{2017})}\BibitemShut
	{NoStop}%
	\bibitem [{\citenamefont {Sato}\ and\ \citenamefont
		{Ando}(2017)}]{Sato2017Rev}%
	\BibitemOpen
	\bibfield  {author} {\bibinfo {author} {\bibfnamefont {M.}~\bibnamefont
			{Sato}}\ and\ \bibinfo {author} {\bibfnamefont {Y.}~\bibnamefont 
			{Ando}},\
	}\bibfield  {title} {\bibinfo {title} {{Topological superconductors: a
				review}},\ }\href {https://doi.org/10.1088/1361-6633/aa6ac7} 
				{\bibfield
		{journal} {\bibinfo  {journal} {Rep. Prog. Phys.}\ }\textbf {\bibinfo
			{volume} {80}},\ \bibinfo {pages} {076501} (\bibinfo {year}
		{2017})}\BibitemShut {NoStop}%
	\bibitem [{\citenamefont {Mourik}\ \emph {et~al.}(2012)\citenamefont 
	{Mourik},
		\citenamefont {Zuo}, \citenamefont {Frolov}, \citenamefont {Plissard},
		\citenamefont {Bakkers},\ and\ \citenamefont 
		{Kouwenhoven}}]{Mourik2012May}%
	\BibitemOpen
	\bibfield  {author} {\bibinfo {author} {\bibfnamefont {V.}~\bibnamefont
			{Mourik}}, \bibinfo {author} {\bibfnamefont {K.}~\bibnamefont 
			{Zuo}},
		\bibinfo {author} {\bibfnamefont {S.~M.}\ \bibnamefont {Frolov}}, 
		\bibinfo
		{author} {\bibfnamefont {S.~R.}\ \bibnamefont {Plissard}}, \bibinfo 
		{author}
		{\bibfnamefont {E.~P. A.~M.}\ \bibnamefont {Bakkers}},\ and\ \bibinfo
		{author} {\bibfnamefont {L.~P.}\ \bibnamefont {Kouwenhoven}},\ 
		}\bibfield
	{title} {\bibinfo {title} {{Signatures of Majorana Fermions in Hybrid
				Superconductor-Semiconductor Nanowire Devices}},\ }\href
	{https://doi.org/10.1126/science.1222360} {\bibfield  {journal} {\bibinfo
			{journal} {Science}\ }\textbf {\bibinfo {volume} {336}},\ \bibinfo 
			{pages}
		{1003} (\bibinfo {year} {2012})}\BibitemShut {NoStop}%
	\bibitem [{\citenamefont {Lutchyn}\ \emph {et~al.}(2010)\citenamefont
		{Lutchyn}, \citenamefont {Sau},\ and\ \citenamefont
		{Das~Sarma}}]{Lutchyn2010Aug}%
	\BibitemOpen
	\bibfield  {author} {\bibinfo {author} {\bibfnamefont {R.~M.}\ \bibnamefont
			{Lutchyn}}, \bibinfo {author} {\bibfnamefont {J.~D.}\ \bibnamefont 
			{Sau}},\
		and\ \bibinfo {author} {\bibfnamefont {S.}~\bibnamefont {Das~Sarma}},\
	}\bibfield  {title} {\bibinfo {title} {{Majorana Fermions and a Topological
				Phase Transition in Semiconductor-Superconductor 
				Heterostructures}},\ }\href
	{https://doi.org/10.1103/PhysRevLett.105.077001} {\bibfield  {journal}
		{\bibinfo  {journal} {Phys. Rev. Lett.}\ }\textbf {\bibinfo {volume} 
		{105}},\
		\bibinfo {pages} {077001} (\bibinfo {year} {2010})}\BibitemShut 
		{NoStop}%
	\bibitem [{\citenamefont {Oreg}\ \emph {et~al.}(2010)\citenamefont {Oreg},
		\citenamefont {Refael},\ and\ \citenamefont {von Oppen}}]{Oreg2010Oct}%
	\BibitemOpen
	\bibfield  {author} {\bibinfo {author} {\bibfnamefont {Y.}~\bibnamefont
			{Oreg}}, \bibinfo {author} {\bibfnamefont {G.}~\bibnamefont 
			{Refael}},\ and\
		\bibinfo {author} {\bibfnamefont {F.}~\bibnamefont {von Oppen}},\ 
		}\bibfield
	{title} {\bibinfo {title} {{Helical Liquids and Majorana Bound States in
				Quantum Wires}},\ }\href 
				{https://doi.org/10.1103/PhysRevLett.105.177002}
	{\bibfield  {journal} {\bibinfo  {journal} {Phys. Rev. Lett.}\ }\textbf
		{\bibinfo {volume} {105}},\ \bibinfo {pages} {177002} (\bibinfo {year}
		{2010})}\BibitemShut {NoStop}%
	\bibitem [{\citenamefont {Lutchyn}\ \emph {et~al.}(2018)\citenamefont
		{Lutchyn}, \citenamefont {Bakkers}, \citenamefont {Kouwenhoven},
		\citenamefont {Krogstrup}, \citenamefont {Marcus},\ and\ \citenamefont
		{Oreg}}]{Lutchyn2018May}%
	\BibitemOpen
	\bibfield  {author} {\bibinfo {author} {\bibfnamefont {R.~M.}\ \bibnamefont
			{Lutchyn}}, \bibinfo {author} {\bibfnamefont {E.~P. A.~M.}\ 
			\bibnamefont
			{Bakkers}}, \bibinfo {author} {\bibfnamefont {L.~P.}\ \bibnamefont
			{Kouwenhoven}}, \bibinfo {author} {\bibfnamefont {P.}~\bibnamefont
			{Krogstrup}}, \bibinfo {author} {\bibfnamefont {C.~M.}\ \bibnamefont
			{Marcus}},\ and\ \bibinfo {author} {\bibfnamefont {Y.}~\bibnamefont 
			{Oreg}},\
	}\bibfield  {title} {\bibinfo {title} {{Majorana zero modes in
				superconductor{\textendash}semiconductor heterostructures}},\ 
				}\href
	{https://doi.org/10.1038/s41578-018-0003-1} {\bibfield  {journal} {\bibinfo
			{journal} {Nat. Rev. Mater.}\ }\textbf {\bibinfo {volume} {3}},\ 
			\bibinfo
		{pages} {52} (\bibinfo {year} {2018})}\BibitemShut {NoStop}%
	\bibitem [{\citenamefont {Prada}\ \emph {et~al.}(2020)\citenamefont {Prada},
		\citenamefont {San-Jose}, \citenamefont {de~Moor}, \citenamefont 
		{Geresdi},
		\citenamefont {Lee}, \citenamefont {Klinovaja}, \citenamefont {Loss},
		\citenamefont {Nyg{\aa}rd}, \citenamefont {Aguado},\ and\ \citenamefont
		{Kouwenhoven}}]{Prada2020Rev}%
	\BibitemOpen
	\bibfield  {author} {\bibinfo {author} {\bibfnamefont {E.}~\bibnamefont
			{Prada}}, \bibinfo {author} {\bibfnamefont {P.}~\bibnamefont 
			{San-Jose}},
		\bibinfo {author} {\bibfnamefont {M.~W.~A.}\ \bibnamefont {de~Moor}},
		\bibinfo {author} {\bibfnamefont {A.}~\bibnamefont {Geresdi}}, \bibinfo
		{author} {\bibfnamefont {E.~J.~H.}\ \bibnamefont {Lee}}, \bibinfo 
		{author}
		{\bibfnamefont {J.}~\bibnamefont {Klinovaja}}, \bibinfo {author}
		{\bibfnamefont {D.}~\bibnamefont {Loss}}, \bibinfo {author} 
		{\bibfnamefont
			{J.}~\bibnamefont {Nyg{\aa}rd}}, \bibinfo {author} {\bibfnamefont
			{R.}~\bibnamefont {Aguado}},\ and\ \bibinfo {author} {\bibfnamefont 
			{L.~P.}\
			\bibnamefont {Kouwenhoven}},\ }\bibfield  {title} {\bibinfo {title} 
			{{From
				Andreev to Majorana bound states in hybrid
				superconductor{\textendash}semiconductor nanowires}},\ }\href
	{https://doi.org/10.1038/s42254-020-0228-y} {\bibfield  {journal} {\bibinfo
			{journal} {Nat. Rev. Phys.}\ }\textbf {\bibinfo {volume} {2}},\ 
			\bibinfo
		{pages} {575} (\bibinfo {year} {2020})}\BibitemShut {NoStop}%
	\bibitem [{\citenamefont {Lee}\ \emph 
	{et~al.}(2013{\natexlab{a}})\citenamefont
		{Lee}, \citenamefont {Jiang}, \citenamefont {Houzet}, \citenamefont 
		{Aguado},
		\citenamefont {Lieber},\ and\ \citenamefont 
		{De~Franceschi}}]{Lee2013Dec}%
	\BibitemOpen
	\bibfield  {author} {\bibinfo {author} {\bibfnamefont {E.~J.~H.}\
			\bibnamefont {Lee}}, \bibinfo {author} {\bibfnamefont 
			{X.}~\bibnamefont
			{Jiang}}, \bibinfo {author} {\bibfnamefont {M.}~\bibnamefont 
			{Houzet}},
		\bibinfo {author} {\bibfnamefont {R.}~\bibnamefont {Aguado}}, \bibinfo
		{author} {\bibfnamefont {C.~M.}\ \bibnamefont {Lieber}},\ and\ \bibinfo
		{author} {\bibfnamefont {S.}~\bibnamefont {De~Franceschi}},\ }\bibfield
	{title} {\bibinfo {title} {{Spin-resolved Andreev levels and parity 
	crossings
				in hybrid superconductor{\textendash}semiconductor 
				nanostructures}},\ }\href
	{https://doi.org/10.1038/nnano.2013.267} {\bibfield  {journal} {\bibinfo
			{journal} {Nat. Nanotechnol.}\ }\textbf {\bibinfo {volume} {9}},\ 
			\bibinfo
		{pages} {79} (\bibinfo {year} {2013}{\natexlab{a}})}\BibitemShut 
		{NoStop}%
	\bibitem [{\citenamefont {Kells}\ \emph {et~al.}(2012)\citenamefont {Kells},
		\citenamefont {Meidan},\ and\ \citenamefont {Brouwer}}]{Kells2012Sep}%
	\BibitemOpen
	\bibfield  {author} {\bibinfo {author} {\bibfnamefont {G.}~\bibnamefont
			{Kells}}, \bibinfo {author} {\bibfnamefont {D.}~\bibnamefont 
			{Meidan}},\ and\
		\bibinfo {author} {\bibfnamefont {P.~W.}\ \bibnamefont {Brouwer}},\
	}\bibfield  {title} {\bibinfo {title} {{Near-zero-energy end states in
				topologically trivial spin-orbit coupled superconducting 
				nanowires with a
				smooth confinement}},\ }\href 
				{https://doi.org/10.1103/PhysRevB.86.100503}
	{\bibfield  {journal} {\bibinfo  {journal} {Phys. Rev. B}\ }\textbf 
	{\bibinfo
			{volume} {86}},\ \bibinfo {pages} {100503(R)} (\bibinfo {year}
		{2012})}\BibitemShut {NoStop}%
	\bibitem [{\citenamefont {Wang}\ \emph {et~al.}(2021)\citenamefont {Wang},
		\citenamefont {Wiebe}, \citenamefont {Zhong}, \citenamefont {Gu},\ and\
		\citenamefont {Wiesendanger}}]{Wang2021Feb}%
	\BibitemOpen
	\bibfield  {author} {\bibinfo {author} {\bibfnamefont {D.}~\bibnamefont
			{Wang}}, \bibinfo {author} {\bibfnamefont {J.}~\bibnamefont 
			{Wiebe}},
		\bibinfo {author} {\bibfnamefont {R.}~\bibnamefont {Zhong}}, \bibinfo
		{author} {\bibfnamefont {G.}~\bibnamefont {Gu}},\ and\ \bibinfo {author}
		{\bibfnamefont {R.}~\bibnamefont {Wiesendanger}},\ }\bibfield  {title}
	{\bibinfo {title} {{Spin-Polarized Yu-Shiba-Rusinov States in an Iron-Based
				Superconductor}},\ }\href 
				{https://doi.org/10.1103/PhysRevLett.126.076802}
	{\bibfield  {journal} {\bibinfo  {journal} {Phys. Rev. Lett.}\ }\textbf
		{\bibinfo {volume} {126}},\ \bibinfo {pages} {076802} (\bibinfo {year}
		{2021})}\BibitemShut {NoStop}%
	\bibitem [{\citenamefont {Pikulin}\ \emph {et~al.}(2012)\citenamefont
		{Pikulin}, \citenamefont {Dahlhaus}, \citenamefont {Wimmer}, 
		\citenamefont
		{Schomerus},\ and\ \citenamefont {Beenakker}}]{Pikulin2012Dec}%
	\BibitemOpen
	\bibfield  {author} {\bibinfo {author} {\bibfnamefont {D.~I.}\ \bibnamefont
			{Pikulin}}, \bibinfo {author} {\bibfnamefont {J.~P.}\ \bibnamefont
			{Dahlhaus}}, \bibinfo {author} {\bibfnamefont {M.}~\bibnamefont 
			{Wimmer}},
		\bibinfo {author} {\bibfnamefont {H.}~\bibnamefont {Schomerus}},\ and\
		\bibinfo {author} {\bibfnamefont {C.~W.~J.}\ \bibnamefont {Beenakker}},\
	}\bibfield  {title} {\bibinfo {title} {{A zero-voltage conductance peak from
				weak antilocalization in a Majorana nanowire}},\ }\href
	{https://doi.org/10.1088/1367-2630/14/12/125011} {\bibfield  {journal}
		{\bibinfo  {journal} {New J. Phys.}\ }\textbf {\bibinfo {volume} {14}},\
		\bibinfo {pages} {125011} (\bibinfo {year} {2012})}\BibitemShut 
		{NoStop}%
	\bibitem [{\citenamefont {Borsoi}\ \emph {et~al.}(2020)\citenamefont 
	{Borsoi},
		\citenamefont {Zuo}, \citenamefont {Gazibegovic}, \citenamefont {Op~het
			Veld}, \citenamefont {Bakkers}, \citenamefont {Kouwenhoven},\ and\
		\citenamefont {Heedt}}]{Borsoi2020Jul}%
	\BibitemOpen
	\bibfield  {author} {\bibinfo {author} {\bibfnamefont {F.}~\bibnamefont
			{Borsoi}}, \bibinfo {author} {\bibfnamefont {K.}~\bibnamefont 
			{Zuo}},
		\bibinfo {author} {\bibfnamefont {S.}~\bibnamefont {Gazibegovic}}, 
		\bibinfo
		{author} {\bibfnamefont {R.~L.~M.}\ \bibnamefont {Op~het Veld}}, 
		\bibinfo
		{author} {\bibfnamefont {E.~P. A.~M.}\ \bibnamefont {Bakkers}}, \bibinfo
		{author} {\bibfnamefont {L.~P.}\ \bibnamefont {Kouwenhoven}},\ and\ 
		\bibinfo
		{author} {\bibfnamefont {S.}~\bibnamefont {Heedt}},\ }\bibfield  {title}
	{\bibinfo {title} {{Transmission phase read-out of a large quantum dot in a
				nanowire interferometer}},\ }\href
	{https://doi.org/10.1038/s41467-020-17461-5} {\bibfield  {journal} {\bibinfo
			{journal} {Nat. Commun.}\ }\textbf {\bibinfo {volume} {11}},\ 
			\bibinfo
		{pages} {1} (\bibinfo {year} {2020})}\BibitemShut {NoStop}%
	\bibitem [{\citenamefont {Whiticar}\ \emph {et~al.}(2020)\citenamefont
		{Whiticar}, \citenamefont {Fornieri}, \citenamefont {O{'}Farrell},
		\citenamefont {Drachmann}, \citenamefont {Wang}, \citenamefont {Thomas},
		\citenamefont {Gronin}, \citenamefont {Kallaher}, \citenamefont 
		{Gardner},
		\citenamefont {Manfra}, \citenamefont {Marcus},\ and\ \citenamefont
		{Nichele}}]{Whiticar2020Jun}%
	\BibitemOpen
	\bibfield  {author} {\bibinfo {author} {\bibfnamefont {A.~M.}\ \bibnamefont
			{Whiticar}}, \bibinfo {author} {\bibfnamefont {A.}~\bibnamefont 
			{Fornieri}},
		\bibinfo {author} {\bibfnamefont {E.~C.~T.}\ \bibnamefont 
		{O{'}Farrell}},
		\bibinfo {author} {\bibfnamefont {A.~C.~C.}\ \bibnamefont {Drachmann}},
		\bibinfo {author} {\bibfnamefont {T.}~\bibnamefont {Wang}}, \bibinfo 
		{author}
		{\bibfnamefont {C.}~\bibnamefont {Thomas}}, \bibinfo {author} 
		{\bibfnamefont
			{S.}~\bibnamefont {Gronin}}, \bibinfo {author} {\bibfnamefont
			{R.}~\bibnamefont {Kallaher}}, \bibinfo {author} {\bibfnamefont 
			{G.~C.}\
			\bibnamefont {Gardner}}, \bibinfo {author} {\bibfnamefont {M.~J.}\
			\bibnamefont {Manfra}}, \bibinfo {author} {\bibfnamefont {C.~M.}\
			\bibnamefont {Marcus}},\ and\ \bibinfo {author} {\bibfnamefont
			{F.}~\bibnamefont {Nichele}},\ }\bibfield  {title} {\bibinfo {title}
		{{Coherent transport through a Majorana island in an
				Aharonov{\textendash}Bohm interferometer}},\ }\href
	{https://doi.org/10.1038/s41467-020-16988-x} {\bibfield  {journal} {\bibinfo
			{journal} {Nat. Commun.}\ }\textbf {\bibinfo {volume} {11}},\ 
			\bibinfo
		{pages} {1} (\bibinfo {year} {2020})}\BibitemShut {NoStop}%
	\bibitem [{\citenamefont {Avila}\ \emph {et~al.}(2019)\citenamefont {Avila},
		\citenamefont {Pe{\ifmmode\tilde{n}\else\~{n}\fi}aranda}, \citenamefont
		{Prada}, \citenamefont {San-Jose},\ and\ \citenamefont
		{Aguado}}]{Avila2019Oct}%
	\BibitemOpen
	\bibfield  {author} {\bibinfo {author} {\bibfnamefont {J.}~\bibnamefont
			{Avila}}, \bibinfo {author} {\bibfnamefont {F.}~\bibnamefont
			{Pe{\ifmmode\tilde{n}\else\~{n}\fi}aranda}}, \bibinfo {author} 
			{\bibfnamefont
			{E.}~\bibnamefont {Prada}}, \bibinfo {author} {\bibfnamefont
			{P.}~\bibnamefont {San-Jose}},\ and\ \bibinfo {author} 
			{\bibfnamefont
			{R.}~\bibnamefont {Aguado}},\ }\bibfield  {title} {\bibinfo {title}
		{{Non-hermitian topology as a unifying framework for the Andreev versus
				Majorana states controversy}},\ }\href
	{https://doi.org/10.1038/s42005-019-0231-8} {\bibfield  {journal} {\bibinfo
			{journal} {Commun. Phys.}\ }\textbf {\bibinfo {volume} {2}},\ 
			\bibinfo
		{pages} {1} (\bibinfo {year} {2019})}\BibitemShut {NoStop}%
	\bibitem [{\citenamefont {{\ifmmode\acute{A}\else\'{A}\fi}vila}\ \emph
		{et~al.}(2020)\citenamefont {{\ifmmode\acute{A}\else\'{A}\fi}vila},
		\citenamefont {Prada}, \citenamefont {San-Jose},\ and\ \citenamefont
		{Aguado}}]{Avila2020Sep}%
	\BibitemOpen
	\bibfield  {author} {\bibinfo {author} {\bibfnamefont {J.}~\bibnamefont
			{{\ifmmode\acute{A}\else\'{A}\fi}vila}}, \bibinfo {author} 
			{\bibfnamefont
			{E.}~\bibnamefont {Prada}}, \bibinfo {author} {\bibfnamefont
			{P.}~\bibnamefont {San-Jose}},\ and\ \bibinfo {author} 
			{\bibfnamefont
			{R.}~\bibnamefont {Aguado}},\ }\bibfield  {title} {\bibinfo {title}
		{{Majorana oscillations and parity crossings in semiconductor 
		nanowire-based
				transmon qubits}},\ }\href 
				{https://doi.org/10.1103/PhysRevResearch.2.033493}
	{\bibfield  {journal} {\bibinfo  {journal} {Phys. Rev. Res.}\ }\textbf
		{\bibinfo {volume} {2}},\ \bibinfo {pages} {033493} (\bibinfo {year}
		{2020})}\BibitemShut {NoStop}%
	\bibitem [{\citenamefont {Salda{\ifmmode\tilde{n}\else\~{n}\fi}a}\ \emph
		{et~al.}(2021)\citenamefont {Salda{\ifmmode\tilde{n}\else\~{n}\fi}a},
		\citenamefont {Vekris}, \citenamefont
		{Pave{\ifmmode\check{s}\else\v{s}\fi}i{\ifmmode\check{c}\else\v{c}\fi}},
		\citenamefont {Krogstrup}, \citenamefont
		{{\ifmmode\check{Z}\else\v{Z}\fi}itko}, \citenamefont 
		{Grove-Rasmussen},\
		and\ \citenamefont {Nyg{\aa}rd}}]{Saldana2021Jan}%
	\BibitemOpen
	\bibfield  {author} {\bibinfo {author} {\bibfnamefont {J.~C.~E.}\
			\bibnamefont {Salda{\ifmmode\tilde{n}\else\~{n}\fi}a}}, \bibinfo 
			{author}
		{\bibfnamefont {A.}~\bibnamefont {Vekris}}, \bibinfo {author} 
		{\bibfnamefont
			{L.}~\bibnamefont
			{Pave{\ifmmode\check{s}\else\v{s}\fi}i{\ifmmode\check{c}\else\v{c}\fi}}},
		\bibinfo {author} {\bibfnamefont {P.}~\bibnamefont {Krogstrup}}, 
		\bibinfo
		{author} {\bibfnamefont {R.}~\bibnamefont
			{{\ifmmode\check{Z}\else\v{Z}\fi}itko}}, \bibinfo {author} 
			{\bibfnamefont
			{K.}~\bibnamefont {Grove-Rasmussen}},\ and\ \bibinfo {author} 
			{\bibfnamefont
			{J.}~\bibnamefont {Nyg{\aa}rd}},\ }\bibfield  {title} {\bibinfo 
			{title}
		{{Bias asymmetric subgap states mimicking Majorana signatures}},\ }\href
	{https://arxiv.org/abs/2101.10794v1} {\bibfield  {journal} {\bibinfo
			{journal} {arXiv}\ } (\bibinfo {year} {2021})},\ \Eprint
	{https://arxiv.org/abs/2101.10794} {2101.10794} \BibitemShut {NoStop}%
	\bibitem [{\citenamefont {Deng}\ \emph {et~al.}(2016)\citenamefont {Deng},
		\citenamefont {Vaitiek{\ifmmode\dot{e}\else\.{e}\fi}nas}, \citenamefont
		{Hansen}, \citenamefont {Danon}, \citenamefont {Leijnse}, \citenamefont
		{Flensberg}, \citenamefont {Nyg{\aa}rd}, \citenamefont {Krogstrup},\ 
		and\
		\citenamefont {Marcus}}]{Deng2016Dec}%
	\BibitemOpen
	\bibfield  {author} {\bibinfo {author} {\bibfnamefont {M.~T.}\ \bibnamefont
			{Deng}}, \bibinfo {author} {\bibfnamefont {S.}~\bibnamefont
			{Vaitiek{\ifmmode\dot{e}\else\.{e}\fi}nas}}, \bibinfo {author} 
			{\bibfnamefont
			{E.~B.}\ \bibnamefont {Hansen}}, \bibinfo {author} {\bibfnamefont
			{J.}~\bibnamefont {Danon}}, \bibinfo {author} {\bibfnamefont
			{M.}~\bibnamefont {Leijnse}}, \bibinfo {author} {\bibfnamefont
			{K.}~\bibnamefont {Flensberg}}, \bibinfo {author} {\bibfnamefont
			{J.}~\bibnamefont {Nyg{\aa}rd}}, \bibinfo {author} {\bibfnamefont
			{P.}~\bibnamefont {Krogstrup}},\ and\ \bibinfo {author} 
			{\bibfnamefont
			{C.~M.}\ \bibnamefont {Marcus}},\ }\bibfield  {title} {\bibinfo 
			{title}
		{{Majorana bound state in a coupled quantum-dot hybrid-nanowire 
		system}},\
	}\href {https://doi.org/10.1126/science.aaf3961} {\bibfield  {journal}
		{\bibinfo  {journal} {Science}\ }\textbf {\bibinfo {volume} {354}},\ 
		\bibinfo
		{pages} {1557} (\bibinfo {year} {2016})}\BibitemShut {NoStop}%
	\bibitem [{\citenamefont {Kondo}(1964)}]{Kondo1964Jul}%
	\BibitemOpen
	\bibfield  {author} {\bibinfo {author} {\bibfnamefont {J.}~\bibnamefont
			{Kondo}},\ }\bibfield  {title} {\bibinfo {title} {{Resistance 
			Minimum in
				Dilute Magnetic Alloys}},\ }\href 
				{https://doi.org/10.1143/PTP.32.37}
	{\bibfield  {journal} {\bibinfo  {journal} {Prog. Theor. Phys.}\ }\textbf
		{\bibinfo {volume} {32}},\ \bibinfo {pages} {37} (\bibinfo {year}
		{1964})}\BibitemShut {NoStop}%
	\bibitem [{\citenamefont {Hewson}(1993)}]{hewson_1993}%
	\BibitemOpen
	\bibfield  {author} {\bibinfo {author} {\bibfnamefont {A.~C.}\ \bibnamefont
			{Hewson}},\ }\href {https://doi.org/10.1017/CBO9780511470752} {\emph
		{\bibinfo {title} {{The Kondo Problem to Heavy Fermions}}}},\ Cambridge
	Studies in Magnetism\ (\bibinfo  {publisher} {Cambridge University Press},\
	\bibinfo {year} {1993})\BibitemShut {NoStop}%
	\bibitem [{\citenamefont {L{\ifmmode\ddot{o}\else\"{o}\fi}hneysen}\ \emph
		{et~al.}(2007)\citenamefont {L{\ifmmode\ddot{o}\else\"{o}\fi}hneysen},
		\citenamefont {Rosch}, \citenamefont {Vojta},\ and\ \citenamefont
		{W{\ifmmode\ddot{o}\else\"{o}\fi}lfle}}]{Lohneysen2007Aug}%
	\BibitemOpen
	\bibfield  {author} {\bibinfo {author} {\bibfnamefont {H.~v.}\ \bibnamefont
			{L{\ifmmode\ddot{o}\else\"{o}\fi}hneysen}}, \bibinfo {author} 
			{\bibfnamefont
			{A.}~\bibnamefont {Rosch}}, \bibinfo {author} {\bibfnamefont
			{M.}~\bibnamefont {Vojta}},\ and\ \bibinfo {author} {\bibfnamefont
			{P.}~\bibnamefont {W{\ifmmode\ddot{o}\else\"{o}\fi}lfle}},\ 
			}\bibfield
	{title} {\bibinfo {title} {{Fermi-liquid instabilities at magnetic quantum
				phase transitions}},\ }\href 
				{https://doi.org/10.1103/RevModPhys.79.1015}
	{\bibfield  {journal} {\bibinfo  {journal} {Rev. Mod. Phys.}\ }\textbf
		{\bibinfo {volume} {79}},\ \bibinfo {pages} {1015} (\bibinfo {year}
		{2007})}\BibitemShut {NoStop}%
	\bibitem [{\citenamefont {Si}\ \emph {et~al.}(2016)\citenamefont {Si},
		\citenamefont {Yu},\ and\ \citenamefont {Abrahams}}]{Si2016Mar}%
	\BibitemOpen
	\bibfield  {author} {\bibinfo {author} {\bibfnamefont {Q.}~\bibnamefont
			{Si}}, \bibinfo {author} {\bibfnamefont {R.}~\bibnamefont {Yu}},\ 
			and\
		\bibinfo {author} {\bibfnamefont {E.}~\bibnamefont {Abrahams}},\ 
		}\bibfield
	{title} {\bibinfo {title} {{High-temperature superconductivity in iron
				pnictides and chalcogenides}},\ }\href
	{https://doi.org/10.1038/natrevmats.2016.17} {\bibfield  {journal} {\bibinfo
			{journal} {Nat. Rev. Mater.}\ }\textbf {\bibinfo {volume} {1}},\ 
			\bibinfo
		{pages} {1} (\bibinfo {year} {2016})}\BibitemShut {NoStop}%
	\bibitem [{\citenamefont {Paschen}\ and\ \citenamefont
		{Si}(2021)}]{Paschen2021Jan}%
	\BibitemOpen
	\bibfield  {author} {\bibinfo {author} {\bibfnamefont {S.}~\bibnamefont
			{Paschen}}\ and\ \bibinfo {author} {\bibfnamefont {Q.}~\bibnamefont 
			{Si}},\
	}\bibfield  {title} {\bibinfo {title} {{Quantum phases driven by strong
				correlations}},\ }\href 
				{https://doi.org/10.1038/s42254-020-00262-6}
	{\bibfield  {journal} {\bibinfo  {journal} {Nat. Rev. Phys.}\ }\textbf
		{\bibinfo {volume} {3}},\ \bibinfo {pages} {9} (\bibinfo {year}
		{2021})}\BibitemShut {NoStop}%
	\bibitem [{\citenamefont {Dzero}\ \emph {et~al.}(2016)\citenamefont {Dzero},
		\citenamefont {Xia}, \citenamefont {Galitski},\ and\ \citenamefont
		{Coleman}}]{Dzero2016Mar}%
	\BibitemOpen
	\bibfield  {author} {\bibinfo {author} {\bibfnamefont {M.}~\bibnamefont
			{Dzero}}, \bibinfo {author} {\bibfnamefont {J.}~\bibnamefont 
			{Xia}}, \bibinfo
		{author} {\bibfnamefont {V.}~\bibnamefont {Galitski}},\ and\ \bibinfo
		{author} {\bibfnamefont {P.}~\bibnamefont {Coleman}},\ }\bibfield  
		{title}
	{\bibinfo {title} {{Topological Kondo Insulators}},\ }\href
	{https://doi.org/10.1146/annurev-conmatphys-031214-014749} {\bibfield
		{journal} {\bibinfo  {journal} {Annu. Rev. Condens. Matter Phys.}\ 
		}\textbf
		{\bibinfo {volume} {7}},\ \bibinfo {pages} {249} (\bibinfo {year}
		{2016})}\BibitemShut {NoStop}%
	\bibitem [{\citenamefont {Lai}\ \emph {et~al.}(2018)\citenamefont {Lai},
		\citenamefont {Grefe}, \citenamefont {Paschen},\ and\ \citenamefont
		{Si}}]{Lai2018Jan}%
	\BibitemOpen
	\bibfield  {author} {\bibinfo {author} {\bibfnamefont {H.-H.}\ \bibnamefont
			{Lai}}, \bibinfo {author} {\bibfnamefont {S.~E.}\ \bibnamefont 
			{Grefe}},
		\bibinfo {author} {\bibfnamefont {S.}~\bibnamefont {Paschen}},\ and\ 
		\bibinfo
		{author} {\bibfnamefont {Q.}~\bibnamefont {Si}},\ }\bibfield  {title}
	{\bibinfo {title} {{Weyl{\textendash}Kondo semimetal in heavy-fermion
				systems}},\ }\href {https://doi.org/10.1073/pnas.1715851115} 
				{\bibfield
		{journal} {\bibinfo  {journal} {Proc. Natl. Acad. Sci. U.S.A.}\ }\textbf
		{\bibinfo {volume} {115}},\ \bibinfo {pages} {93} (\bibinfo {year}
		{2018})}\BibitemShut {NoStop}%
	\bibitem [{\citenamefont {Cheng}\ \emph {et~al.}(2014)\citenamefont {Cheng},
		\citenamefont {Becker}, \citenamefont {Bauer},\ and\ \citenamefont
		{Lutchyn}}]{Cheng2014Sep}%
	\BibitemOpen
	\bibfield  {author} {\bibinfo {author} {\bibfnamefont {M.}~\bibnamefont
			{Cheng}}, \bibinfo {author} {\bibfnamefont {M.}~\bibnamefont 
			{Becker}},
		\bibinfo {author} {\bibfnamefont {B.}~\bibnamefont {Bauer}},\ and\ 
		\bibinfo
		{author} {\bibfnamefont {R.~M.}\ \bibnamefont {Lutchyn}},\ }\bibfield
	{title} {\bibinfo {title} {{Interplay between Kondo and Majorana 
	Interactions
				in Quantum Dots}},\ }\href 
				{https://doi.org/10.1103/PhysRevX.4.031051}
	{\bibfield  {journal} {\bibinfo  {journal} {Phys. Rev. X}\ }\textbf 
	{\bibinfo
			{volume} {4}},\ \bibinfo {pages} {031051} (\bibinfo {year}
		{2014})}\BibitemShut {NoStop}%
	\bibitem [{\citenamefont {Silva}\ \emph {et~al.}(2020)\citenamefont {Silva},
		\citenamefont {da~Silva},\ and\ \citenamefont {Vernek}}]{Vernek2019}%
	\BibitemOpen
	\bibfield  {author} {\bibinfo {author} {\bibfnamefont {J.~F.}\ \bibnamefont
			{Silva}}, \bibinfo {author} {\bibfnamefont {L.~G. G. V.~D.}\ 
			\bibnamefont
			{da~Silva}},\ and\ \bibinfo {author} {\bibfnamefont 
			{E.}~\bibnamefont
			{Vernek}},\ }\bibfield  {title} {\bibinfo {title} {{Robustness of 
			the Kondo
				effect in a quantum dot coupled to Majorana zero modes}},\ 
				}\href
	{https://doi.org/10.1103/PhysRevB.101.075428} {\bibfield  {journal} 
	{\bibinfo
			{journal} {Phys. Rev. B}\ }\textbf {\bibinfo {volume} {101}},\ 
			\bibinfo
		{pages} {075428} (\bibinfo {year} {2020})}\BibitemShut {NoStop}%
	\bibitem [{\citenamefont {van Beek}\ and\ \citenamefont
		{Braunecker}(2016)}]{vanBeek2016Sep}%
	\BibitemOpen
	\bibfield  {author} {\bibinfo {author} {\bibfnamefont {I.~J.}\ \bibnamefont
			{van Beek}}\ and\ \bibinfo {author} {\bibfnamefont {B.}~\bibnamefont
			{Braunecker}},\ }\bibfield  {title} {\bibinfo {title} {{Non-Kondo 
			many-body
				physics in a Majorana-based Kondo type system}},\ }\href
	{https://doi.org/10.1103/PhysRevB.94.115416} {\bibfield  {journal} {\bibinfo
			{journal} {Phys. Rev. B}\ }\textbf {\bibinfo {volume} {94}},\ 
			\bibinfo
		{pages} {115416} (\bibinfo {year} {2016})}\BibitemShut {NoStop}%
	\bibitem [{\citenamefont {B{\ifmmode\acute{e}\else\'{e}\fi}ri}\ and\
		\citenamefont {Cooper}(2012)}]{Beri2012Oct}%
	\BibitemOpen
	\bibfield  {author} {\bibinfo {author} {\bibfnamefont {B.}~\bibnamefont
			{B{\ifmmode\acute{e}\else\'{e}\fi}ri}}\ and\ \bibinfo {author} 
			{\bibfnamefont
			{N.~R.}\ \bibnamefont {Cooper}},\ }\bibfield  {title} {\bibinfo 
			{title}
		{{Topological Kondo Effect with Majorana Fermions}},\ }\href
	{https://doi.org/10.1103/PhysRevLett.109.156803} {\bibfield  {journal}
		{\bibinfo  {journal} {Phys. Rev. Lett.}\ }\textbf {\bibinfo {volume} 
		{109}},\
		\bibinfo {pages} {156803} (\bibinfo {year} {2012})}\BibitemShut 
		{NoStop}%
	\bibitem [{\citenamefont {Herviou}\ \emph {et~al.}(2016)\citenamefont
		{Herviou}, \citenamefont {Le~Hur},\ and\ \citenamefont
		{Mora}}]{Herviou2016Dec}%
	\BibitemOpen
	\bibfield  {author} {\bibinfo {author} {\bibfnamefont {L.}~\bibnamefont
			{Herviou}}, \bibinfo {author} {\bibfnamefont {K.}~\bibnamefont 
			{Le~Hur}},\
		and\ \bibinfo {author} {\bibfnamefont {C.}~\bibnamefont {Mora}},\ 
		}\bibfield
	{title} {\bibinfo {title} {{Many-terminal Majorana island: From topological
				to multichannel Kondo model}},\ }\href
	{https://doi.org/10.1103/PhysRevB.94.235102} {\bibfield  {journal} {\bibinfo
			{journal} {Phys. Rev. B}\ }\textbf {\bibinfo {volume} {94}},\ 
			\bibinfo
		{pages} {235102} (\bibinfo {year} {2016})}\BibitemShut {NoStop}%
	\bibitem [{\citenamefont {Kornja{\ifmmode\check{c}\else\v{c}\fi}a}\ \emph
		{et~al.}(2021)\citenamefont {Kornja{\ifmmode\check{c}\else\v{c}\fi}a},
		\citenamefont {Quito},\ and\ \citenamefont {Flint}}]{Kornjaca2021Apr}%
	\BibitemOpen
	\bibfield  {author} {\bibinfo {author} {\bibfnamefont {M.}~\bibnamefont
			{Kornja{\ifmmode\check{c}\else\v{c}\fi}a}}, \bibinfo {author} 
			{\bibfnamefont
			{V.~L.}\ \bibnamefont {Quito}},\ and\ \bibinfo {author} 
			{\bibfnamefont
			{R.}~\bibnamefont {Flint}},\ }\bibfield  {title} {\bibinfo {title} 
			{{Mobile
				Majorana zero-modes in two-channel Kondo insulators}},\ }\href
	{https://arxiv.org/abs/2104.11173v1} {\bibfield  {journal} {\bibinfo
			{journal} {arXiv}\ } (\bibinfo {year} {2021})},\ \Eprint
	{https://arxiv.org/abs/2104.11173} {2104.11173} \BibitemShut {NoStop}%
	\bibitem [{\citenamefont {Lee}\ \emph 
	{et~al.}(2013{\natexlab{b}})\citenamefont
		{Lee}, \citenamefont {Lim},\ and\ \citenamefont
		{L{\ifmmode\acute{o}\else\'{o}\fi}pez}}]{Lee2013Jun}%
	\BibitemOpen
	\bibfield  {author} {\bibinfo {author} {\bibfnamefont {M.}~\bibnamefont
			{Lee}}, \bibinfo {author} {\bibfnamefont {J.~S.}\ \bibnamefont 
			{Lim}},\ and\
		\bibinfo {author} {\bibfnamefont {R.}~\bibnamefont
			{L{\ifmmode\acute{o}\else\'{o}\fi}pez}},\ }\bibfield  {title} 
			{\bibinfo
		{title} {{Kondo effect in a quantum dot side-coupled to a topological
				superconductor}},\ }\href 
				{https://doi.org/10.1103/PhysRevB.87.241402}
	{\bibfield  {journal} {\bibinfo  {journal} {Phys. Rev. B}\ }\textbf 
	{\bibinfo
			{volume} {87}},\ \bibinfo {pages} {241402(R)} (\bibinfo {year}
		{2013}{\natexlab{b}})}\BibitemShut {NoStop}%
	\bibitem [{\citenamefont {Ruiz-Tijerina}\ \emph {et~al.}(2015)\citenamefont
		{Ruiz-Tijerina}, \citenamefont {Vernek}, \citenamefont {Dias~da 
		Silva},\ and\
		\citenamefont {Egues}}]{Ruiz-Tijerina2015Mar}%
	\BibitemOpen
	\bibfield  {author} {\bibinfo {author} {\bibfnamefont {D.~A.}\ \bibnamefont
			{Ruiz-Tijerina}}, \bibinfo {author} {\bibfnamefont {E.}~\bibnamefont
			{Vernek}}, \bibinfo {author} {\bibfnamefont {L.~G. G.~V.}\ 
			\bibnamefont
			{Dias~da Silva}},\ and\ \bibinfo {author} {\bibfnamefont {J.~C.}\
			\bibnamefont {Egues}},\ }\bibfield  {title} {\bibinfo {title} 
			{{Interaction
				effects on a Majorana zero mode leaking into a quantum dot}},\ 
				}\href
	{https://doi.org/10.1103/PhysRevB.91.115435} {\bibfield  {journal} {\bibinfo
			{journal} {Phys. Rev. B}\ }\textbf {\bibinfo {volume} {91}},\ 
			\bibinfo
		{pages} {115435} (\bibinfo {year} {2015})}\BibitemShut {NoStop}%
	\bibitem [{\citenamefont {Weymann}\ and\ \citenamefont
		{W{\ifmmode\acute{o}\else\'{o}\fi}jcik}(2017)}]{Weymann2017Apr}%
	\BibitemOpen
	\bibfield  {author} {\bibinfo {author} {\bibfnamefont {I.}~\bibnamefont
			{Weymann}}\ and\ \bibinfo {author} {\bibfnamefont {K.~P.}\ 
			\bibnamefont
			{W{\ifmmode\acute{o}\else\'{o}\fi}jcik}},\ }\bibfield  {title} 
			{\bibinfo
		{title} {{Transport properties of a hybrid Majorana wire-quantum dot 
		system
				with ferromagnetic contacts}},\ }\href
	{https://doi.org/10.1103/PhysRevB.95.155427} {\bibfield  {journal} {\bibinfo
			{journal} {Phys. Rev. B}\ }\textbf {\bibinfo {volume} {95}},\ 
			\bibinfo
		{pages} {155427} (\bibinfo {year} {2017})}\BibitemShut {NoStop}%
	\bibitem [{\citenamefont {Flensberg}(2010)}]{Flensberg2010Nov}%
	\BibitemOpen
	\bibfield  {author} {\bibinfo {author} {\bibfnamefont {K.}~\bibnamefont
			{Flensberg}},\ }\bibfield  {title} {\bibinfo {title} {{Tunneling
				characteristics of a chain of Majorana bound states}},\ }\href
	{https://doi.org/10.1103/PhysRevB.82.180516} {\bibfield  {journal} {\bibinfo
			{journal} {Phys. Rev. B}\ }\textbf {\bibinfo {volume} {82}},\ 
			\bibinfo
		{pages} {180516(R)} (\bibinfo {year} {2010})}\BibitemShut {NoStop}%
	\bibitem [{\citenamefont {Vernek}\ \emph {et~al.}(2014)\citenamefont 
	{Vernek},
		\citenamefont {Penteado}, \citenamefont {Seridonio},\ and\ \citenamefont
		{Egues}}]{Vernek2014Apr}%
	\BibitemOpen
	\bibfield  {author} {\bibinfo {author} {\bibfnamefont {E.}~\bibnamefont
			{Vernek}}, \bibinfo {author} {\bibfnamefont {P.~H.}\ \bibnamefont
			{Penteado}}, \bibinfo {author} {\bibfnamefont {A.~C.}\ \bibnamefont
			{Seridonio}},\ and\ \bibinfo {author} {\bibfnamefont {J.~C.}\ 
			\bibnamefont
			{Egues}},\ }\bibfield  {title} {\bibinfo {title} {{Subtle leakage 
			of a
				Majorana mode into a quantum dot}},\ }\href
	{https://doi.org/10.1103/PhysRevB.89.165314} {\bibfield  {journal} {\bibinfo
			{journal} {Phys. Rev. B}\ }\textbf {\bibinfo {volume} {89}},\ 
			\bibinfo
		{pages} {165314} (\bibinfo {year} {2014})}\BibitemShut {NoStop}%
	\bibitem [{\citenamefont {Weymann}\ \emph {et~al.}(2020)\citenamefont
		{Weymann}, \citenamefont {W{\ifmmode\acute{o}\else\'{o}\fi}jcik},\ and\
		\citenamefont {Majek}}]{Weymann2020Jan}%
	\BibitemOpen
	\bibfield  {author} {\bibinfo {author} {\bibfnamefont {I.}~\bibnamefont
			{Weymann}}, \bibinfo {author} {\bibfnamefont {K.~P.}\ \bibnamefont
			{W{\ifmmode\acute{o}\else\'{o}\fi}jcik}},\ and\ \bibinfo {author}
		{\bibfnamefont {P.}~\bibnamefont {Majek}},\ }\bibfield  {title} 
		{\bibinfo
		{title} {{Majorana-Kondo interplay in T-shaped double quantum dots}},\ 
		}\href
	{https://doi.org/10.1103/PhysRevB.101.235404} {\bibfield  {journal} 
	{\bibinfo
			{journal} {Phys. Rev. B}\ }\textbf {\bibinfo {volume} {101}},\ 
			\bibinfo
		{pages} {235404} (\bibinfo {year} {2020})}\BibitemShut {NoStop}%
	\bibitem [{\citenamefont {Pustilnik}\ and\ \citenamefont
		{Glazman}(2001)}]{Pustilnik2001Nov}%
	\BibitemOpen
	\bibfield  {author} {\bibinfo {author} {\bibfnamefont {M.}~\bibnamefont
			{Pustilnik}}\ and\ \bibinfo {author} {\bibfnamefont {L.~I.}\ 
			\bibnamefont
			{Glazman}},\ }\bibfield  {title} {\bibinfo {title} {{Kondo Effect 
			in Real
				Quantum Dots}},\ }\href 
				{https://doi.org/10.1103/PhysRevLett.87.216601}
	{\bibfield  {journal} {\bibinfo  {journal} {Phys. Rev. Lett.}\ }\textbf
		{\bibinfo {volume} {87}},\ \bibinfo {pages} {216601} (\bibinfo {year}
		{2001})}\BibitemShut {NoStop}%
	\bibitem [{\citenamefont {Cornaglia}\ and\ \citenamefont
		{Grempel}(2005)}]{Cornaglia2005Feb}%
	\BibitemOpen
	\bibfield  {author} {\bibinfo {author} {\bibfnamefont {P.~S.}\ \bibnamefont
			{Cornaglia}}\ and\ \bibinfo {author} {\bibfnamefont {D.~R.}\ 
			\bibnamefont
			{Grempel}},\ }\bibfield  {title} {\bibinfo {title} {{Strongly 
			correlated
				regimes in a double quantum dot device}},\ }\href
	{https://doi.org/10.1103/PhysRevB.71.075305} {\bibfield  {journal} {\bibinfo
			{journal} {Phys. Rev. B}\ }\textbf {\bibinfo {volume} {71}},\ 
			\bibinfo
		{pages} {075305} (\bibinfo {year} {2005})}\BibitemShut {NoStop}%
	\bibitem [{\citenamefont {Sasaki}\ \emph {et~al.}(2009)\citenamefont 
	{Sasaki},
		\citenamefont {Tamura}, \citenamefont {Akazaki},\ and\ \citenamefont
		{Fujisawa}}]{Sasaki2009Dec}%
	\BibitemOpen
	\bibfield  {author} {\bibinfo {author} {\bibfnamefont {S.}~\bibnamefont
			{Sasaki}}, \bibinfo {author} {\bibfnamefont {H.}~\bibnamefont 
			{Tamura}},
		\bibinfo {author} {\bibfnamefont {T.}~\bibnamefont {Akazaki}},\ and\ 
		\bibinfo
		{author} {\bibfnamefont {T.}~\bibnamefont {Fujisawa}},\ }\bibfield  
		{title}
	{\bibinfo {title} {{Fano-Kondo Interplay in a Side-Coupled Double Quantum
				Dot}},\ }\href {https://doi.org/10.1103/PhysRevLett.103.266806} 
				{\bibfield
		{journal} {\bibinfo  {journal} {Phys. Rev. Lett.}\ }\textbf {\bibinfo
			{volume} {103}},\ \bibinfo {pages} {266806} (\bibinfo {year}
		{2009})}\BibitemShut {NoStop}%
	\bibitem [{\citenamefont {W{\ifmmode\acute{o}\else\'{o}\fi}jcik}\ and\
		\citenamefont {Weymann}(2015)}]{Wojcik2015Apr}%
	\BibitemOpen
	\bibfield  {author} {\bibinfo {author} {\bibfnamefont {K.~P.}\ \bibnamefont
			{W{\ifmmode\acute{o}\else\'{o}\fi}jcik}}\ and\ \bibinfo {author}
		{\bibfnamefont {I.}~\bibnamefont {Weymann}},\ }\bibfield  {title} 
		{\bibinfo
		{title} {{Two-stage Kondo effect in T-shaped double quantum dots with
				ferromagnetic leads}},\ }\href 
				{https://doi.org/10.1103/PhysRevB.91.134422}
	{\bibfield  {journal} {\bibinfo  {journal} {Phys. Rev. B}\ }\textbf 
	{\bibinfo
			{volume} {91}},\ \bibinfo {pages} {134422} (\bibinfo {year}
		{2015})}\BibitemShut {NoStop}%
	\bibitem [{\citenamefont {Guo}\ \emph {et~al.}(2021)\citenamefont {Guo},
		\citenamefont {Zhu}, \citenamefont {Zhou}, \citenamefont {Yu}, 
		\citenamefont
		{Lu},\ and\ \citenamefont {Liang}}]{Guo2021Mar}%
	\BibitemOpen
	\bibfield  {author} {\bibinfo {author} {\bibfnamefont {X.}~\bibnamefont
			{Guo}}, \bibinfo {author} {\bibfnamefont {Q.}~\bibnamefont {Zhu}}, 
			\bibinfo
		{author} {\bibfnamefont {L.}~\bibnamefont {Zhou}}, \bibinfo {author}
		{\bibfnamefont {W.}~\bibnamefont {Yu}}, \bibinfo {author} {\bibfnamefont
			{W.}~\bibnamefont {Lu}},\ and\ \bibinfo {author} {\bibfnamefont
			{W.}~\bibnamefont {Liang}},\ }\bibfield  {title} {\bibinfo {title}
		{{Evolution and universality of two-stage Kondo effect in single 
		manganese
				phthalocyanine molecule transistors}},\ }\href
	{https://doi.org/10.1038/s41467-021-21492-x} {\bibfield  {journal} {\bibinfo
			{journal} {Nat. Commun.}\ }\textbf {\bibinfo {volume} {12}},\ 
			\bibinfo
		{pages} {1} (\bibinfo {year} {2021})}\BibitemShut {NoStop}%
	\bibitem [{\citenamefont {L{\ifmmode\acute{o}\else\'{o}\fi}pez}\ \emph
		{et~al.}(2014)\citenamefont {L{\ifmmode\acute{o}\else\'{o}\fi}pez},
		\citenamefont {Lee}, \citenamefont {Serra},\ and\ \citenamefont
		{Lim}}]{Lopez2014May}%
	\BibitemOpen
	\bibfield  {author} {\bibinfo {author} {\bibfnamefont {R.}~\bibnamefont
			{L{\ifmmode\acute{o}\else\'{o}\fi}pez}}, \bibinfo {author} 
			{\bibfnamefont
			{M.}~\bibnamefont {Lee}}, \bibinfo {author} {\bibfnamefont 
			{L.}~\bibnamefont
			{Serra}},\ and\ \bibinfo {author} {\bibfnamefont {J.~S.}\ 
			\bibnamefont
			{Lim}},\ }\bibfield  {title} {\bibinfo {title} {{Thermoelectrical 
			detection
				of Majorana states}},\ }\href 
				{https://doi.org/10.1103/PhysRevB.89.205418}
	{\bibfield  {journal} {\bibinfo  {journal} {Phys. Rev. B}\ }\textbf 
	{\bibinfo
			{volume} {89}},\ \bibinfo {pages} {205418} (\bibinfo {year}
		{2014})}\BibitemShut {NoStop}%
	\bibitem [{\citenamefont {Weymann}(2017)}]{Weymann2017Jan}%
	\BibitemOpen
	\bibfield  {author} {\bibinfo {author} {\bibfnamefont {I.}~\bibnamefont
			{Weymann}},\ }\bibfield  {title} {\bibinfo {title} {{Spin Seebeck 
			effect in
				quantum dot side-coupled to topological superconductor}},\ 
				}\href
	{https://doi.org/10.1088/1361-648x/aa5526} {\bibfield  {journal} {\bibinfo
			{journal} {J. Phys.: Condens. Matter}\ }\textbf {\bibinfo {volume} 
			{29}},\
		\bibinfo {pages} {095301} (\bibinfo {year} {2017})}\BibitemShut 
		{NoStop}%
	\bibitem [{\citenamefont {Costi}\ and\ \citenamefont
		{Zlati{\ifmmode\acute{c}\else\'{c}\fi}}(2010)}]{CostiZlatic}%
	\BibitemOpen
	\bibfield  {author} {\bibinfo {author} {\bibfnamefont {T.~A.}\ \bibnamefont
			{Costi}}\ and\ \bibinfo {author} {\bibfnamefont {V.}~\bibnamefont
			{Zlati{\ifmmode\acute{c}\else\'{c}\fi}}},\ }\bibfield  {title} 
			{\bibinfo
		{title} {{Thermoelectric transport through strongly correlated quantum
				dots}},\ }\href {https://doi.org/10.1103/PhysRevB.81.235127} 
				{\bibfield
		{journal} {\bibinfo  {journal} {Phys. Rev. B}\ }\textbf {\bibinfo 
		{volume}
			{81}},\ \bibinfo {pages} {235127} (\bibinfo {year} 
			{2010})}\BibitemShut
	{NoStop}%
	\bibitem [{\citenamefont {Dutta}\ \emph {et~al.}(2019)\citenamefont {Dutta},
		\citenamefont {Majidi}, \citenamefont {García~Corral}, \citenamefont
		{Erdman}, \citenamefont {Florens}, \citenamefont {Costi}, \citenamefont
		{Courtois},\ and\ \citenamefont {Winkelmann}}]{Dutta2018Dec}%
	\BibitemOpen
	\bibfield  {author} {\bibinfo {author} {\bibfnamefont {B.}~\bibnamefont
			{Dutta}}, \bibinfo {author} {\bibfnamefont {D.}~\bibnamefont 
			{Majidi}},
		\bibinfo {author} {\bibfnamefont {A.}~\bibnamefont {García~Corral}},
		\bibinfo {author} {\bibfnamefont {P.~A.}\ \bibnamefont {Erdman}}, 
		\bibinfo
		{author} {\bibfnamefont {S.}~\bibnamefont {Florens}}, \bibinfo {author}
		{\bibfnamefont {T.~A.}\ \bibnamefont {Costi}}, \bibinfo {author}
		{\bibfnamefont {H.}~\bibnamefont {Courtois}},\ and\ \bibinfo {author}
		{\bibfnamefont {C.~B.}\ \bibnamefont {Winkelmann}},\ }\bibfield  {title}
	{\bibinfo {title} {Direct probe of the seebeck coefficient in a
			kondo-correlated single-quantum-dot transistor},\ }\href
	{https://doi.org/10.1021/acs.nanolett.8b04398} {\bibfield  {journal}
		{\bibinfo  {journal} {Nano Letters}\ }\textbf {\bibinfo {volume} {19}},\
		\bibinfo {pages} {506} (\bibinfo {year} {2019})},\ \Eprint
	{https://arxiv.org/abs/https://doi.org/10.1021/acs.nanolett.8b04398}
	{https://doi.org/10.1021/acs.nanolett.8b04398} \BibitemShut {NoStop}%
	\bibitem [{\citenamefont {Svilans}\ \emph {et~al.}(2018)\citenamefont
		{Svilans}, \citenamefont {Josefsson}, \citenamefont {Burke}, 
		\citenamefont
		{Fahlvik}, \citenamefont {Thelander}, \citenamefont {Linke},\ and\
		\citenamefont {Leijnse}}]{Svilans2018Nov}%
	\BibitemOpen
	\bibfield  {author} {\bibinfo {author} {\bibfnamefont {A.}~\bibnamefont
			{Svilans}}, \bibinfo {author} {\bibfnamefont {M.}~\bibnamefont 
			{Josefsson}},
		\bibinfo {author} {\bibfnamefont {A.~M.}\ \bibnamefont {Burke}}, 
		\bibinfo
		{author} {\bibfnamefont {S.}~\bibnamefont {Fahlvik}}, \bibinfo {author}
		{\bibfnamefont {C.}~\bibnamefont {Thelander}}, \bibinfo {author}
		{\bibfnamefont {H.}~\bibnamefont {Linke}},\ and\ \bibinfo {author}
		{\bibfnamefont {M.}~\bibnamefont {Leijnse}},\ }\bibfield  {title} 
		{\bibinfo
		{title} {Thermoelectric characterization of the kondo resonance in 
		nanowire
			quantum dots},\ }\href 
			{https://doi.org/10.1103/PhysRevLett.121.206801}
	{\bibfield  {journal} {\bibinfo  {journal} {Phys. Rev. Lett.}\ }\textbf
		{\bibinfo {volume} {121}},\ \bibinfo {pages} {206801} (\bibinfo {year}
		{2018})}\BibitemShut {NoStop}%
	\bibitem [{\citenamefont {Bauer}\ \emph {et~al.}(2012)\citenamefont {Bauer},
		\citenamefont {Saitoh},\ and\ \citenamefont {van Wees}}]{Bauer2012May}%
	\BibitemOpen
	\bibfield  {author} {\bibinfo {author} {\bibfnamefont {G.~E.~W.}\
			\bibnamefont {Bauer}}, \bibinfo {author} {\bibfnamefont 
			{E.}~\bibnamefont
			{Saitoh}},\ and\ \bibinfo {author} {\bibfnamefont {B.~J.}\ 
			\bibnamefont {van
				Wees}},\ }\bibfield  {title} {\bibinfo {title} {{Spin 
				caloritronics}},\
	}\href {https://doi.org/10.1038/nmat3301} {\bibfield  {journal} {\bibinfo
			{journal} {Nat. Mater.}\ }\textbf {\bibinfo {volume} {11}},\ 
			\bibinfo {pages}
		{391} (\bibinfo {year} {2012})}\BibitemShut {NoStop}%
	\bibitem [{\citenamefont {Gong}\ \emph {et~al.}(2014)\citenamefont {Gong},
		\citenamefont {Zhang}, \citenamefont {Li}, \citenamefont {Yi},\ and\
		\citenamefont {Zheng}}]{Gong2014Jun}%
	\BibitemOpen
	\bibfield  {author} {\bibinfo {author} {\bibfnamefont {W.-J.}\ \bibnamefont
			{Gong}}, \bibinfo {author} {\bibfnamefont {S.-F.}\ \bibnamefont 
			{Zhang}},
		\bibinfo {author} {\bibfnamefont {Z.-C.}\ \bibnamefont {Li}}, \bibinfo
		{author} {\bibfnamefont {G.}~\bibnamefont {Yi}},\ and\ \bibinfo {author}
		{\bibfnamefont {Y.-S.}\ \bibnamefont {Zheng}},\ }\bibfield  {title} 
		{\bibinfo
		{title} {{Detection of a Majorana fermion zero mode by a T-shaped 
		quantum-dot
				structure}},\ }\href 
				{https://doi.org/10.1103/PhysRevB.89.245413} {\bibfield
		{journal} {\bibinfo  {journal} {Phys. Rev. B}\ }\textbf {\bibinfo 
		{volume}
			{89}},\ \bibinfo {pages} {245413} (\bibinfo {year} 
			{2014})}\BibitemShut
	{NoStop}%
	\bibitem [{\citenamefont {Ramos-Andrade}\ \emph {et~al.}(2016)\citenamefont
		{Ramos-Andrade}, \citenamefont {{\ifmmode\acute{A}\else\'{A}\fi}valos
			Ovando}, \citenamefont {Orellana},\ and\ \citenamefont
		{Ulloa}}]{Ramos-Andrade2016Oct}%
	\BibitemOpen
	\bibfield  {author} {\bibinfo {author} {\bibfnamefont {J.~P.}\ \bibnamefont
			{Ramos-Andrade}}, \bibinfo {author} {\bibfnamefont {O.}~\bibnamefont
			{{\ifmmode\acute{A}\else\'{A}\fi}valos Ovando}}, \bibinfo {author}
		{\bibfnamefont {P.~A.}\ \bibnamefont {Orellana}},\ and\ \bibinfo 
		{author}
		{\bibfnamefont {S.~E.}\ \bibnamefont {Ulloa}},\ }\bibfield  {title} 
		{\bibinfo
		{title} {{Thermoelectric transport through Majorana bound states and
				violation of Wiedemann-Franz law}},\ }\href
	{https://doi.org/10.1103/PhysRevB.94.155436} {\bibfield  {journal} {\bibinfo
			{journal} {Phys. Rev. B}\ }\textbf {\bibinfo {volume} {94}},\ 
			\bibinfo
		{pages} {155436} (\bibinfo {year} {2016})}\BibitemShut {NoStop}%
	\bibitem [{\citenamefont {Buccheri}\ \emph {et~al.}(2021)\citenamefont
		{Buccheri}, \citenamefont {Nava}, \citenamefont {Egger}, \citenamefont
		{Sodano},\ and\ \citenamefont {Giuliano}}]{Buccheri2021Aug}%
	\BibitemOpen
	\bibfield  {author} {\bibinfo {author} {\bibfnamefont {F.}~\bibnamefont
			{Buccheri}}, \bibinfo {author} {\bibfnamefont {A.}~\bibnamefont 
			{Nava}},
		\bibinfo {author} {\bibfnamefont {R.}~\bibnamefont {Egger}}, \bibinfo
		{author} {\bibfnamefont {P.}~\bibnamefont {Sodano}},\ and\ \bibinfo 
		{author}
		{\bibfnamefont {D.}~\bibnamefont {Giuliano}},\ }\bibfield  {title} 
		{\bibinfo
		{title} {{Violation of the Wiedemann-Franz law in the Topological Kondo
				model}},\ }\href {https://arxiv.org/abs/2108.04156v1} 
				{\bibfield  {journal}
		{\bibinfo  {journal} {arXiv}\ } (\bibinfo {year} {2021})},\ \Eprint
	{https://arxiv.org/abs/2108.04156} {2108.04156} \BibitemShut {NoStop}%
	\bibitem [{\citenamefont {Hou}\ \emph {et~al.}(2013)\citenamefont {Hou},
		\citenamefont {Shtengel},\ and\ \citenamefont {Refael}}]{Hou2013Aug}%
	\BibitemOpen
	\bibfield  {author} {\bibinfo {author} {\bibfnamefont {C.-Y.}\ \bibnamefont
			{Hou}}, \bibinfo {author} {\bibfnamefont {K.}~\bibnamefont 
			{Shtengel}},\ and\
		\bibinfo {author} {\bibfnamefont {G.}~\bibnamefont {Refael}},\ 
		}\bibfield
	{title} {\bibinfo {title} {{Thermopower and Mott formula for a Majorana edge
				state}},\ }\href {https://doi.org/10.1103/PhysRevB.88.075304} 
				{\bibfield
		{journal} {\bibinfo  {journal} {Phys. Rev. B}\ }\textbf {\bibinfo 
		{volume}
			{88}},\ \bibinfo {pages} {075304} (\bibinfo {year} 
			{2013})}\BibitemShut
	{NoStop}%
	\bibitem [{\citenamefont {Leijnse}(2014)}]{Leijnse2014Jan}%
	\BibitemOpen
	\bibfield  {author} {\bibinfo {author} {\bibfnamefont {M.}~\bibnamefont
			{Leijnse}},\ }\bibfield  {title} {\bibinfo {title} {{Thermoelectric
				signatures of a Majorana bound state coupled to a quantum 
				dot}},\ }\href
	{https://doi.org/10.1088/1367-2630/16/1/015029} {\bibfield  {journal}
		{\bibinfo  {journal} {New J. Phys.}\ }\textbf {\bibinfo {volume} {16}},\
		\bibinfo {pages} {015029} (\bibinfo {year} {2014})}\BibitemShut 
		{NoStop}%
	\bibitem [{\citenamefont {Valentini}\ \emph {et~al.}(2015)\citenamefont
		{Valentini}, \citenamefont {Fazio}, \citenamefont {Giovannetti},\ and\
		\citenamefont {Taddei}}]{Valentini2015Jan}%
	\BibitemOpen
	\bibfield  {author} {\bibinfo {author} {\bibfnamefont {S.}~\bibnamefont
			{Valentini}}, \bibinfo {author} {\bibfnamefont {R.}~\bibnamefont 
			{Fazio}},
		\bibinfo {author} {\bibfnamefont {V.}~\bibnamefont {Giovannetti}},\ and\
		\bibinfo {author} {\bibfnamefont {F.}~\bibnamefont {Taddei}},\ 
		}\bibfield
	{title} {\bibinfo {title} {{Thermopower of three-terminal topological
				superconducting systems}},\ }\href
	{https://doi.org/10.1103/PhysRevB.91.045430} {\bibfield  {journal} {\bibinfo
			{journal} {Phys. Rev. B}\ }\textbf {\bibinfo {volume} {91}},\ 
			\bibinfo
		{pages} {045430} (\bibinfo {year} {2015})}\BibitemShut {NoStop}%
	\bibitem [{\citenamefont {Smirnov}(2018)}]{Smirnov2018Apr}%
	\BibitemOpen
	\bibfield  {author} {\bibinfo {author} {\bibfnamefont {S.}~\bibnamefont
			{Smirnov}},\ }\bibfield  {title} {\bibinfo {title} {{Universal 
			Majorana
				thermoelectric noise}},\ }\href 
				{https://doi.org/10.1103/PhysRevB.97.165434}
	{\bibfield  {journal} {\bibinfo  {journal} {Phys. Rev. B}\ }\textbf 
	{\bibinfo
			{volume} {97}},\ \bibinfo {pages} {165434} (\bibinfo {year}
		{2018})}\BibitemShut {NoStop}%
	\bibitem [{\citenamefont {Wang}\ \emph {et~al.}(2019)\citenamefont {Wang},
		\citenamefont {Zhang}, \citenamefont {Han}, \citenamefont {Yi},\ and\
		\citenamefont {Gong}}]{Wang2019May}%
	\BibitemOpen
	\bibfield  {author} {\bibinfo {author} {\bibfnamefont {X.-Q.}\ \bibnamefont
			{Wang}}, \bibinfo {author} {\bibfnamefont {S.-F.}\ \bibnamefont 
			{Zhang}},
		\bibinfo {author} {\bibfnamefont {Y.}~\bibnamefont {Han}}, \bibinfo 
		{author}
		{\bibfnamefont {G.-Y.}\ \bibnamefont {Yi}},\ and\ \bibinfo {author}
		{\bibfnamefont {W.-J.}\ \bibnamefont {Gong}},\ }\bibfield  {title} 
		{\bibinfo
		{title} {{Efficient enhancement of the thermoelectric effect due to the
				Majorana zero modes coupled to one quantum-dot system}},\ }\href
	{https://doi.org/10.1103/PhysRevB.99.195424} {\bibfield  {journal} {\bibinfo
			{journal} {Phys. Rev. B}\ }\textbf {\bibinfo {volume} {99}},\ 
			\bibinfo
		{pages} {195424} (\bibinfo {year} {2019})}\BibitemShut {NoStop}%
	\bibitem [{\citenamefont {Smirnov}(2020)}]{Smirnov2020Mar}%
	\BibitemOpen
	\bibfield  {author} {\bibinfo {author} {\bibfnamefont {S.}~\bibnamefont
			{Smirnov}},\ }\bibfield  {title} {\bibinfo {title} {{Dual Majorana
				universality in thermally induced nonequilibrium}},\ }\href
	{https://doi.org/10.1103/PhysRevB.101.125417} {\bibfield  {journal} 
	{\bibinfo
			{journal} {Phys. Rev. B}\ }\textbf {\bibinfo {volume} {101}},\ 
			\bibinfo
		{pages} {125417} (\bibinfo {year} {2020})}\BibitemShut {NoStop}%
	\bibitem [{\citenamefont {W{\ifmmode\acute{o}\else\'{o}\fi}jcik}\ and\
		\citenamefont {Weymann}(2016)}]{Wojcik2016Feb}%
	\BibitemOpen
	\bibfield  {author} {\bibinfo {author} {\bibfnamefont {K.~P.}\ \bibnamefont
			{W{\ifmmode\acute{o}\else\'{o}\fi}jcik}}\ and\ \bibinfo {author}
		{\bibfnamefont {I.}~\bibnamefont {Weymann}},\ }\bibfield  {title} 
		{\bibinfo
		{title} {{Thermopower of strongly correlated T-shaped double quantum 
		dots}},\
	}\href {https://doi.org/10.1103/PhysRevB.93.085428} {\bibfield  {journal}
		{\bibinfo  {journal} {Phys. Rev. B}\ }\textbf {\bibinfo {volume} {93}},\
		\bibinfo {pages} {085428} (\bibinfo {year} {2016})}\BibitemShut 
		{NoStop}%
	\bibitem [{\citenamefont {Sherman}\ \emph {et~al.}(2017)\citenamefont
		{Sherman}, \citenamefont {Yodh}, \citenamefont {Albrecht}, \citenamefont
		{Nyg{\aa}rd}, \citenamefont {Krogstrup},\ and\ \citenamefont
		{Marcus}}]{Sherman2017Mar}%
	\BibitemOpen
	\bibfield  {author} {\bibinfo {author} {\bibfnamefont {D.}~\bibnamefont
			{Sherman}}, \bibinfo {author} {\bibfnamefont {J.~S.}\ \bibnamefont 
			{Yodh}},
		\bibinfo {author} {\bibfnamefont {S.~M.}\ \bibnamefont {Albrecht}}, 
		\bibinfo
		{author} {\bibfnamefont {J.}~\bibnamefont {Nyg{\aa}rd}}, \bibinfo 
		{author}
		{\bibfnamefont {P.}~\bibnamefont {Krogstrup}},\ and\ \bibinfo {author}
		{\bibfnamefont {C.~M.}\ \bibnamefont {Marcus}},\ }\bibfield  {title}
	{\bibinfo {title} {{Normal, superconducting and topological regimes of 
	hybrid
				double quantum dots}},\ }\href 
				{https://doi.org/10.1038/nnano.2016.227}
	{\bibfield  {journal} {\bibinfo  {journal} {Nat. Nanotechnol.}\ }\textbf
		{\bibinfo {volume} {12}},\ \bibinfo {pages} {212} (\bibinfo {year}
		{2017})}\BibitemShut {NoStop}%
	\bibitem [{\citenamefont {V{\ifmmode\ddot{a}\else\"{a}\fi}yrynen}\ \emph
		{et~al.}(2020)\citenamefont {V{\ifmmode\ddot{a}\else\"{a}\fi}yrynen},
		\citenamefont {Pikulin},\ and\ \citenamefont 
		{Lutchyn}}]{Vayrynen2020Oct}%
	\BibitemOpen
	\bibfield  {author} {\bibinfo {author} {\bibfnamefont {J.~I.}\ \bibnamefont
			{V{\ifmmode\ddot{a}\else\"{a}\fi}yrynen}}, \bibinfo {author} 
			{\bibfnamefont
			{D.~I.}\ \bibnamefont {Pikulin}},\ and\ \bibinfo {author} 
			{\bibfnamefont
			{R.~M.}\ \bibnamefont {Lutchyn}},\ }\bibfield  {title} {\bibinfo 
			{title}
		{{Majorana signatures in charge transport through a topological
				superconducting double-island system}},\ }\href
	{https://arxiv.org/abs/2010.05963v1} {\bibfield  {journal} {\bibinfo
			{journal} {arXiv}\ } (\bibinfo {year} {2020})},\ \Eprint
	{https://arxiv.org/abs/2010.05963} {2010.05963} \BibitemShut {NoStop}%
	\bibitem [{\citenamefont {Carrad}\ \emph {et~al.}(2020)\citenamefont 
	{Carrad},
		\citenamefont {Bjergfelt}, \citenamefont {Kanne}, \citenamefont 
		{Aagesen},
		\citenamefont {Krizek}, \citenamefont {Fiordaliso}, \citenamefont 
		{Johnson},
		\citenamefont {Nyg{\aa}rd},\ and\ \citenamefont 
		{Jespersen}}]{Carrad2020Jun}%
	\BibitemOpen
	\bibfield  {author} {\bibinfo {author} {\bibfnamefont {D.~J.}\ \bibnamefont
			{Carrad}}, \bibinfo {author} {\bibfnamefont {M.}~\bibnamefont 
			{Bjergfelt}},
		\bibinfo {author} {\bibfnamefont {T.}~\bibnamefont {Kanne}}, \bibinfo
		{author} {\bibfnamefont {M.}~\bibnamefont {Aagesen}}, \bibinfo {author}
		{\bibfnamefont {F.}~\bibnamefont {Krizek}}, \bibinfo {author} 
		{\bibfnamefont
			{E.~M.}\ \bibnamefont {Fiordaliso}}, \bibinfo {author} 
			{\bibfnamefont
			{E.}~\bibnamefont {Johnson}}, \bibinfo {author} {\bibfnamefont
			{J.}~\bibnamefont {Nyg{\aa}rd}},\ and\ \bibinfo {author} 
			{\bibfnamefont
			{T.~S.}\ \bibnamefont {Jespersen}},\ }\bibfield  {title} {\bibinfo 
			{title}
		{{Shadow Epitaxy for In Situ Growth of Generic 
		Semiconductor/Superconductor
				Hybrids}},\ }\href {https://doi.org/10.1002/adma.201908411} 
				{\bibfield
		{journal} {\bibinfo  {journal} {Adv. Mater.}\ }\textbf {\bibinfo 
		{volume}
			{32}},\ \bibinfo {pages} {1908411} (\bibinfo {year} 
			{2020})}\BibitemShut
	{NoStop}%
	\bibitem [{\citenamefont {Heedt}\ \emph {et~al.}(2020)\citenamefont {Heedt},
		\citenamefont {Quintero-P{\ifmmode\acute{e}\else\'{e}\fi}rez}, 
		\citenamefont
		{Borsoi}, \citenamefont {Fursina}, \citenamefont {van Loo}, 
		\citenamefont
		{Mazur}, \citenamefont {Nowak}, \citenamefont {Ammerlaan}, \citenamefont
		{Li}, \citenamefont {Korneychuk}, \citenamefont {Shen}, \citenamefont 
		{van~de
			Poll}, \citenamefont {Badawy}, \citenamefont {Gazibegovic}, 
			\citenamefont
		{van Hoogdalem}, \citenamefont {Bakkers},\ and\ \citenamefont
		{Kouwenhoven}}]{Heedt2020Jul}%
	\BibitemOpen
	\bibfield  {author} {\bibinfo {author} {\bibfnamefont {S.}~\bibnamefont
			{Heedt}}, \bibinfo {author} {\bibfnamefont {M.}~\bibnamefont
			{Quintero-P{\ifmmode\acute{e}\else\'{e}\fi}rez}}, \bibinfo {author}
		{\bibfnamefont {F.}~\bibnamefont {Borsoi}}, \bibinfo {author} 
		{\bibfnamefont
			{A.}~\bibnamefont {Fursina}}, \bibinfo {author} {\bibfnamefont
			{N.}~\bibnamefont {van Loo}}, \bibinfo {author} {\bibfnamefont 
			{G.~P.}\
			\bibnamefont {Mazur}}, \bibinfo {author} {\bibfnamefont {M.~P.}\ 
			\bibnamefont
			{Nowak}}, \bibinfo {author} {\bibfnamefont {M.}~\bibnamefont 
			{Ammerlaan}},
		\bibinfo {author} {\bibfnamefont {K.}~\bibnamefont {Li}}, \bibinfo 
		{author}
		{\bibfnamefont {S.}~\bibnamefont {Korneychuk}}, \bibinfo {author}
		{\bibfnamefont {J.}~\bibnamefont {Shen}}, \bibinfo {author} 
		{\bibfnamefont
			{M.~A.~Y.}\ \bibnamefont {van~de Poll}}, \bibinfo {author} 
			{\bibfnamefont
			{G.}~\bibnamefont {Badawy}}, \bibinfo {author} {\bibfnamefont
			{S.}~\bibnamefont {Gazibegovic}}, \bibinfo {author} {\bibfnamefont
			{K.}~\bibnamefont {van Hoogdalem}}, \bibinfo {author} 
			{\bibfnamefont {E.~P.
				A.~M.}\ \bibnamefont {Bakkers}},\ and\ \bibinfo {author} 
				{\bibfnamefont
			{L.~P.}\ \bibnamefont {Kouwenhoven}},\ }\bibfield  {title} 
			{\bibinfo {title}
		{{Shadow-wall lithography of ballistic superconductor-semiconductor 
		quantum
				devices}},\ }\href {https://arxiv.org/abs/2007.14383v1} 
				{\bibfield  {journal}
		{\bibinfo  {journal} {arXiv}\ } (\bibinfo {year} {2020})},\ \Eprint
	{https://arxiv.org/abs/2007.14383} {2007.14383} \BibitemShut {NoStop}%
	\bibitem [{\citenamefont {Kanne}\ \emph {et~al.}(2020)\citenamefont {Kanne},
		\citenamefont {Marnauza}, \citenamefont {Olsteins}, \citenamefont 
		{Carrad},
		\citenamefont {Sestoft}, \citenamefont {de~Bruijckere}, \citenamefont 
		{Zeng},
		\citenamefont {Johnson}, \citenamefont {Olsson}, \citenamefont
		{Grove-Rasmussen},\ and\ \citenamefont {Nyg{\aa}rd}}]{Kanne2020Feb}%
	\BibitemOpen
	\bibfield  {author} {\bibinfo {author} {\bibfnamefont {T.}~\bibnamefont
			{Kanne}}, \bibinfo {author} {\bibfnamefont {M.}~\bibnamefont 
			{Marnauza}},
		\bibinfo {author} {\bibfnamefont {D.}~\bibnamefont {Olsteins}}, \bibinfo
		{author} {\bibfnamefont {D.~J.}\ \bibnamefont {Carrad}}, \bibinfo 
		{author}
		{\bibfnamefont {J.~E.}\ \bibnamefont {Sestoft}}, \bibinfo {author}
		{\bibfnamefont {J.}~\bibnamefont {de~Bruijckere}}, \bibinfo {author}
		{\bibfnamefont {L.}~\bibnamefont {Zeng}}, \bibinfo {author} 
		{\bibfnamefont
			{E.}~\bibnamefont {Johnson}}, \bibinfo {author} {\bibfnamefont
			{E.}~\bibnamefont {Olsson}}, \bibinfo {author} {\bibfnamefont
			{K.}~\bibnamefont {Grove-Rasmussen}},\ and\ \bibinfo {author} 
			{\bibfnamefont
			{J.}~\bibnamefont {Nyg{\aa}rd}},\ }\bibfield  {title} {\bibinfo 
			{title}
		{{Epitaxial Pb on InAs nanowires}},\ }\href
	{https://arxiv.org/abs/2002.11641v1} {\bibfield  {journal} {\bibinfo
			{journal} {arXiv}\ } (\bibinfo {year} {2020})},\ \Eprint
	{https://arxiv.org/abs/2002.11641} {2002.11641} \BibitemShut {NoStop}%
	\bibitem [{\citenamefont {M{\ifmmode\ddot{u}\else\"{u}\fi}nning}\ \emph
		{et~al.}(2021)\citenamefont {M{\ifmmode\ddot{u}\else\"{u}\fi}nning},
		\citenamefont {Breunig}, \citenamefont {Legg}, \citenamefont {Roitsch},
		\citenamefont {Fan}, \citenamefont
		{R{\ifmmode\ddot{o}\else\"{o}\fi}{\ss}ler}, \citenamefont {Rosch},\ and\
		\citenamefont {Ando}}]{Munning2021Feb}%
	\BibitemOpen
	\bibfield  {author} {\bibinfo {author} {\bibfnamefont {F.}~\bibnamefont
			{M{\ifmmode\ddot{u}\else\"{u}\fi}nning}}, \bibinfo {author} 
			{\bibfnamefont
			{O.}~\bibnamefont {Breunig}}, \bibinfo {author} {\bibfnamefont 
			{H.~F.}\
			\bibnamefont {Legg}}, \bibinfo {author} {\bibfnamefont 
			{S.}~\bibnamefont
			{Roitsch}}, \bibinfo {author} {\bibfnamefont {D.}~\bibnamefont 
			{Fan}},
		\bibinfo {author} {\bibfnamefont {M.}~\bibnamefont
			{R{\ifmmode\ddot{o}\else\"{o}\fi}{\ss}ler}}, \bibinfo {author} 
			{\bibfnamefont
			{A.}~\bibnamefont {Rosch}},\ and\ \bibinfo {author} {\bibfnamefont
			{Y.}~\bibnamefont {Ando}},\ }\bibfield  {title} {\bibinfo {title} 
			{{Quantum
				confinement of the Dirac surface states in topological-insulator
				nanowires}},\ }\href 
				{https://doi.org/10.1038/s41467-021-21230-3} {\bibfield
		{journal} {\bibinfo  {journal} {Nat. Commun.}\ }\textbf {\bibinfo 
		{volume}
			{12}},\ \bibinfo {pages} {1} (\bibinfo {year} {2021})}\BibitemShut 
			{NoStop}%
	\bibitem [{\citenamefont {Desjardins}\ \emph {et~al.}(2019)\citenamefont
		{Desjardins}, \citenamefont {Contamin}, \citenamefont {Delbecq},
		\citenamefont {Dartiailh}, \citenamefont {Bruhat}, \citenamefont 
		{Cubaynes},
		\citenamefont {Viennot}, \citenamefont {Mallet}, \citenamefont {Rohart},
		\citenamefont {Thiaville}, \citenamefont {Cottet},\ and\ \citenamefont
		{Kontos}}]{Desjardins2019Oct}%
	\BibitemOpen
	\bibfield  {author} {\bibinfo {author} {\bibfnamefont {M.~M.}\ \bibnamefont
			{Desjardins}}, \bibinfo {author} {\bibfnamefont {L.~C.}\ 
			\bibnamefont
			{Contamin}}, \bibinfo {author} {\bibfnamefont {M.~R.}\ \bibnamefont
			{Delbecq}}, \bibinfo {author} {\bibfnamefont {M.~C.}\ \bibnamefont
			{Dartiailh}}, \bibinfo {author} {\bibfnamefont {L.~E.}\ \bibnamefont
			{Bruhat}}, \bibinfo {author} {\bibfnamefont {T.}~\bibnamefont 
			{Cubaynes}},
		\bibinfo {author} {\bibfnamefont {J.~J.}\ \bibnamefont {Viennot}}, 
		\bibinfo
		{author} {\bibfnamefont {F.}~\bibnamefont {Mallet}}, \bibinfo {author}
		{\bibfnamefont {S.}~\bibnamefont {Rohart}}, \bibinfo {author} 
		{\bibfnamefont
			{A.}~\bibnamefont {Thiaville}}, \bibinfo {author} {\bibfnamefont
			{A.}~\bibnamefont {Cottet}},\ and\ \bibinfo {author} {\bibfnamefont
			{T.}~\bibnamefont {Kontos}},\ }\bibfield  {title} {\bibinfo {title}
		{{Synthetic spin{\textendash}orbit interaction for Majorana devices}},\
	}\href {https://doi.org/10.1038/s41563-019-0457-6} {\bibfield  {journal}
		{\bibinfo  {journal} {Nat. Mater.}\ }\textbf {\bibinfo {volume} {18}},\
		\bibinfo {pages} {1060} (\bibinfo {year} {2019})}\BibitemShut {NoStop}%
	\bibitem [{\citenamefont {Delfanazari}\ \emph {et~al.}(2020)\citenamefont
		{Delfanazari}, \citenamefont {Serra}, \citenamefont {Ma}, \citenamefont
		{Puddy}, \citenamefont {Yi}, \citenamefont {Cao}, \citenamefont {Gul},
		\citenamefont {Farrer}, \citenamefont {Ritchie}, \citenamefont {Joyce},
		\citenamefont {Kelly},\ and\ \citenamefont 
		{Smith}}]{Delfanazari2020Jul}%
	\BibitemOpen
	\bibfield  {author} {\bibinfo {author} {\bibfnamefont {K.}~\bibnamefont
			{Delfanazari}}, \bibinfo {author} {\bibfnamefont {L.}~\bibnamefont 
			{Serra}},
		\bibinfo {author} {\bibfnamefont {P.}~\bibnamefont {Ma}}, \bibinfo 
		{author}
		{\bibfnamefont {R.~K.}\ \bibnamefont {Puddy}}, \bibinfo {author}
		{\bibfnamefont {T.}~\bibnamefont {Yi}}, \bibinfo {author} {\bibfnamefont
			{M.}~\bibnamefont {Cao}}, \bibinfo {author} {\bibfnamefont 
			{Y.}~\bibnamefont
			{Gul}}, \bibinfo {author} {\bibfnamefont {I.}~\bibnamefont 
			{Farrer}},
		\bibinfo {author} {\bibfnamefont {D.~A.}\ \bibnamefont {Ritchie}}, 
		\bibinfo
		{author} {\bibfnamefont {H.~J.}\ \bibnamefont {Joyce}}, \bibinfo 
		{author}
		{\bibfnamefont {M.~J.}\ \bibnamefont {Kelly}},\ and\ \bibinfo {author}
		{\bibfnamefont {C.~G.}\ \bibnamefont {Smith}},\ }\bibfield  {title} 
		{\bibinfo
		{title} {{Experimental evidence for topological phases in the
				magnetoconductance of 2DEG-based hybrid junctions}},\ }\href
	{https://arxiv.org/abs/2007.02057v2} {\bibfield  {journal} {\bibinfo
			{journal} {arXiv}\ } (\bibinfo {year} {2020})},\ \Eprint
	{https://arxiv.org/abs/2007.02057} {2007.02057} \BibitemShut {NoStop}%
	\bibitem [{\citenamefont {Liu}\ and\ \citenamefont
		{Baranger}(2011)}]{Liu2011Nov}%
	\BibitemOpen
	\bibfield  {author} {\bibinfo {author} {\bibfnamefont {D.~E.}\ \bibnamefont
			{Liu}}\ and\ \bibinfo {author} {\bibfnamefont {H.~U.}\ \bibnamefont
			{Baranger}},\ }\bibfield  {title} {\bibinfo {title} {{Detecting a
				Majorana-fermion zero mode using a quantum dot}},\ }\href
	{https://doi.org/10.1103/PhysRevB.84.201308} {\bibfield  {journal} {\bibinfo
			{journal} {Phys. Rev. B}\ }\textbf {\bibinfo {volume} {84}},\ 
			\bibinfo
		{pages} {201308(R)} (\bibinfo {year} {2011})}\BibitemShut {NoStop}%
	\bibitem [{\citenamefont {Albrecht}\ \emph {et~al.}(2016)\citenamefont
		{Albrecht}, \citenamefont {Higginbotham}, \citenamefont {Madsen},
		\citenamefont {Kuemmeth}, \citenamefont {Jespersen}, \citenamefont
		{Nyg{\aa}rd}, \citenamefont {Krogstrup},\ and\ \citenamefont
		{Marcus}}]{Albrecht2016Mar}%
	\BibitemOpen
	\bibfield  {author} {\bibinfo {author} {\bibfnamefont {S.~M.}\ \bibnamefont
			{Albrecht}}, \bibinfo {author} {\bibfnamefont {A.~P.}\ \bibnamefont
			{Higginbotham}}, \bibinfo {author} {\bibfnamefont {M.}~\bibnamefont
			{Madsen}}, \bibinfo {author} {\bibfnamefont {F.}~\bibnamefont 
			{Kuemmeth}},
		\bibinfo {author} {\bibfnamefont {T.~S.}\ \bibnamefont {Jespersen}}, 
		\bibinfo
		{author} {\bibfnamefont {J.}~\bibnamefont {Nyg{\aa}rd}}, \bibinfo 
		{author}
		{\bibfnamefont {P.}~\bibnamefont {Krogstrup}},\ and\ \bibinfo {author}
		{\bibfnamefont {C.~M.}\ \bibnamefont {Marcus}},\ }\bibfield  {title}
	{\bibinfo {title} {{Exponential protection of zero modes in Majorana
				islands}},\ }\href {https://doi.org/10.1038/nature17162} 
				{\bibfield
		{journal} {\bibinfo  {journal} {Nature}\ }\textbf {\bibinfo {volume} 
		{531}},\
		\bibinfo {pages} {206} (\bibinfo {year} {2016})}\BibitemShut {NoStop}%
	\bibitem [{\citenamefont {Benenti}\ \emph {et~al.}(2017)\citenamefont
		{Benenti}, \citenamefont {Casati}, \citenamefont {Saito},\ and\ 
		\citenamefont
		{Whitney}}]{Benenti2017Jun}%
	\BibitemOpen
	\bibfield  {author} {\bibinfo {author} {\bibfnamefont {G.}~\bibnamefont
			{Benenti}}, \bibinfo {author} {\bibfnamefont {G.}~\bibnamefont 
			{Casati}},
		\bibinfo {author} {\bibfnamefont {K.}~\bibnamefont {Saito}},\ and\ 
		\bibinfo
		{author} {\bibfnamefont {R.~S.}\ \bibnamefont {Whitney}},\ }\bibfield
	{title} {\bibinfo {title} {{Fundamental aspects of steady-state conversion 
	of
				heat to work at the nanoscale}},\ }\href
	{https://doi.org/10.1016/j.physrep.2017.05.008} {\bibfield  {journal}
		{\bibinfo  {journal} {Phys. Rep.}\ }\textbf {\bibinfo {volume} {694}},\
		\bibinfo {pages} {1} (\bibinfo {year} {2017})}\BibitemShut {NoStop}%
	\bibitem [{\citenamefont {Barnard}(1972)}]{barnard1972thermoelectricity}%
	\BibitemOpen
	\bibfield  {author} {\bibinfo {author} {\bibfnamefont {R.}~\bibnamefont
			{Barnard}},\ }\href {https://books.google.pl/books?id=CayRQAAACAAJ} 
			{\emph
		{\bibinfo {title} {Thermoelectricity in Metals and Alloys}}}\ (\bibinfo
	{publisher} {Taylor \& Francis},\ \bibinfo {year} {1972})\BibitemShut
	{NoStop}%
	\bibitem [{\citenamefont {Dias~da Silva}\ \emph {et~al.}(2013)\citenamefont
		{Dias~da Silva}, \citenamefont {Vernek}, \citenamefont {Ingersent},
		\citenamefont {Sandler},\ and\ \citenamefont 
		{Ulloa}}]{DiasdaSilva2013May}%
	\BibitemOpen
	\bibfield  {author} {\bibinfo {author} {\bibfnamefont {L.~G. G.~V.}\
			\bibnamefont {Dias~da Silva}}, \bibinfo {author} {\bibfnamefont
			{E.}~\bibnamefont {Vernek}}, \bibinfo {author} {\bibfnamefont
			{K.}~\bibnamefont {Ingersent}}, \bibinfo {author} {\bibfnamefont
			{N.}~\bibnamefont {Sandler}},\ and\ \bibinfo {author} 
			{\bibfnamefont {S.~E.}\
			\bibnamefont {Ulloa}},\ }\bibfield  {title} {\bibinfo {title}
		{{Spin-polarized conductance in double quantum dots: Interplay of Kondo,
				Zeeman, and interference effects}},\ }\href
	{https://doi.org/10.1103/PhysRevB.87.205313} {\bibfield  {journal} {\bibinfo
			{journal} {Phys. Rev. B}\ }\textbf {\bibinfo {volume} {87}},\ 
			\bibinfo
		{pages} {205313} (\bibinfo {year} {2013})}\BibitemShut {NoStop}%
	\bibitem [{\citenamefont {W{\ifmmode\acute{o}\else\'{o}\fi}jcik}\ and\
		\citenamefont {Weymann}(2014)}]{Wojcik2014Sep}%
	\BibitemOpen
	\bibfield  {author} {\bibinfo {author} {\bibfnamefont {K.~P.}\ \bibnamefont
			{W{\ifmmode\acute{o}\else\'{o}\fi}jcik}}\ and\ \bibinfo {author}
		{\bibfnamefont {I.}~\bibnamefont {Weymann}},\ }\bibfield  {title} 
		{\bibinfo
		{title} {{Perfect spin polarization in T-shaped double quantum dots due 
		to
				the spin-dependent Fano effect}},\ }\href
	{https://doi.org/10.1103/PhysRevB.90.115308} {\bibfield  {journal} {\bibinfo
			{journal} {Phys. Rev. B}\ }\textbf {\bibinfo {volume} {90}},\ 
			\bibinfo
		{pages} {115308} (\bibinfo {year} {2014})}\BibitemShut {NoStop}%
	\bibitem [{\citenamefont {Wilson}(1975)}]{Wilson1975Oct}%
	\BibitemOpen
	\bibfield  {author} {\bibinfo {author} {\bibfnamefont {K.~G.}\ \bibnamefont
			{Wilson}},\ }\bibfield  {title} {\bibinfo {title} {{The 
			renormalization
				group: Critical phenomena and the Kondo problem}},\ }\href
	{https://doi.org/10.1103/RevModPhys.47.773} {\bibfield  {journal} {\bibinfo
			{journal} {Rev. Mod. Phys.}\ }\textbf {\bibinfo {volume} {47}},\ 
			\bibinfo
		{pages} {773} (\bibinfo {year} {1975})}\BibitemShut {NoStop}%
	\bibitem [{\citenamefont {Bulla}\ \emph {et~al.}(2008)\citenamefont {Bulla},
		\citenamefont {Costi},\ and\ \citenamefont {Pruschke}}]{Bulla2008Apr}%
	\BibitemOpen
	\bibfield  {author} {\bibinfo {author} {\bibfnamefont {R.}~\bibnamefont
			{Bulla}}, \bibinfo {author} {\bibfnamefont {T.~A.}\ \bibnamefont 
			{Costi}},\
		and\ \bibinfo {author} {\bibfnamefont {T.}~\bibnamefont {Pruschke}},\
	}\bibfield  {title} {\bibinfo {title} {{Numerical renormalization group
				method for quantum impurity systems}},\ }\href
	{https://doi.org/10.1103/RevModPhys.80.395} {\bibfield  {journal} {\bibinfo
			{journal} {Rev. Mod. Phys.}\ }\textbf {\bibinfo {volume} {80}},\ 
			\bibinfo
		{pages} {395} (\bibinfo {year} {2008})}\BibitemShut {NoStop}%
	\bibitem [{\citenamefont {Legeza}\ \emph {et~al.}(2008)\citenamefont 
	{Legeza},
		\citenamefont {Moca}, \citenamefont {T\'{o}th}, \citenamefont 
		{Weymann},\
		and\ \citenamefont {Zar\'{a}nd}}]{NRG_code}%
	\BibitemOpen
	\bibfield  {author} {\bibinfo {author} {\bibfnamefont {O.}~\bibnamefont
			{Legeza}}, \bibinfo {author} {\bibfnamefont {C.~P.}\ \bibnamefont 
			{Moca}},
		\bibinfo {author} {\bibfnamefont {A.~I.}\ \bibnamefont {T\'{o}th}}, 
		\bibinfo
		{author} {\bibfnamefont {I.}~\bibnamefont {Weymann}},\ and\ \bibinfo 
		{author}
		{\bibfnamefont {G.}~\bibnamefont {Zar\'{a}nd}},\ }\href
	{http://arxiv.org/abs/0809.3143} {\bibinfo {title} {{Manual for the Flexible
				DM-NRG code}}},\ \bibinfo {howpublished} {arXiv:0809.3143v1} 
				(\bibinfo {year}
	{2008}),\ \bibinfo {note} {(the open access Flexible DM-NRG Budapest code is
		available at
		\href{http://www.phy.bme.hu/\~dmnrg/}{http:/\!/www.phy.bme.hu/\textasciitilde{}dmnrg/}}\BibitemShut
	{NoStop}%
	\bibitem [{\citenamefont {Anders}\ and\ \citenamefont
		{Schiller}(2005)}]{Anders2005}%
	\BibitemOpen
	\bibfield  {author} {\bibinfo {author} {\bibfnamefont {F.~B.}\ \bibnamefont
			{Anders}}\ and\ \bibinfo {author} {\bibfnamefont {A.}~\bibnamefont
			{Schiller}},\ }\bibfield  {title} {\bibinfo {title} {Real-time 
			dynamics in
			quantum-impurity systems: A time-dependent numerical 
			renormalization-group
			approach},\ }\href {https://doi.org/10.1103/PhysRevLett.95.196801} 
			{\bibfield
		{journal} {\bibinfo  {journal} {Phys. Rev. Lett.}\ }\textbf {\bibinfo
			{volume} {95}},\ \bibinfo {pages} {196801} (\bibinfo {year}
		{2005})}\BibitemShut {NoStop}%
	\bibitem [{\citenamefont {Mahan}\ and\ \citenamefont 
	{Sofo}(1996)}]{MahanSofo}%
	\BibitemOpen
	\bibfield  {author} {\bibinfo {author} {\bibfnamefont {G.~D.}\ \bibnamefont
			{Mahan}}\ and\ \bibinfo {author} {\bibfnamefont {J.~O.}\ 
			\bibnamefont
			{Sofo}},\ }\bibfield  {title} {\bibinfo {title} {{The best 
			thermoelectric}},\
	}\href {https://doi.org/10.1073/pnas.93.15.7436} {\bibfield  {journal}
		{\bibinfo  {journal} {Proc. Natl. Acad. Sci. U.S.A.}\ }\textbf {\bibinfo
			{volume} {93}},\ \bibinfo {pages} {7436} (\bibinfo {year}
		{1996})}\BibitemShut {NoStop}%
	\bibitem [{\citenamefont {Haldane}(1978)}]{Haldane_Phys.Rev.Lett.40/1978}%
	\BibitemOpen
	\bibfield  {author} {\bibinfo {author} {\bibfnamefont {F.}~\bibnamefont
			{Haldane}},\ }\bibfield  {title} {\bibinfo {title} {{Scaling theory 
			of the
				asymmetric Anderson model}},\ }\href@noop {} {\bibfield  
				{journal} {\bibinfo
			{journal} {Phys. Rev. Lett.}\ }\textbf {\bibinfo {volume} {40}},\ 
			\bibinfo
		{pages} {416} (\bibinfo {year} {1978})}\BibitemShut {NoStop}%
	\bibitem [{\citenamefont {Franz}\ and\ \citenamefont
		{Wiedemann}(1853)}]{Franz1853Jan}%
	\BibitemOpen
	\bibfield  {author} {\bibinfo {author} {\bibfnamefont {R.}~\bibnamefont
			{Franz}}\ and\ \bibinfo {author} {\bibfnamefont {G.}~\bibnamefont
			{Wiedemann}},\ }\bibfield  {title} {\bibinfo {title} {{Ueber die
				W{\ifmmode\ddot{a}\else\"{a}\fi}rme-Leitungsf{\ifmmode\ddot{a}\else\"{a}\fi}higkeit
				der Metalle}},\ }\href 
				{https://doi.org/10.1002/andp.18531650802} {\bibfield
		{journal} {\bibinfo  {journal} {Ann. Phys.}\ }\textbf {\bibinfo {volume}
			{165}},\ \bibinfo {pages} {497} (\bibinfo {year} 
			{1853})}\BibitemShut
	{NoStop}%
	\bibitem [{\citenamefont {Giuliano}\ \emph {et~al.}(2021)\citenamefont
		{Giuliano}, \citenamefont {Nava}, \citenamefont {Egger}, \citenamefont
		{Sodano},\ and\ \citenamefont {Buccheri}}]{Giuliano2021Aug}%
	\BibitemOpen
	\bibfield  {author} {\bibinfo {author} {\bibfnamefont {D.}~\bibnamefont
			{Giuliano}}, \bibinfo {author} {\bibfnamefont {A.}~\bibnamefont 
			{Nava}},
		\bibinfo {author} {\bibfnamefont {R.}~\bibnamefont {Egger}}, \bibinfo
		{author} {\bibfnamefont {P.}~\bibnamefont {Sodano}},\ and\ \bibinfo 
		{author}
		{\bibfnamefont {F.}~\bibnamefont {Buccheri}},\ }\bibfield  {title} 
		{\bibinfo
		{title} {{Multi-particle scattering and breakdown of the 
		Wiedemann-Franz law
				at a junction of N interacting quantum wires}},\ }\href
	{https://arxiv.org/abs/2108.04149v2} {\bibfield  {journal} {\bibinfo
			{journal} {arXiv}\ } (\bibinfo {year} {2021})},\ \Eprint
	{https://arxiv.org/abs/2108.04149} {2108.04149} \BibitemShut {NoStop}%
	\bibitem [{\citenamefont {Weymann}(2011)}]{Weymann2011Mar}%
	\BibitemOpen
	\bibfield  {author} {\bibinfo {author} {\bibfnamefont {I.}~\bibnamefont
			{Weymann}},\ }\bibfield  {title} {\bibinfo {title} 
			{{Finite-temperature
				spintronic transport through Kondo quantum dots: Numerical 
				renormalization
				group study}},\ }\href 
				{https://doi.org/10.1103/PhysRevB.83.113306}
	{\bibfield  {journal} {\bibinfo  {journal} {Phys. Rev. B}\ }\textbf 
	{\bibinfo
			{volume} {83}},\ \bibinfo {pages} {113306} (\bibinfo {year}
		{2011})}\BibitemShut {NoStop}%
	\bibitem [{\citenamefont {Costi}(2000)}]{Costi2000Aug}%
	\BibitemOpen
	\bibfield  {author} {\bibinfo {author} {\bibfnamefont {T.~A.}\ \bibnamefont
			{Costi}},\ }\bibfield  {title} {\bibinfo {title} {{Kondo Effect in 
			a Magnetic
				Field and the Magnetoresistivity of Kondo Alloys}},\ }\href
	{https://doi.org/10.1103/PhysRevLett.85.1504} {\bibfield  {journal} 
	{\bibinfo
			{journal} {Phys. Rev. Lett.}\ }\textbf {\bibinfo {volume} {85}},\ 
			\bibinfo
		{pages} {1504} (\bibinfo {year} {2000})}\BibitemShut {NoStop}%
	\bibitem [{\citenamefont {Martinek}\ \emph {et~al.}(2005)\citenamefont
		{Martinek}, \citenamefont {Sindel}, \citenamefont {Borda}, \citenamefont
		{Barna{\ifmmode\acute{s}\else\'{s}\fi}}, \citenamefont {Bulla}, 
		\citenamefont
		{K{\ifmmode\ddot{o}\else\"{o}\fi}nig}, \citenamefont
		{Sch{\ifmmode\ddot{o}\else\"{o}\fi}n}, \citenamefont {Maekawa},\ and\
		\citenamefont {von Delft}}]{MartinekEx}%
	\BibitemOpen
	\bibfield  {author} {\bibinfo {author} {\bibfnamefont {J.}~\bibnamefont
			{Martinek}}, \bibinfo {author} {\bibfnamefont {M.}~\bibnamefont 
			{Sindel}},
		\bibinfo {author} {\bibfnamefont {L.}~\bibnamefont {Borda}}, \bibinfo
		{author} {\bibfnamefont {J.}~\bibnamefont
			{Barna{\ifmmode\acute{s}\else\'{s}\fi}}}, \bibinfo {author} 
			{\bibfnamefont
			{R.}~\bibnamefont {Bulla}}, \bibinfo {author} {\bibfnamefont
			{J.}~\bibnamefont {K{\ifmmode\ddot{o}\else\"{o}\fi}nig}}, \bibinfo 
			{author}
		{\bibfnamefont {G.}~\bibnamefont {Sch{\ifmmode\ddot{o}\else\"{o}\fi}n}},
		\bibinfo {author} {\bibfnamefont {S.}~\bibnamefont {Maekawa}},\ and\ 
		\bibinfo
		{author} {\bibfnamefont {J.}~\bibnamefont {von Delft}},\ }\bibfield  
		{title}
	{\bibinfo {title} {{Gate-controlled spin splitting in quantum dots with
				ferromagnetic leads in the Kondo regime}},\ }\href
	{https://doi.org/10.1103/PhysRevB.72.121302} {\bibfield  {journal} {\bibinfo
			{journal} {Phys. Rev. B}\ }\textbf {\bibinfo {volume} {72}},\ 
			\bibinfo
		{pages} {121302} (\bibinfo {year} {2005})}\BibitemShut {NoStop}%
	\bibitem [{\citenamefont {Tooski}\ \emph {et~al.}(2014)\citenamefont 
	{Tooski},
		\citenamefont {Ram{\ifmmode\check{s}\else\v{s}\fi}ak}, \citenamefont
		{Bu{\l}ka},\ and\ \citenamefont
		{{\ifmmode\check{Z}\else\v{Z}\fi}itko}}]{BulkaZitko}%
	\BibitemOpen
	\bibfield  {author} {\bibinfo {author} {\bibfnamefont {S.~B.}\ \bibnamefont
			{Tooski}}, \bibinfo {author} {\bibfnamefont {A.}~\bibnamefont
			{Ram{\ifmmode\check{s}\else\v{s}\fi}ak}}, \bibinfo {author} 
			{\bibfnamefont
			{B.~R.}\ \bibnamefont {Bu{\l}ka}},\ and\ \bibinfo {author} 
			{\bibfnamefont
			{R.}~\bibnamefont {{\ifmmode\check{Z}\else\v{Z}\fi}itko}},\ 
			}\bibfield
	{title} {\bibinfo {title} {{Effect of assisted hopping on thermopower in an
				interacting quantum dot}},\ }\href
	{https://doi.org/10.1088/1367-2630/16/5/055001} {\bibfield  {journal}
		{\bibinfo  {journal} {New J. Phys.}\ }\textbf {\bibinfo {volume} {16}},\
		\bibinfo {pages} {055001} (\bibinfo {year} {2014})}\BibitemShut 
		{NoStop}%
	\bibitem [{\citenamefont {Weymann}\ and\ \citenamefont
		{Barna{\ifmmode\acute{s}\else\'{s}\fi}}(2013)}]{Weymann2013Aug}%
	\BibitemOpen
	\bibfield  {author} {\bibinfo {author} {\bibfnamefont {I.}~\bibnamefont
			{Weymann}}\ and\ \bibinfo {author} {\bibfnamefont {J.}~\bibnamefont
			{Barna{\ifmmode\acute{s}\else\'{s}\fi}}},\ }\bibfield  {title} 
			{\bibinfo
		{title} {{Spin thermoelectric effects in Kondo quantum dots coupled to
				ferromagnetic leads}},\ }\href 
				{https://doi.org/10.1103/PhysRevB.88.085313}
	{\bibfield  {journal} {\bibinfo  {journal} {Phys. Rev. B}\ }\textbf 
	{\bibinfo
			{volume} {88}},\ \bibinfo {pages} {085313} (\bibinfo {year}
		{2013})}\BibitemShut {NoStop}%
	\bibitem [{\citenamefont {{\ifmmode\acute{S}\else\'{S}\fi}wirkowicz}\ \emph
		{et~al.}(2009)\citenamefont {{\ifmmode\acute{S}\else\'{S}\fi}wirkowicz},
		\citenamefont {Wierzbicki},\ and\ \citenamefont
		{Barna{\ifmmode\acute{s}\else\'{s}\fi}}}]{Swirkowicz2009Nov}%
	\BibitemOpen
	\bibfield  {author} {\bibinfo {author} {\bibfnamefont {R.}~\bibnamefont
			{{\ifmmode\acute{S}\else\'{S}\fi}wirkowicz}}, \bibinfo {author}
		{\bibfnamefont {M.}~\bibnamefont {Wierzbicki}},\ and\ \bibinfo {author}
		{\bibfnamefont {J.}~\bibnamefont 
		{Barna{\ifmmode\acute{s}\else\'{s}\fi}}},\
	}\bibfield  {title} {\bibinfo {title} {{Thermoelectric effects in transport
				through quantum dots attached to ferromagnetic leads with 
				noncollinear
				magnetic moments}},\ }\href 
				{https://doi.org/10.1103/PhysRevB.80.195409}
	{\bibfield  {journal} {\bibinfo  {journal} {Phys. Rev. B}\ }\textbf 
	{\bibinfo
			{volume} {80}},\ \bibinfo {pages} {195409} (\bibinfo {year}
		{2009})}\BibitemShut {NoStop}%
	\bibitem [{\citenamefont {Misiorny}\ and\ \citenamefont
		{Barna{\ifmmode\acute{s}\else\'{s}\fi}}(2015)}]{Misiorny2015Apr}%
	\BibitemOpen
	\bibfield  {author} {\bibinfo {author} {\bibfnamefont {M.}~\bibnamefont
			{Misiorny}}\ and\ \bibinfo {author} {\bibfnamefont {J.}~\bibnamefont
			{Barna{\ifmmode\acute{s}\else\'{s}\fi}}},\ }\bibfield  {title} 
			{\bibinfo
		{title} {{Effect of magnetic anisotropy on spin-dependent thermoelectric
				effects in nanoscopic systems}},\ }\href
	{https://doi.org/10.1103/PhysRevB.91.155426} {\bibfield  {journal} {\bibinfo
			{journal} {Phys. Rev. B}\ }\textbf {\bibinfo {volume} {91}},\ 
			\bibinfo
		{pages} {155426} (\bibinfo {year} {2015})}\BibitemShut {NoStop}%
	\bibitem [{\citenamefont {Wiesendanger}(2009)}]{Wiesendanger2009Nov}%
	\BibitemOpen
	\bibfield  {author} {\bibinfo {author} {\bibfnamefont {R.}~\bibnamefont
			{Wiesendanger}},\ }\bibfield  {title} {\bibinfo {title} {{Spin 
			mapping at the
				nanoscale and atomic scale}},\ }\href
	{https://doi.org/10.1103/RevModPhys.81.1495} {\bibfield  {journal} {\bibinfo
			{journal} {Rev. Mod. Phys.}\ }\textbf {\bibinfo {volume} {81}},\ 
			\bibinfo
		{pages} {1495} (\bibinfo {year} {2009})}\BibitemShut {NoStop}%
	\bibitem [{\citenamefont
		{{\ifmmode\check{Z}\else\v{Z}\fi}uti{\ifmmode\acute{c}\else\'{c}\fi}}\ 
		\emph
		{et~al.}(2004)\citenamefont
		{{\ifmmode\check{Z}\else\v{Z}\fi}uti{\ifmmode\acute{c}\else\'{c}\fi}},
		\citenamefont {Fabian},\ and\ \citenamefont {Das~Sarma}}]{Zutic2004Apr}%
	\BibitemOpen
	\bibfield  {author} {\bibinfo {author} {\bibfnamefont {I.}~\bibnamefont
			{{\ifmmode\check{Z}\else\v{Z}\fi}uti{\ifmmode\acute{c}\else\'{c}\fi}}},
		\bibinfo {author} {\bibfnamefont {J.}~\bibnamefont {Fabian}},\ and\ 
		\bibinfo
		{author} {\bibfnamefont {S.}~\bibnamefont {Das~Sarma}},\ }\bibfield  
		{title}
	{\bibinfo {title} {{Spintronics: Fundamentals and applications}},\ }\href
	{https://doi.org/10.1103/RevModPhys.76.323} {\bibfield  {journal} {\bibinfo
			{journal} {Rev. Mod. Phys.}\ }\textbf {\bibinfo {volume} {76}},\ 
			\bibinfo
		{pages} {323} (\bibinfo {year} {2004})}\BibitemShut {NoStop}%
\end{thebibliography}

%


\end{document}